\documentclass[10pt,reprint,prx,aps,amsmath,amssymb,floatfix,superscriptaddress,longbibliography]{revtex4-2}

\usepackage[utf8]{inputenc}
\usepackage{lipsum}

\usepackage{amsmath, amssymb,amsfonts,amsthm,mathtools,nicefrac}
\usepackage{xspace,graphicx,relsize,bm,xcolor}
\usepackage{soul} 

\usepackage{libertinus}  
\usepackage[zerostyle=a,scaled=.97]{newtxtt}  
\usepackage[T1]{fontenc}
\usepackage[libertine]{newtxmath}  
\usepackage{dsfont}  
\usepackage{anyfontsize}

\usepackage{url}
\usepackage[linktocpage=true]{hyperref}
\usepackage[singlespacing]{setspace}
\definecolor{linkcol}{rgb}{0.0,0.55,0.7}
\definecolor{citecol}{rgb}{0.0, 0.6, 0.45}
\definecolor{urlcol}{rgb}{0.7, 0.0, 0.55}
\hypersetup{
    colorlinks,
    linkcolor={linkcol},
    citecolor={citecol},
    urlcolor={urlcol},
    pdftitle = {Efficient distributed inner product estimation},
    pdfauthor = {
      Marcel Hinsche,
      Marios Ioannou,
      Jose Carrasco
    },
    pdfsubject = {Quantum computation, quantum information},
    pdfkeywords = {
        Quantum verification,
        Pauli sampling,
   }
}

\usepackage{tabularx}
\usepackage{mleftright}
\usepackage{tipa}
\usepackage{multirow}
\usepackage{physics}  

\usepackage{subfiles} 

\usepackage{algorithm}
\usepackage{algpseudocode}

\usepackage{multirow}

\usepackage{pifont}

\usepackage{float}

\usepackage{chngpage}

\usepackage[capitalize]{cleveref}  
\Crefname{equation}{Eq.}{Eqs.}
\crefformat{footnote}{#2\footnotemark[#1]#3}

\def\01{\{0,1\}}

\DeclareMathOperator*{\Ex}{\mathbf{E}}
\let\Pr\relax
\DeclareMathOperator*{\Pr}{\mathbf{Pr}}

\newcommand{\eps}{\epsilon}

\newcommand{\A}{\ensuremath{\mathcal{A}}}

\newcommand{\E}{\ensuremath{\mathcal{E}}}

\newcommand{\poly}{\mathrm{poly}}

\renewcommand{\norm}[1]{\lVert{#1}\rVert}

\DeclarePairedDelimiter\lrangle{\langle}{\rangle}
\makeatletter
\let\oldlrangle\lrangle
\def\lrangle{\@ifstar{\oldlrangle}{\oldlrangle*}}
\newcommand{\multiline}[1]{%
  \begin{tabularx}{\dimexpr\linewidth-\ALG@thistlm}[t]{@{}X@{}}
    #1
  \end{tabularx}
}
\makeatother

\def\pmix{p_{\mathrm{mix}}}

\newtheoremstyle{mydefinitionsty}
{10pt}
{10pt}
{}
{}
{}
{}
{.5em}
{\textbf{\thmname{#1}~\thmnumber{#2}:  }\thmnote{(#3)}}
\theoremstyle{mydefinitionsty}
\newtheorem{definition}{Definition}
\newtheorem{remark}{Remark}

\newtheoremstyle{mythmsty}
{10pt}
{10pt}
{\itshape}
{}
{}
{}
{.5em}
{\textbf{\thmname{#1}~\thmnumber{#2}:  }\thmnote{(#3)}}
\theoremstyle{mythmsty}

\newtheorem{theorem}{Theorem}
\newtheorem{lemma}{Lemma}

\newtheorem{corollary}{Corollary}

 \definecolor{marcelorange}{rgb}{0.8, 0.43, 0}

\allowdisplaybreaks

\begin{document}

\title{Efficient distributed inner product estimation via Pauli sampling}

\newcommand\CoAuthorMark{\footnotemark[\arabic{footnote}]}

\author{M. Hinsche\thanks{
Contributed equally. \\
Corresponding authors: \{\href{mailto:m.hinsche@fu-berlin.de}{m.hinsche}, \href{mailto:marios.ionannou@fu-berlin.de}{marios.ioannou}, \href{mailto:jose.carrasco@fu-berlin.de}{jose.carrasco}\}@fu-berlin.de}}
\affiliation{Dahlem Center for Complex Quantum Systems, Freie Universit\"at Berlin, Berlin, Germany}
\author{M. Ioannou\protect\CoAuthorMark}
\affiliation{Dahlem Center for Complex Quantum Systems, Freie Universit\"at Berlin, Berlin, Germany}
\author{S. Jerbi}
\affiliation{Dahlem Center for Complex Quantum Systems, Freie Universit\"at Berlin, Berlin, Germany}
\author{L. Leone}
\affiliation{Dahlem Center for Complex Quantum Systems, Freie Universit\"at Berlin, Berlin, Germany}
\author{J. Eisert}
\affiliation{Dahlem Center for Complex Quantum Systems, Freie Universit\"at Berlin, Berlin, Germany}
\affiliation{Helmholtz-Zentrum Berlin für Materialien und Energie, Berlin, Germany}
\author{J.~Carrasco\protect\CoAuthorMark}
\affiliation{Dahlem Center for Complex Quantum Systems, Freie Universit\"at Berlin, Berlin, Germany}

\begin{abstract}
Cross-platform verification is the task of comparing the output states produced by different physical platforms using solely local quantum operations and classical communication. While protocols have previously been suggested for this task, their exponential sample complexity renders them unpractical even for intermediate-scale
quantum
systems. In this work, we propose a novel protocol for this task based on \emph{Pauli sampling}, a subroutine which generates Paulis distributed according to their weight in the expansion of a quantum state in the Pauli basis. We show that our protocols for both Pauli sampling and cross-platform verification are efficient for pure states with low magic and entanglement (i.e., of the order $O(\log n)$). Conversely, we show super-polynomial lower bounds on the complexity of both tasks for states with $\omega(\log (n))$ magic and entanglement. Interestingly, when considering states with real amplitudes the requirements of our protocol for cross-platform verification can be significantly weakened.
\end{abstract}
\maketitle

\section{Introduction}
Numerous institutions are currently engaged in the development of small and intermediate-sized quantum computers, each utilizing diverse physical platforms. The benchmarking of these devices is imperative for technological progress \cite{eisertQuantumCertificationBenchmarking2020}. However, as we advance to ever larger-sized devices, their individual benchmarking based on certification methods, such as tomography \cite{odonnellEfficientQuantumTomography2016} and fidelity estimation \cite{dasilvaPracticalCharacterizationQuantum2011,flammiaDirectFidelityEstimation2011b}, becomes increasingly impractical due to the exponential 
or at least infeasible scaling of resources required.
In some instances, a classical simulation of the quantum device is within reach, but also here the scaling of resources is generically inhibitive. 
An alternative strategy involves \emph{directly comparing} different quantum devices, bypassing classical benchmarks altogether. This approach gives rise to a central task known as \emph{cross-platform verification}, which aims at estimating the fidelity of states prepared on two quantum computers in distinct laboratories, potentially operating on different physical platforms.

Establishing a high quality quantum channel between distant laboratories operating on different physical platforms is currently beyond technological reach. This limitation emphasizes the importance of cross-platform verification protocols that do not rely on quantum communication between laboratories \cite{elbenCrossPlatformVerificationIntermediate2020, anshuDistributedQuantumInner2022,carrascoTheoreticalExperimentalPerspectives2021,qianMultimodalDeepRepresentation2023,zhengCrossPlatformComparisonArbitrary2024}. The first protocol in this vein has been proposed in 
Ref.\ \cite{elbenCrossPlatformVerificationIntermediate2020} and was experimentally implemented in Ref.\ \cite{zhuCrossPlatformComparisonArbitrary2022}. It is based on randomized measurements coordinated between the two parties, that is, the two parties agree on the measurement bases before data collection.  Despite being less expensive than individual state tomography, this protocol is not scalable even to intermediate-sized systems, as the required number of state copies grows exponentially with the system size 
$n$. Recently, this scalability challenge was further highlighted by Ref.\ \cite{anshuDistributedQuantumInner2022} which studied the task at the core of cross-platform verification: estimating the overlap $\tr(\rho\sigma)$ for quantum states $\rho$ and $\sigma$ prepared in separate experimental platforms without quantum communication. They refer to this task as distributed quantum inner product estimation and show a $\Omega(\sqrt{d})$ lower bound on the required number of copies of $\rho$ and $\sigma$, indicating unavoidable exponential scaling in general. Conversely, efficient protocols for the task require further assumptions on the input states $\rho$ and $\sigma$.

This motivates the main question this work addresses:

\begin{center}
\emph{Which assumptions on $\rho,\sigma$ allow for an efficient distributed estimation of $\tr(\rho\sigma)$?}
\end{center}

In this work, we approach this question from different angles: On the one hand, we propose new algorithms for the distributed estimation of $\tr(\rho\sigma)$ and demonstrate regimes where these algorithms run efficiently. On the other hand, we show that the task remains hard even if we severely restrict the class of input states. Interestingly, both our hardness and easiness results connect to commonly studied properties of states, namely their magic and their multipartite entanglement.

At the core of our proposed algorithms is a sub-routine that we refer to as \textit{Pauli sampling}. This sub-routine operates on copies of an unknown state $\rho$ and aims to approximately sample Pauli strings with probability proportional to their weight $\alpha_{\rho}(P)^2$ in the decomposition $\rho = 2^{-n}\sum_P \alpha_{\rho}(P) P$. We construct an algorithm for this sub-routine by combining tools from two distinct fields. Firstly, our algorithm incorporates Bell measurements across $\rho^{\otimes2}$, a measurement scheme that has attracted recent attention from a quantum learning viewpoint due to its rich information content \cite{montanaroLearningStabilizerStates2017,grossSchurWeylDualityClifford2021,huangQuantumAdvantageLearning2022b,hangleiterBellSamplingQuantum2023,grewalEfficientLearningQuantum2024,grewalImprovedStabilizerEstimation2024a,gutierrezSimpleAlgorithmsTest2024}. Secondly, the algorithm processes this data classically, drawing on concepts from the field of classical simulation of quantum circuits \cite{bravyiHowSimulateQuantum2022,bremnerAchievingQuantumSupremacy2017a,cliffordClassicalComplexityBoson2018}. Our Pauli sampling sub-routine has applications beyond inner product estimation, for instance in the context of learning states prepared from Clifford and few $T$-gates \cite{leoneLearningTdopedStabilizer2024, grewalEfficientLearningQuantum2024, chiaEfficientLearningDoped2024}.

\subsection{Distributed inner product estimation}
\begin{figure*}[t]
     \centering
  \includegraphics[width=0.6\textwidth]{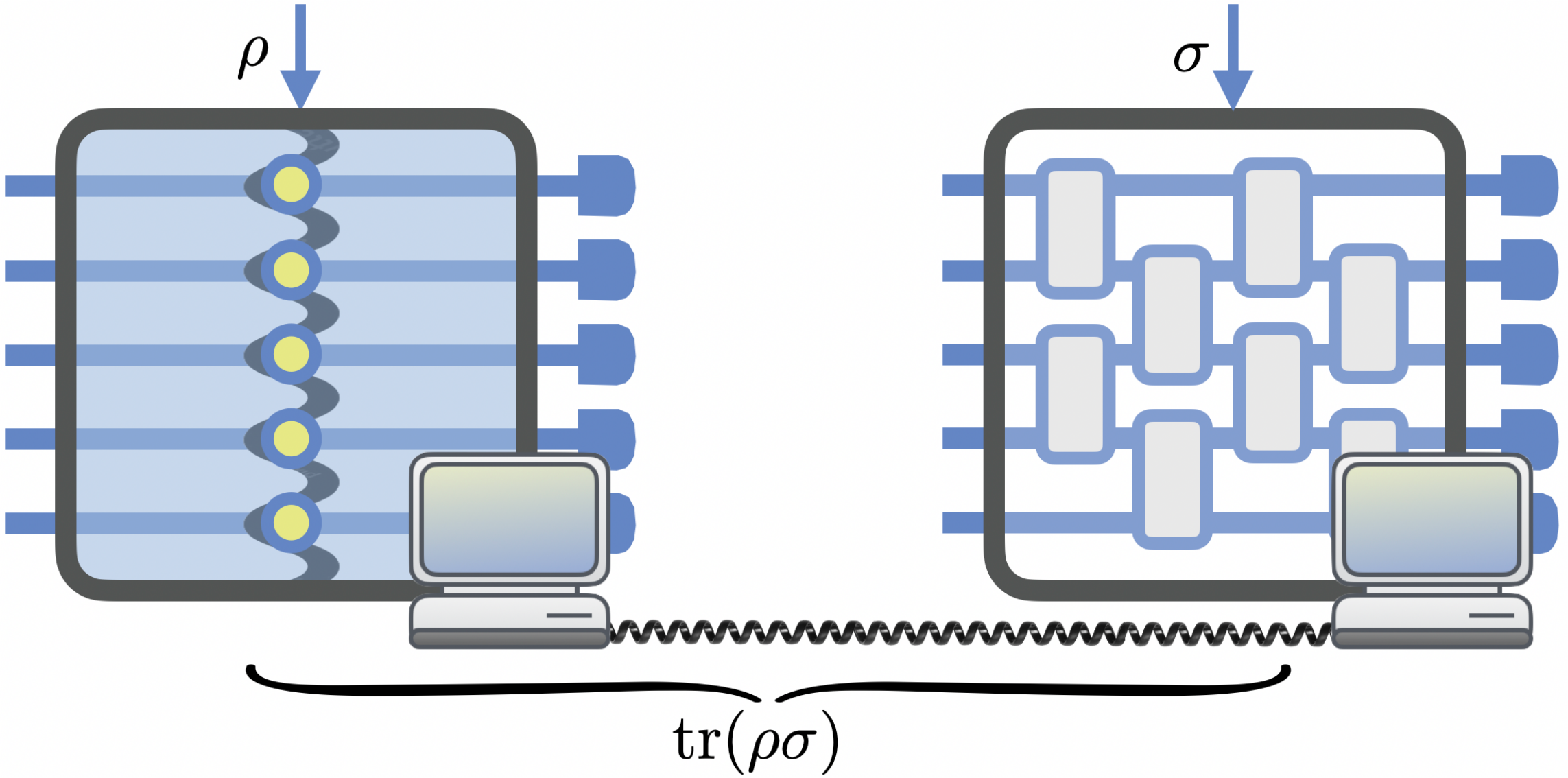}
    \caption{Sketch of the setting of distributed inner product estimation: Alice and Bob each have access to many copies of their respective states $\rho$ and $\sigma$. Their goal is to obtain an estimate of $\tr(\rho \sigma)$ based on local quantum operations and classical communication, commonly abbreviated as LOCC. The two implementations may but do not have to be realized based on the same kind of physical architecture.}
    \label{fig:distributed_inner_product_estimation}
 \end{figure*}

We start by specifying the task studied in this work. We largely follow the formulation of 
Ref.\ \cite{anshuDistributedQuantumInner2022}
and consider the task of \textit{distributed quantum inner product estimation} (IP). This task involves two parties, Alice and Bob. A sketch of the setting is provided in \Cref{fig:distributed_inner_product_estimation}.

\begin{definition}[Inner product estimation, IP]\label{def:IP}
Alice is given $k$ copies of an unknown state $\rho$ and Bob is given $k'$ copies of an unknown state $\sigma$. Their goal is to estimate the overlap $\trace\left(\rho\sigma\right)$ up to some desired additive error
$\epsilon\in\left(0,1\right)$ with success probability at least $2/3$ using only 
\emph{local quantum operations and classical communication} (LOCC).
They are not allowed to have any quantum communication.
\end{definition}

Firstly, note that this task is interesting only in the distributed setting: If Alice and Bob could use quantum communication, then they could run a SWAP test to efficiently estimate the overlap $\trace\left(\rho\sigma\right)$. 
In the distributed setting with no quantum communication, a reasonable strategy is for Alice and Bob to coordinate the bases in which they measure their individual states. This way, they can correlate the classical data they collect. In Ref.\ \cite{elbenCrossPlatformVerificationIntermediate2020}, the authors proposed a protocol based on such coordinated measurements. In their protocol, the measurement bases are selected by drawing local random unitaries. More recently, Anshu \emph{et al.} \cite{anshuDistributedQuantumInner2022} proved that, in general, at least $\Omega(\sqrt{d})$ copies of the states $\rho,\sigma$ acting on $\mathbb{C}^d$ are necessary to solve IP. They further presented a variant of the approach laid out in
Ref.\ \cite{elbenCrossPlatformVerificationIntermediate2020} using coordinated measurement bases selected via drawing fully Haar random unitaries. In both approaches, the random measurement bases are not tailored to the states $\rho,\sigma$. In our approach, we will instead rely on Pauli measurements that we tailor to the input states. In light of the lower bound proven by
Ref.\ \cite{anshuDistributedQuantumInner2022}, this approach can only improve over fully random measurements under certain restrictions on the input states.

How can we select Pauli measurements that are tailored to the input states $\rho,\sigma$? This question motivates the study of the task we call \textit{Pauli sampling}.

\subsection{Pauli sampling}
Consider a state $\rho$ expanded in the $n$-qubit Pauli basis as
\begin{equation}
    \rho = \frac{1}{2^n}\sum_P \alpha_{\rho}(P) P.
    \label{eq:pauli-decomposition}
\end{equation}
When correctly normalized, the squared coefficients $\alpha_{\rho}(P)^2 = \trace\left(\rho P\right)^2$ form a probability distribution $p_{\rho}$ over the set $\{I,X,Y,Z\}^n$, i.e., set of phaseless $n$-qubit Pauli strings. We refer to this distribution $p_{\rho}$ as the \textit{Pauli distribution}. 
We study in detail the following task that forms the core sub-routine of our protocols for IP.

\begin{definition}[Pauli sampling]\label{def:pauli-sampling}
Given access to copies of an unknown state $\rho$, sample from the \textit{Pauli distribution}
\begin{equation}
p_{\rho}\left(P\right)=\frac{1}{2^{n}}\,\frac{\trace\left(\rho P\right)^{2}}{\trace\left(\rho^{2}\right)}\label{eq:pauli_distribution}.
\end{equation}
\end{definition}
We are also content with sampling from a distribution that approximates $p_{\rho}$ up to error $\Delta$ in total variation distance. Importantly, Pauli sampling generates Pauli strings with probability proportional to their weight in the decomposition of the state (c.f. \Cref{eq:pauli-decomposition}). Thus, it is well-suited to tailoring Pauli measurements to the input states $\rho,\sigma$ in inner product estimation.

In general, no efficient quantum algorithm exists to directly sample from $p_{\rho}$ given access to an unknown state $\rho$. Instead, one must measure copies of 
$\rho$ to obtain classical data and then post-process it to generate samples.
However, if $\rho$ is a stabilizer state or a real state (with real computational basis amplitudes), one can sample directly from $p_\rho$ using Bell sampling or Bell difference sampling \cite{montanaroLearningStabilizerStates2017, grossSchurWeylDualityClifford2021}.

The Pauli distribution $p_{\rho}$ of a state $\rho$ is intimately connected to the non-stabilizerness also known as magic of the state $\rho$. In particular, recently, stabilizer Rényi entropies have been introduced as a measure of magic \cite{leoneStabilizerEnyiEntropy2022}. For pure states, they are defined as the Rényi entropies of the Pauli distribution $p_\rho$ (up to a normalization). Interestingly, in Ref.\ \cite{leoneNonstabilizernessDeterminingHardness2023}, the complexity of \emph{direct fidelity estimation} (DFE) was found to scale exponentially with the stabilizer entropy of the target state. As explained in detail in \Cref{subsec:algorithm_ip}, we find a similar connection between the complexity of distributed inner product estimation and the stabilizer entropy of the input states.

Lastly, we note that Pauli sampling has appeared in a different context in several works \cite{lamiNonstabilizernessPerfectPauli2023, haugStabilizerEntropiesNonstabilizerness2023b, tarabungaManybodyMagicPauliMarkov2023} trying to quantify non-stabilizerness in tensor network states such as matrix product states (MPS). However, these papers rely on a classical description of $\rho$ in the form of a tensor network while we work in the scenario where the state $\rho$ is not characterized yet.

\section{Main results}
In this section, we discuss our novel algorithms for distributed inner product estimation and Pauli sampling as well as our complementary no-go results. These results connect to two key characteristics of input states $\rho,\sigma$: their magic (or non-stabilizerness) and their entanglement. We quantify magic through a magic measure $M(\rho)$ such as the stabilizer entropies \cite{leoneStabilizerEnyiEntropy2022} (see \Cref{ssec:preliminaries-measures-magic-entanglement} for more details) and we quantify entanglement in terms of the $\alpha$-Rényi entropies of entanglement $\mathcal{S}_\alpha(\rho_{A:B})$.

\subsection{Pauli sampling}
The first main contribution of this work is the study of Pauli sampling and its resource requirements. More concretely, our focus is on characterizing the sample complexity (the required number of copies of the unknown input state $\rho$) and computational complexity across classes of input states. Note that our findings predominantly apply to classes of pure states.

\paragraph{No-go:}
We now state our first main result, a no-go theorem for efficient Pauli sampling:

\begin{theorem}[Approximate Pauli sampling no-go -- informal version of \Cref{th:hardnesspauli}]\label{ithm:main-result-Pauli-sampling-no-go}
There is no sample-efficient algorithm for approximate Pauli sampling with respect to the class of all pure 
quantum states. Moreover, let $\mathcal{C}$ be the class of $n$-qubit pure states $\rho$ such that $M(\rho) < f(n)$ and $\mathcal{S}_{\alpha}(\rho_{A:B}) < g(n)$ for $0\leq\alpha\leq\infty$ and for all bi-partitions $(A:B)$ where $f(n), g(n) = \omega(\log n)$. Then, there is no sample-efficient algorithm for approximate Pauli sampling up to any constant TV distance $\Delta\leq 1/3$ with respect to $\mathcal{C}$.
\end{theorem}

Intuitively, \Cref{ithm:main-result-Pauli-sampling-no-go} reveals that Pauli sampling is challenging not only for general states but also remains difficult when we restrict the class of input states to states that have magic and entanglement only slightly larger than $\log(n)$.

We obtain this result by connecting to so-called \emph{pseudo-random states} (PRS). 
This concept and the first constructions have been introduced 
in Ref.\  \cite{jiPseudorandomQuantumStates2018}. More recently, PRS constructions that feature tunable entanglement \cite{aaronsonQuantumPseudoentanglement2024} and tunable magic \cite{guPseudomagicQuantumStates2024} have been discovered. Our no-go results are based on these tunable PRS constructions. Concretely, we obtain \Cref{ithm:main-result-Pauli-sampling-no-go} by demonstrating a reduction between Pauli sampling and imaginarity testing, that is, testing if a state is either real or highly imaginary. We then make use of a proof strategy from Ref.\ \cite{haugPseudorandomUnitariesAre2023} which showed that imaginarity testing requires at least $\Omega\left(2^{n/2}\right)$ copies based on an ensemble of real pseudo-random states. We apply this strategy to the tunable PRS constructions mentioned above.

\paragraph{Algorithm:}
Turning to positive results, in \Cref{ssec:ancestral-sampling-algorithm}, we propose an algorithm for approximate Pauli sampling. To summarize, our algorithm applies the ancestral sampling method---a standard tool in classical simulation of quantum computation \cite{bravyiHowSimulateQuantum2022,bremnerAchievingQuantumSupremacy2017a,cliffordClassicalComplexityBoson2018}---to the Pauli distribution $p_\rho$. In ancestral sampling, a distribution $p$ over $\{0,1\}^n$
is decomposed via the chain rule,
\begin{equation}
    p(x) = p(x_1)p(x_2 | x_1) \cdots p(x_n | x_1, \dots, x_{n-1})\,,
\end{equation}
and a sample is generated bit by bit
 by sampling from the corresponding conditional/marginal distributions. Here, we represent $n$-qubit Pauli operators (from $\{I,X,Y,Z\}^n$) as bit-strings in $\{0,1\}^{2n}$ and generate them qubit by qubit. Instead of computing the required marginals of $p_\rho$ from a classical state description of $\rho$, we estimate them from Bell measurement data.
Specifically, the marginals of the Pauli distribution $p_\rho$ for a pure state $\rho$ are given by
\begin{equation}
p_{\rho}\left(x_1, \dots, x_k \right) =
2^{-k}\, \trace\big(P_{x_{1}}^{\otimes2} \otimes \cdots\otimes  P_{x_{k}}^{\otimes2}\otimes \mathbb{S}^{\otimes n-k}\rho^{\otimes2}\big) \,,
\end{equation}
where the observables---tensor products of two-copy single-qubit Paulis $P_i^{\otimes2}$ and SWAP operators $\mathbb{S}$---are estimated from the Bell measurement data. 

We underscore the algorithm’s practicality by noting it relies solely on Bell measurement data and classical post-processing. Our analysis (see \Cref{ssec:efficiently-samplable-states}) builds a connection—flatness of the Pauli distribution reflects the state's magic, while the size of its marginals corresponds to entanglement across cuts—and, through this connection, we are able to prove the following guarantee:
\begin{theorem}[Efficient approximate Pauli sampling -- informal version of \cref{corollar:sampleable_states}]\label{ithm:main-result-pauli-sampling-states}
Let $\mathcal{C}$ be the class of $n$-qubit pure states $\rho$ with $M(\rho) = O(\log (n))$ and $\mathcal{S}_0(\rho_{A:B}) = O(\log (n))$ for all bi-partitions $(A:B)$. Then, our algorithm for approximate Pauli sampling with respect to $\mathcal{C}$ is both sample- and computationally efficient.

\end{theorem}
\Cref{ithm:main-result-pauli-sampling-states} shows that approximate Pauli sampling via our algorithm is efficient when we limit both magic and entanglement sufficiently, namely to at most $O(\log n)$. As explained in more detail in \Cref{ssec:efficiently-samplable-states}, if we only require sample efficiency, our algorithm can cover an extended class of states for which bounded entanglement is only needed for a subset of bi-partitions instead of all bi-partitions.

We further emphasize that our algorithm can be applied to any state, including mixed states. Though \cref{ithm:main-result-pauli-sampling-states} applies only to a specific set of pure states, the most general result we obtain (see \cref{thm:sampling_conditions_marginals}) characterizes the complexity of Pauli sampling with respect to properties of the underlying Pauli distribution. This characterization also holds for mixed states. However, in the mixed state case, we are not yet able to connect the relevant properties of the Pauli distribution to mixed state entanglement measures \cite{plenioIntroductionEntanglementMeasures2007} and magic measures \cite{liuManybodyQuantumMagic2022c, leoneStabilizerEntropiesAre2024}.

\subsection{Distributed inner product estimation}
As the second main contribution of this work, we study which restricted classes of states allow for efficient distributed inner product estimation, i.e., estimation of $\trace\left(\rho\sigma\right)$. Similarly to our results for Pauli sampling above, we again provide both positive and negative results.
\paragraph{No-go:}
We start by presenting a no-go theorem. Note that a first no-go was demonstrated in Ref.\ \cite{anshuDistributedQuantumInner2022}, namely that there is no sample-efficient algorithm for IP with respect to the class of all pure states. Here, we use the proof strategy 
from Ref.\  \cite{anshuDistributedQuantumInner2022} and combine it with the tunable PRS constructions 
\cite{aaronsonQuantumPseudoentanglement2024,guPseudomagicQuantumStates2024} also used to prove \Cref{ithm:main-result-Pauli-sampling-no-go}, in order to show that this no-go holds even for a much more restricted class of states. Note that the IP problem has two input states, $\rho$ and $\sigma$. In the following informal theorem statements, when referring to a single class of states $\mathcal{C}$, we will always assume that both $\rho$ and $\sigma$ belong to that class $\mathcal{C}$.

\begin{theorem}[IP no-go -- informal version of \Cref{thm:IP*-lower-bound}]\label{ithm:main-result-IP-no-go}
Let $\mathcal{C}$ be the class $n$-qubit pure states $\rho$ such that $M(\rho) < f(n)$ and $\mathcal{S}_{\alpha}(\rho_{A:B}) < g(n)$ for $0\leq\alpha\leq\infty$ and for all bi-partitions $(A:B)$ where $f(n), g(n) = \omega(\log n)$. Then, there is no sample-efficient algorithm for IP up to error $\epsilon\in(0,1)$ with respect to $\mathcal{C}$.
\end{theorem}

\paragraph{Algorithm:}
Turning to positive results, in \Cref{subsec:algorithm_ip}, we propose two novel protocols for IP that employ coordinated Pauli basis measurements selected via the Pauli sampling sub-routine. We refer to them as the \textit{symmetric} and \textit{asymmetric} protocol. In the symmetric protocol, both Alice and Bob perform Pauli sampling according to $p_\rho$ and $p_\sigma$, respectively. This way, they obtain a list of Pauli strings distributed according to $\pmix=\frac{1}{2}\left(p_\rho + p_\sigma \right)$. Then, both parties measure the sampled Pauli operators on their respective states in order to estimate $\trace\left(\rho P\right), \trace\left(\sigma P\right)$. Lastly, by combining these estimates via classical communication, they can Monte Carlo estimate $\trace\left(\rho\sigma\right)$ via the rewriting
\begin{equation}
    \left(1+ \trace\left(\rho\sigma\right) \right)=\underset{\pmix}{\mathbb{E}} \left[G\left(\trace(\rho P),\trace(\sigma P)
\right)\right] \, ,
\label{eq:symmetric-expectation}
\end{equation}
where $
 G\left(x,y
\right) := \frac{(x+ y)^2}{ x^2 +y^2}
$.
The asymmetric protocol proceeds in a similar way, however only one party performs Pauli sampling and the estimator is given by
\begin{equation}
    \trace\left(\rho\sigma \right)=\underset{p_\rho}{\mathbb{E}} \left[\frac{\trace(\sigma P)}{\trace(\rho P)}\right] \, .
\label{eq:asymmetric-expectation}
\end{equation}
We give a detailed description of both protocols in \Cref{subsec:algorithm_ip}.

By analyzing the symmetric and asymmetric protocol, we find that their efficiency is linked to the magic of the input states. In particular, we are able to show the following guarantee:

\begin{theorem}[Efficient IP via Pauli sampling  -- informal version of \Cref{cor:main-result-IP-low-magic-symmetric}]\label{ithm:main-result-IP}
Let $\mathcal{C}$ be the class of $n$-qubit pure states with $M(\rho) = O(\log (n))$. Then, assuming access to an oracle for approximate Pauli sampling up to total variation distance $\Delta$, there exists an efficient algorithm for IP up to error $\epsilon+2\Delta$ with respect to $\mathcal{C}$. 
\end{theorem}
As remarked above, \Cref{ithm:main-result-IP} assumes that both inputs $\rho,\sigma$ to IP belong to the class $\mathcal{C}$. Furthermore, we point out that in \Cref{ithm:main-result-IP} the total variation distance error $\Delta$ from the approximate Pauli sampling subroutine directly enters the total estimation error as $\epsilon + 2\Delta$. Here, $\epsilon$ is an additional contribution due to a standard finite sample error. Importantly, the IP algorithm has control over the value of $\Delta$ and can always suppress it by investing more resources.

\Cref{ithm:main-result-IP} tells us that distributed inner product estimation via Pauli sampling can be performed efficiently if we limit ourselves to states with magic of at most $O(\log n)$. Examples of low magic states are very common; they include eigenstates of so-called (perturbed) stabilizer Hamiltonians as shown in Ref.~\cite{guDopedStabilizerStates2024} such as, e.g., the toric code Hamiltonian. Concretely, it has been proven that these Hamiltonians admit eigenstates with bounded magic $M(\rho)$ making them interesting candidates for the application of our algorithm.

By replacing the oracle access in \Cref{ithm:main-result-IP} with concrete Pauli sampling methods, we obtain guarantees on the complexity of IP for two classes of states. First, plugging in our proposed Pauli sampling algorithm and its performance guarantee in \Cref{ithm:main-result-pauli-sampling-states}, we obtain the following corollary:

\begin{corollary}[Efficient IP for low magic and entanglement states  -- informal]\label{cor:main-result-IP}
Let $\mathcal{C}$ be the class $n$-qubit pure states $\rho$ with $M(\rho) = O(\log (n))$ and $\mathcal{S}_0(\rho_{A:B}) = O(\log (n))$ for all bi-partitions $(A:B)$. Then, there exists an efficient algorithm for IP up to error $\eps$ with respect to $\mathcal{C}$.
\end{corollary}

Second, by using the fact that Bell sampling corresponds to Pauli sampling for real states \cite{montanaroLearningStabilizerStates2017} (states with real amplitudes with respect to the computational basis), we can get rid of the restriction of low entanglement and obtain the following corollary:

\begin{corollary}[Efficient IP for real low magic states -- informal]\label{cor:main-result-IP-real-states}
Let $\mathcal{C}$ be the class of real $n$-qubit pure states with $M(\rho) = O(\log (n))$. Then, there exists an efficient algorithm for IP up to error $\eps$ with respect to $\mathcal{C}$.
\end{corollary}

In particular, we note that the class of real states includes arbitrary eigenstates of any real Hamiltonian. Note that the overwhelming majority of many-body Hamiltonians considered in practice have real-valued coefficients with respect to the computational basis \cite{sachdevQuantumPhaseTransitions2011}. This includes, by construction, all stoquastic Hamiltonians \cite{bravyiComplexityStoquasticLocal2008}, but more importantly also most well-known spin Hamiltonians. Further examples include Hamiltonians corresponding to CSS codes, such as the toric code, and Ryberg atom arrays \cite{tengLearningTopologicalStates2024}. Hence, this result applies to a broad class of states with a rich structure that has been studied in the literature.

Finally, we remark that our two protocols each
have advantages that are unique to them: When using the \textit{asymmetric} protocol, only one of the two input states $\rho,\sigma$ has to be belong to the restricted class of states $\mathcal{C}$ with low magic and entanglement whereas the other state is essentially unconstrained. On the other hand, for our {\em symmetric} protocol, the conditions on low magic and entanglement of the previous theorems can be weakened. Thus, to the best of our knowledge, the previous results provide a way to perform distributed inner product estimation that goes beyond any existing methods.

\section{Numerical experiments}
To support our theoretical findings, in this section, we present numerical results from the implementation of our symmetric and asymmetric protocols. For simplicity, we focus on the case where $\rho=\sigma$ such that the true value of the inner product is known to be $\trace(\rho\sigma)=1$. This choice not only makes it easier to assess the deviation of the protocol estimates from the true value but is also practically relevant for the case when two devices prepare the same state.

Our choice of the input state $\rho$ is closely linked to our main theoretical result, \Cref{ithm:main-result-IP}, which shows that the sample complexity of our protocols scales exponentially with the amount of magic in the input states (see also the formal statements in \cref{cor:main-result-IP-low-magic-symmetric} and \cref{cor:main-result-IP-low-magic-asymmetric}). To validate this, we numerically study our protocols on an ensemble of $n$-qubit states with tunable magic. Specifically, we draw the state $\rho$ randomly from the ensemble of $t$-doped $n$-qubit state vectors given by
\begin{equation}
   C (\ket{T}^{\otimes{t}}\otimes \ket{0}^{\otimes n-t})\,,
\end{equation}
 where $C$ is a random global Clifford unitary and $0\leq t\leq n$ and $\ket{T}=T\ket{+}$ is the standard single-qubit magic state. In particular, its Pauli decomposition is
\begin{equation}
\ket{T}\bra{T} = \tfrac{1}{2}\left(I+ (X+Y)/\sqrt{2}\right).
\end{equation}
This ensemble provides an excellent benchmark because it is both generic and highly entangled due to the application of the random Clifford unitary. Moreover, since stabilizer entropy is additive under tensor products and invariant under Clifford operations, we can compute it analytically. In particular, from the Pauli decomposition of the $\ket{T}$-state, we can find that 
\begin{align}
     M_0(\ket{T}) &= \log_2(3)-2 \approx 0.585  \,,\\
     M_1(\ket{T}) &= \tfrac{1}{2} \,.
\end{align}
and $M_0(\ket{T}^{\otimes t}) = t \cdot M_0(\ket{T})$ and similarly for $M_1$. Hence, the number $t$ of $T$-gates in the state preparation corresponds directly to the stabilizer entropy of the state $\rho$.

We now describe our numerical experiments:
To evaluate the estimator in \Cref{eq:symmetric-expectation} for the symmetric protocol and the estimator \Cref{eq:asymmetric-expectation} (c.f. \Cref{eq:f-n1-n2-asym} for a more detailed version) for the asymmetric protocol, we sampled $N_1=1000$ Pauli operators $P\in \{I,X,Y,Z\}^n$ from the Pauli distribution $p_\rho$ and then estimated $\tr(\rho P)$ for each of the sampled Paulis $P$ using $N_2$ state preparations of $\rho,\sigma$ in the corresponding Pauli basis. The total number of shots is hence $2N_1 N_2$.

\begin{figure*}[t]
    \centering
    \includegraphics[]{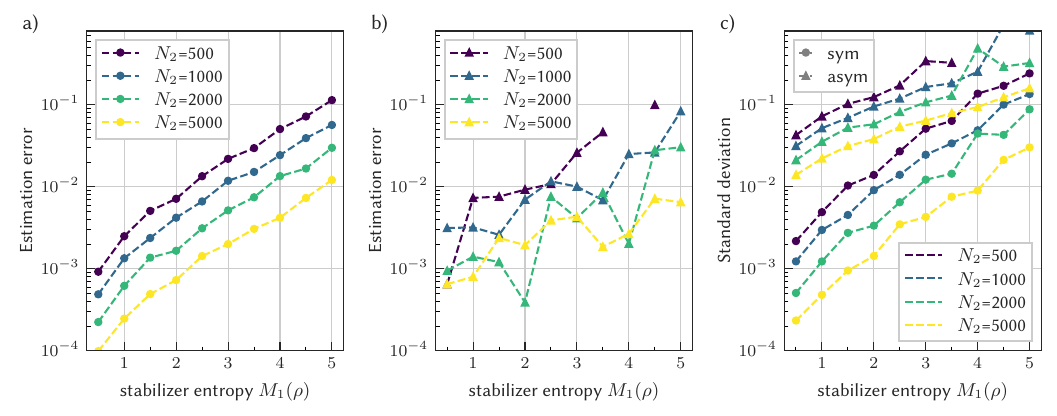}
    \caption{Estimation error and standard deviation of the symmetric and asymmetric protocol as a function of the stabilizer entropy $M_1(\rho)$ and the number of shots per Pauli $N_2$. Here, $\rho$ is drawn from the ensemble $C (\ket{T}^{\otimes{t}}\otimes \ket{0}^{\otimes n-t})$ with increasing $t$ corresponding to increasing values of $M_1(\rho)$, the other parameters in the protocol are fixed to $N_1=1000$ Pauli samples and the number of qubits is $n=10$. a) Estimation error of the symmetric protocol. b) Estimation error of the asymmetric protocol. c) Standard deviations of the estimators of the symmetric and asymmetric protocol, respectively.}   \label{fig:performance_sym_vs_asym}
\end{figure*}

In \Cref{fig:performance_sym_vs_asym}, we present the estimation error, i.e. the deviation of the estimate output by the protocol from the true value of $\tr(\rho \sigma)$, of the symmetric and asymmetric protocol as a function of number of $T$-gates (serving as a proxy for the amount of magic) and $N_2$. Alongside, we also compare the the standard deviation of the estimators. Our numerical results align well with the theoretical analysis, confirming that the estimation error grows exponentially with the amount of magic but can be mitigated by increasing the number of shots in both protocols. Furthermore, when comparing the two approaches, the symmetric estimator demonstrates significantly lower standard deviation, indicating more stable behavior. This advantage arises from the fact that the symmetric estimator is inherently bounded, whereas the asymmetric estimator is not.

\section{Comparison to learning-based approaches}
The key challenge in cross-device verification is determining whether two quantum devices prepare the same state without requiring direct quantum communication between them. One potential approach is to first learn a classical representation of the state produced by each device and then compute $\trace(\rho\sigma)$ based on these descriptions. However, in general, learning a full classical description scales exponentially with the number of qubits, making this approach infeasible. To mitigate this, methods such as Matrix Product State (MPS) tomography~\cite{cramerEfficientQuantumState2010a,lamiNonstabilizernessPerfectPauli2023} and learning-based techniques for stabilizer and $t$-doped stabilizer states~\cite{leoneLearningTdopedStabilizer2024, grewalEfficientLearningQuantum2024} provide efficient ways to infer compressed classical representations of quantum states for verification. However, these methods rely on stringent structural assumptions about the input states 
$\rho,\sigma$, which are often violated under realistic, imperfect state preparations, limiting their practical applicability.

Our approach offers an alternative that bypasses the need for full state reconstruction. Instead of first approximating a classical description, we directly estimate $\trace(\rho \sigma)$ from measurement data, significantly reducing resource overhead. Since our method does not depend on any particular structural constraints, it is robustly applicable even when $\rho,\sigma$
do not conform to specific assumptions, making it more resilient to experimental imperfections. In fact, our Pauli-based protocols enable efficient cross-device verification even for certain highly entangled, highly doped states, as demonstrated in~\cite{DVGC25}. In contrast, MPS-based and 
$t$-doped learning methods are prohibitively expensive in these regimes of large entanglement and doping via non-Clifford gates. Our findings establish our approach as a robust and scalable tool for quantum verification, significantly broadening its applicability beyond traditional learning-based techniques.

\section{Discussion and future work}

In this work, we propose algorithms for distributed inner product estimation that employ Pauli basis measurements selected via
the Pauli sampling sub-routine. In contrast to previous proposals \cite{elbenCrossPlatformVerificationIntermediate2020,anshuDistributedQuantumInner2022}, Pauli measurements are not only experimentally much more feasible but we  also show how to tailor them to the input states in order to avoid previous no-go results. Our work thus opens up the possibility of scalable, distributed cross-platform verification on near-term devices. Furthermore, our results on Pauli sampling are relevant to other applications such as quantum learning tasks related to ($t$-doped)-stabilizer states \cite{leoneLearningTdopedStabilizer2024} and experimental magic estimation \cite{haugEfficientQuantumAlgorithms2024}. Our  results also reveal intimate connections to magic and entanglement, fundamental resources for quantum computation whose interplay determines the performance of our Pauli sampling algorithm.

Our work opens several avenues for future research. For instance, one question is whether the relationship we discovered between entanglement and magic—particularly in \cref{lem:bounding_marginals_with_entanglement} and \cref{lem:bounding_probabilities_with_M0,lem:bounding-probabilities-via-M1}—can be extended to mixed states. Another open issue is finding an efficient method to identify “heavy” Pauli operators of a general state $\rho$ whose squared expectation value $\trace(\rho P)^2$ is nearly maximal. Although Bell measurement data enables us to estimate all these values using only 
$O(n/\epsilon^2)$ 
samples, no efficient algorithm is known for selecting the high-value operators. Moreover, recent work \cite{gongSampleComplexityPurity2024a,arunachalamGeneralizedInnerProduct2025a} has examined the complexity of inner product estimation in scenarios where Alice and Bob share a limited amount of quantum information—a setting analogous to having a low-capacity quantum channel. Extending our protocol to this intermediate regime is another intriguing challenge. Finally, a promising direction is to investigate distributed inner product estimation in other systems, such as qudits, or even bosonic and fermionic systems.

\appendix

\section{Preliminaries}

We denote $\left[n\right]=\left\{ 1,\dots,n\right\} $ and the total variation distance between two probability distributions $p,q$ by $\left\Vert p - q\right\Vert_\mathrm{TV}$. We denote drawing a sample $x$ according to a distribution $p$ by $x\sim p$. Further, we denote the probability weight of a distribution on a set $S$ by $p(S)=\sum_{x \in S}p(x)$. Finally, we use $\log$ to denote the logarithm to base 2 and $\ln$ to denote the natural logarithm. 

\subsection{Bell basis and Bell sampling}\label{sec:bell_basis_bell_sampling}

The single-qubit Pauli matrices are denoted by $\left\{ I,X,Y,Z\right\} $.
Denoting with $\ket{\Omega}:=\frac{1}{\sqrt{2}}\left(\ket{0,0}+\ket{1,1}\right)$
a fiducial maximally entangled state vector,
the 2-qubit
Bell basis is given by the four Bell state vectors $\left\{ \ket{\Phi^{+}},\ket{\Phi^{-}},\ket{\Psi^{+}},\ket{\Psi^{-}}\right\} $
which can expressed in terms of the single-qubit Pauli matrices as
\begin{alignat}{2}\label{eq:Bell-states}
\ket{\Phi^{+}}= & \left(I\otimes I\right) & \ket{\Omega}=\frac{1}{\sqrt{2}}\left(\ket{0,0}+\ket{1,1}\right),\\
\ket{\Phi^{-}}= & \left(I\otimes Z\right) & \ket{\Omega}=\frac{1}{\sqrt{2}} \left(\ket{0,0}-\ket{1,1} \right),\\
\ket{\Psi^{+}}= & \left(I\otimes X\right) & \ket{\Omega}=\frac{1}{\sqrt{2}}\left(\ket{0,1}+\ket{1,0}\right),\\
\ket{\Psi^{-}}= & i\left(I\otimes Y\right) & \ket{\Omega}=\frac{1}{\sqrt{2}} \left( \ket{0,1}-\ket{1,0}\right).
\end{alignat}

The swap operator between two qubits can be expressed
in terms of the single-qubit Pauli matrices as
\begin{equation}
\mathbb{S}=\frac{1}{2}\sum_{P\in\left\{ I,X,Y,Z\right\} }P^{\otimes2}\label{eq:swap_pauli_identity}.
\end{equation}
The four Bell states are eigenstates with eigenvalues
in $\left\{ -1,1\right\} $ of $P\otimes P$ for any single-qubit
Pauli matrix $P$ as well as of the swap operator $\mathbb{S}$.

Let us now generalize the above definitions and observations to $n$
qubits. We will label (phaseless) Pauli operators by bit-strings of
length $2n$ as follows. Let $x=\left(x_{1,}\dots,x_{n}\right)=\left(v_{1},w_{1},\dots,v_{n},w_{n}\right)\in\01^{2n}$,
i.e., $x_{i}=\left(v_{i},w_{i}\right)$. We then define
\begin{equation}
\label{eq:pauli-labelling}
P_{x}=i^{v\cdot w}\left(X^{v_{1}}Z^{w_{1}}\right)\otimes\cdots\otimes\left(X^{v_{n}}Z^{w_{n}}\right)=iX^{v}Z^{w} \, ,
\end{equation}
where the inner product $v\cdot w$ of the $n$-bit strings $v,w$
on the phase in front of the tensor product is by convention meant
to be an integer. Furthermore, we define the generalized $2n$-qubit
canonical Bell state as follows: Let 
\begin{equation}\label{eq:2n-Bell-state}
\ket{\Phi_{n}^{+}}=2^{-n/2}\sum_{v\in\left\{ 0,1\right\} ^{n}}\ket{v, v}
\end{equation}
(by abuse of notation we will often drop the subscript $n$). Then,
the $2n$-qubit Bell basis is given by
\begin{equation}
\left\{ \ket{P_{x}}:=\left(P_{x}\otimes I\right)\ket{\Phi^{+}}\,|\,x\in\01^{2n}\right\} .
\end{equation}
It is an orthonormal basis of $\mathbb{C}^{2^{n}}\otimes\mathbb{C}^{2^{n}}$. By virtue of the tensor product structure, it is the eigenbasis of
any $2n$-qubit Pauli string $P_x^{\otimes2}$ as well as of the operator
$\mathbb{\mathbb{S}}^{\otimes n}$ as well as of strings containing
both $P^{\otimes2}$ and $\mathbb{S}$ factors, 
e.g., 
$X^{\otimes2}\otimes \mathbb{S}^{\otimes n-1}$. In Appendix \ref{highdim}, we comment on generalizations to other higher dimensional systems.

By Bell sampling from a state $\rho$, we mean measuring $\rho^{\otimes2}$
in the $2n$-qubit Bell basis with the outcome of the measurement
being a bit-string $x$ of length $2n$ that we can relate to the Pauli
operator $P_{x}$. To be specific, to implement this measurement,
one measures each qubit pair $\left(i,i+n\right)$ for $i\in\left[n\right]$
in the two-qubit Bell basis defined above.

\subsection{Pauli distribution and its marginals}

Next, we define the \emph{Pauli distribution} of a quantum state
$\rho$ to be the distribution over $x\in \01^{2n}$ with the
probabilities given by
\begin{equation}
p_{\rho}\left(x\right)=\frac{1}{2^{n}}\,\frac{\trace\left(\rho P_{x}\right)^{2}}{\trace\left(\rho^{2}\right)}\label{eq:definition_pauli_distribution}.
\end{equation}
In the following, we will often use the shorthand notation $\alpha_{\rho}(x):=\trace\left(\rho P_{x}\right)$.
In the case of a pure state $\rho=\ket{\psi}\bra{\psi}$, \Cref{eq:definition_pauli_distribution} simplifies
to
\begin{equation}
p_{\psi}\left(x\right)=\frac{1}{2^{n}}\trace\left(P_{x}\ket{\psi}\bra{\psi}\right)^{2}=\frac{\alpha_{\psi}(x)^2}{2^n} \, .
\end{equation}

Let us consider taking marginals of the Pauli distribution $p_\rho$. Let $A\subseteq\left[n\right]$ and let $B$ be the complement of $A$ in $[n]$. We can associate the subset $A$
and its corresponding sub-string $x_{A} \in \{0,1\}^m$ of length $m=2\left|A\right|$ with a marginal 
of the Pauli distribution as 
\begin{equation}
     p_{\rho}\left(x_{A}\right) = \sum_{x_{B}} p_{\rho}\left(x\right)=\frac{1}{2^{n} \trace\left(\rho^{2}\right)}\sum_{x_{B}}\trace\left(P_{x_{A}}^{\otimes2}\otimes P_{x_{B}}^{\otimes2}\rho^{\otimes 2} \right)\,.
\end{equation}
Then, using
$\trace\left(P_{x} \rho\right)^{2}=\trace\left(P_{x}^{\otimes2}\rho^{\otimes2}\right)$,
we can further express such marginals $p_{\rho}\left(x_{A}\right)$
in terms of the swap operator $\mathbb{S}$ as 
\begin{equation}
    \label{eq:marginals_with_swap}
 p_{\rho}\left(x_{A}\right) =\frac{1}{2^{\left|A\right|} \tr (\rho^2)}\trace\left(P_{x_{A}}^{\otimes2}\otimes\mathbb{S}_{B}\, \rho^{\otimes 2} \right) \, .
\end{equation}
For instance,
if $A=\left\{ 1,\dots,k\right\} $, then the corresponding marginal
is given by
\begin{align}
&p_{\rho}\left(x_{A}\right)=p_{\rho}\left(x_{1},\dots,x_{k}\right)  =\sum_{x_{k+1}}\dots\sum_{x_{n}}p_{\rho}\left(x_1,\dots,x_n\right) \\
\nonumber
&=\frac{1}{2^{k} \trace\left(\rho^{2}\right)}\trace\left(P_{x_{1}}^{\otimes2}\otimes \cdots \otimes P_{x_{k}}^{\otimes2} \otimes \mathbb{S}^{\otimes n-k } \rho^{\otimes 2} \right) \, .
\end{align}
We note that, in the case of a state vector $\ket{\psi}$, we can further rewrite \Cref{eq:marginals_with_swap} in terms of the
reduced state $\rho_{A}=\trace_{B}\left(\ket{\psi}\bra{\psi}\right)$ of $\ket{\psi}\bra{\psi}$ as
\begin{align}
\label{eq:pure-state-marginals}
 p_{\psi}\left(x_{A}\right)& =\frac{1}{2^{\left|A\right|}}\trace\left(\ket{\psi}\bra{\psi}\left(P_{A}\otimes I_{B}\right)\ket{\psi}\bra{\psi}\left(P_{A}\otimes I_{B}\right)\right)\\
 & =\frac{1}{2^{\left|A\right|}}\trace\left(\rho_{A}P_{A}\rho_{A}P_{A}\right) \, .
  \nonumber
\end{align}

By \Cref{eq:marginals_with_swap} and \Cref{eq:pure-state-marginals}, it is clear that marginals $p(x_A)$ correspond to Pauli-strings $P_{x_A}$ of length $|A|$. Note that $x_{A}$ are bit-strings of even length and thus we will in some instances refer to these as ``even'' marginals of $p_{\rho}$.

When viewing $p_\rho$ as a distribution over $\01^{2n}$, however, we also have ``odd'' marginals of $p_{\rho}$ corresponding to sub-strings $x_m\in \{0,1\}^m$ for odd $m$. These do not directly correspond to Pauli-strings.
For instance, consider the marginal 
\begin{eqnarray}
    p_\rho(x_1,\dots,x_{k-1},\,v_k) &=& p_\rho((v_1,w_1), \dots (v_{k-1},w_{k-1}) ,v_{k})\nonumber\\
    & =& p_\rho(v_1,w_1, \dots ,v_{k})
\end{eqnarray}
for some $k\in[n]$ (refer to \Cref{eq:pauli-labelling} for this notation). This marginal is given by
\begin{eqnarray}
   && p_\rho(v_1,w_1, \dots ,v_{k})= \sum_{w_k}\sum_{x_{k+1}}\dots\sum_{x_{n}} \\    
   &\times&p_{\rho}\left( (v_1,w_1), \dots, (v_{k-1},w_{k-1}) ,(v_{k},w_k),\, x_{k+1},\dots, x_n \right) \,.
   \nonumber
\end{eqnarray}
This odd marginal can be conveniently rewritten as the sum of two even marginals as
\begin{eqnarray}
       p_\rho(v_1,w_1, \dots ,v_{k}) &=&p_\rho(v_1,w_1, \dots ,v_{k},w_k=0)\nonumber\\
       &+&p_\rho(v_1,w_1, \dots ,v_{k},w_k=1) \,.
\end{eqnarray}

\subsection{Measures of magic and entanglement}\label{ssec:preliminaries-measures-magic-entanglement}

In this section, we define the measures of magic, which quantity the nonstabilizerness of states, and measures of entanglement used in this work.
The \emph{Rényi entropy of a distribution} $p$ over $\{0,1\}^n$ is defined as
\begin{equation}
    H_{\alpha}(p) = \frac{1}{1-\alpha} \log\left(\sum_{x} p(x)^\alpha \right) .
\end{equation}
The limiting value of $H_{\alpha}(p)$ for $\alpha\to 1$ is the Shannon entropy
\begin{equation}
    H_1(p) = -\sum_x p(x) \log p(x) \,.
\end{equation}
This definition of the classical Rényi entropy can be used to define both entanglement and magic measures. 
In particular, letting $(A:B)$ be a bi-partition of the $n$ qubits. Then, the \emph{Rényi entanglement entropies} $\mathcal{S}_{\alpha}$ are defined as the Rényi entropy of the reduced density matrix
\begin{equation}
    \mathcal{S}_{\alpha}\left(\rho_{A:B}\right) = \frac{1}{1-\alpha} \log\left(\tr \rho_A^\alpha \right) = H_\alpha(\vec{\lambda})
\end{equation}
where $\rho_A=\sum_j \lambda_j \ket{\psi_j}\bra{\psi_j}$. The Rényi entanglement entropies measure bi-partite 
pure state entanglement over the cut $(A:B)$. Another important way to quantify entanglement in pure states is through the Schmidt rank: Any state vector $\ket{\psi}$ can be written in terms of the 
Schmidt decomposition as
\begin{equation}\label{eq:Shmidt-decomposition}
    \ket{\psi}= \sum_{i=1}^{r}\sqrt{\lambda_{i}}\ket{i_{A}}\ket{i_{B}}
\end{equation}
where $r$ is the known as the Schmidt rank of $\ket{\psi}$ with respect to the cut $(A,B)$. For pure states, the Schmidt rank is related to the \emph{Rényi-0 entanglement entropy} via
\begin{equation}
\mathcal{S}_{0}\left(\rho_{A:B}\right) =\log (r).
\end{equation}

The \emph{stabilizer (Rényi) entropies} (SEs) \cite{leoneStabilizerEnyiEntropy2022} are (up to normalization) defined as the Rényi entropy of the corresponding Pauli distribution. They can be used to quantify the magic of a pure state~\cite{leoneStabilizerEntropiesAre2024}. In particular, for a pure $n$-qubit state $\psi$, its $\alpha$-stabilizer entropy $M_\alpha(\psi)$ is given by
\begin{equation}
    M_\alpha(\psi) = H_\alpha (p_\psi) - n \,. 
\end{equation}
Here, the normalization is chosen such that for a stabilizer state  $M_\alpha(\psi)=0$. Note that $M_\alpha(\psi)$ takes values in $\left[0, n\right]$.

Let us remark here that $M_0$ is not a robust measure of magic in the context of our work. Concretely, there exist pure states $\psi$ that are very close to stabilizer states in trace distance but they exhibit almost maximal $M_0(\psi)$. So, while these states should be regarded as low-magic from the perspective of higher order stabilizer entropies $M_{\alpha}$ with $\alpha>1$ \cite{haugStabilizerEntropiesNonstabilizerness2023b} or other magic monotones such as min-relative entropy of magic \cite{liuManybodyQuantumMagic2022c}, according to $M_{0}\left(\psi\right)$, they are as magical as it gets. 
The different notions of magic captured by stabilizer entropies with $\alpha<1$ and $\alpha>1$ have also recently been discussed in Ref.\ \cite{haugProbingQuantumComplexity2024}.

Nevertheless, $M_0$ captures well the magic of some important classes of states, such as the class of t-doped stabilizer states \cite{leoneLearningTdopedStabilizer2024}. These are states obtained from computational basis states by Clifford circuits doped with a finite number $t$ of non-Clifford gates. This is because any such $t$-doped stabilizer state (where $t$ is the number of single qubit non-Clifford gates) satisfies $ M_{0}\left(\psi\right)\leq2t$.

\subsection{Relating properties of the Pauli distribution to magic and entanglement}\label{subsec:relating-pauli-distribution-to-entanglement-magic}
In this section, we collect a few Lemmata connecting properties of the Pauli distribution $p_{\rho}$ to magic and entanglement properties of the state $\rho$. These will be used notably in \cref{ssec:efficiently-samplable-states} to obtain sufficient conditions on an unknown state for efficient Pauli sampling.

We start by relating the entanglement across a cut $(A:B)$ in a state $\rho$ to the size of the corresponding marginals of $p_{\rho}$.
Recall from \Cref{eq:pure-state-marginals}, that for pure states $\rho=\ket{\psi}\bra{\psi}$, the marginals of the Pauli distribution are given by $ p_{\psi}\left(x_{A}\right)
=\frac{1}{2^{\left|A\right|}}\trace\left(\rho_{A}P_{A}\rho_{A}P_{A}\right) $. In the following lemma, we show that we can provide lower bounds on these marginals in terms of the Schmidt rank of the corresponding bipartition. After completing this work, we learned that an equivalent result was independently derived in Ref. \cite{lamiLearningStabilizerGroup2024} in the context of matrix product states and their stabilizer group.

\begin{lemma}[Bounding marginals via Schmidt rank]\label{lem:bounding_marginals_with_entanglement}
Let $(A:B)$ be a bipartition of the set of $n$ qubits. Let $\rho=\ket{\psi}\bra{\psi}$ be a pure state and let $\rho_{A}$ be its reduced
state on the subsystem $A$. Let $r$ be the Schmidt rank of $\rho$ corresponding to
the bipartition $(A:B)$, then for any $n$-qubit Pauli
string $P=P_A\otimes P_B$, it holds that
\begin{equation}
p_{\psi}(x_A)=\frac{\trace\left(\rho_{A}P_{A}\rho_{A}P_{A}\right)}{2^{|A|}}\ge\frac{\trace\left(\rho P\right)^{2}}{2^{|A|}r} \, .
\end{equation}
\end{lemma}

\begin{proof}
Start by Schmidt decomposing the state vector $\ket{\psi}=\sum_{i=1}^{r}\sqrt{\lambda_{i}}\ket{i_{A}}\ket{i_{B}}$
in terms of the Schmidt coefficients $\sqrt{\lambda_{i}}$. The reduced
state is then given by $\rho_{A}=\sum_{i=1}^{r}\lambda_{i}\ket{i_{A}}\bra{i_{A}}$.
Using this expression for the reduced state $\rho_{A}$, we find
\begin{equation}
\trace\left(\rho_{A}P_{A}\rho_{A}P_{A}\right)=\sum_{i=1}^{r}\sum_{j=1}^{r}\lambda_{i}\lambda_{j}\left|\bra{i_{A}}P_{A}\ket{j_{A}}\right|^{2}\,.
\end{equation}
On the other hand, expanding $\trace\left(\rho P\right)$ in terms
of the Schmidt decomposition, we find
\begin{eqnarray}
\trace\left(\rho P\right)&=&\trace\left(\rho P_{A}\otimes P_{B}\right)\nonumber \\
&=&\sum_{i=1}^{r}\sum_{j=1}^{r}\sqrt{\lambda_{i}\lambda_{j}}\bra{i_{A}}P_{A}\ket{j_{A}}\bra{i_{B}}P_{B}\ket{j_{B}}\,.
\end{eqnarray}
Using the Cauchy-Schwarz inequality, we find
\begin{align}
\trace\left(\rho P\right)^2 
&  \leq \left(\sum_{i,j=1}^{r} \lambda_{i}\lambda_{j}\left|\bra{i_{A}}P_{A}\ket{j_{A}}\right|^2\right) \left(\sum_{i,j=1}^{r} \left|\bra{i_{B}}P_{B}\ket{j_{B}}\right|^2\right) \,,\nonumber \\
&= \trace\left(\rho_{A}P_{A}\rho_{A}P_{A}\right) \left(\sum_{i,j=1}^{r} \left|\bra{i_{B}}P_{B}\ket{j_{B}}\right|^2\right)\, .
\end{align}
Now, it remains to upper bound $\sum_{ij}^{r}\left|\bra{i_{B}}P_{B}\ket{j_{B}}\right|^{2}.$ To this end, note that since $P_{B}$ has eigenvalues bounded by 1 we have that
\begin{equation}
    \sum_{ij}^{r}\left|\bra{i_{B}}P_{B}\ket{j_{B}}\right|^{2}	\leq\sum_{ij}^{r}\left|\bra{i_{B}}I_{B}\ket{j_{B}}\right|^{2} =\sum_{ij}^{r}\delta_{ij}=r,
\end{equation}
where we have used orthonormality of the Schmidt vectors $\ket{i_{B}}$. This completes the proof.
\end{proof}

Next, we consider the connection of the Pauli distribution $p_\rho$ to the magic of the state $\rho$ as measured by its stabilizer entropy. In particular, we are interested in the average size of $\alpha_\rho(x)^2=\trace\left(\rho P_x\right)^2$ for $x\sim p_{\rho}$. To this end, we start with the following definition:

\begin{definition}\label{def:cdf}
Let $\rho$ be an $n$-qubit state and let $\tau \in \left[0,1\right]$. Denote $\alpha_\rho(x)^2=\trace\left(\rho P_x\right)^2$. Then, we define the \emph{cumulative distribution function} (CDF) associated to $\rho$ as 
\begin{eqnarray}\label{eq:cdf}
          F_\rho(\tau)&:=&\Pr_{x\sim p_\rho}[\alpha_\rho(x)^2<\tau]\\
          &=&\sum_{\{x:\, \alpha_\rho(x)^2<\tau\}}p_\rho(x)\nonumber\\
          &=&\sum_{\{x:\,\alpha_\rho(x)^2<\tau\}}\frac{\alpha_\rho(x)^2}{2^n\tr\left(\rho^2\right)}\,.\nonumber
\end{eqnarray}
\end{definition}

Note that 
\begin{equation}
     \Pr_{x\sim p_{\rho}}\left(\alpha_\rho(x)^2\geq\tau\right) = 1- F_{\rho}(\tau) \,.
\end{equation}
We can now relate the CDF to the stabilizer entropies $M_{\alpha}$ for different $\alpha$. For $\alpha=0$, the corresponding stabilizer entropy $M_0(\psi)$ is related to the size of the support of the Pauli distribution $p_\psi$ via
\begin{equation}
    M_0(\psi)=\log |\{x: p_{\psi}(x)>0\}| -n = \log| \{x:\alpha_\psi(x)^2\neq0\}|-n \,.
\end{equation}
From this relation, we directly obtain the following lemma:

\begin{lemma}[Bounding CDF in terms of $M_0$]\label{lem:bounding_probabilities_with_M0}
For any pure state $\psi$ with stabilizer entropy $M_{0}\left(\psi\right)$ and $0<\tau\leq1$,
it holds that
\begin{equation}
    F_{\psi}(\tau)  \leq 2^{M_0(\psi)}\tau.
\end{equation}
\end{lemma}
\begin{proof}
We have that
\begin{eqnarray}
    F_{\psi}(\tau) &=& \sum_{\{x:\alpha_\rho(x)^2\leq \tau\}}\frac{\alpha_\psi(x)^2}{2^n}\\
    &\le& \frac{\tau}{2^n}\sum_{\{x:\alpha_\rho(x)^2\le\tau\}}1\le\frac{\tau}{2^n}\sum_{\{x:\alpha_\rho(x)^2\neq0\}}1
    \nonumber\\
    &=&\frac{\tau}{2^n}2^{n+M_0(\psi)}=2^{M_0(\psi)}\tau .\nonumber
\end{eqnarray}
\end{proof}

Similarly, we can also obtain a bound on $F_{\psi}(\tau)$ in terms of $M_1(\psi)$ by means of Markov's inequality:

\begin{lemma}[Bounding CDF in terms of $M_1$]\label{lem:bounding-probabilities-via-M1}
For any pure state $\psi$ with stabilizer entropy $M_{1}\left(\psi\right)$ and $0<\tau\leq1$,
it holds that
\begin{equation}
    F_{\psi}(\tau)  \leq \frac{M_{1}\left(\psi\right)}{\log(1/\tau)}.
\end{equation}
\end{lemma}

\begin{proof}
    The proof uses Markov's inequality, $\Pr\left(X\geq a\right)\leq\frac{\mathbb{E}\left[X\right]}{a}$, which holds for any non-negative random variable $X$ and $a>0$. Now, in our case, let 
    \begin{equation}
    X:=\log\left(\frac{1}{\trace\left(P_{x}\psi\right)^{2}}\right)
    \end{equation}
    where $x$ is drawn according to $p_{\psi}$. The bound $\trace\left(P_{x}\psi\right)^{2}\leq1$ implies that $X\geq0$. Then 
    \begin{equation}
\mathbb{E}\left[X\right]=\sum_{x}p_{\psi}\left(x\right)\log\left(\frac{1}{\trace\left(P_{x}\psi\right)^{2}}\right)=H_{1}\left(p_{\psi}\right)-n=M_{1}\left(\psi\right).
    \end{equation}
    So, using Markov's inequality, we find that for any $a>0$, 
    \begin{equation}
        \Pr_{x\sim p_{\psi}}\left(\trace\left(P_{x}\psi\right)^{2}\leq2^{-a}\right)\leq\frac{M_{1}\left(\psi\right)}{a}\, ,
    \end{equation}
    or equivalently, letting $\tau=2^{-a}$,
    \begin{equation}
        \Pr_{x\sim p_{\psi}}\left(\alpha_{\psi}(x)^{2}\leq \tau \right)\leq- \frac{M_{1}\left(\psi\right)}{\log(\tau)} \,.
    \end{equation}
\end{proof}

\subsection{Statistical versus computational indistinguishability}\label{ssec:statistical-computational-indistinguishability}
In this section, we provide a brief overview of the concepts of statistical indistinguishability and computational indistinguishability, which are frequently employed throughout the manuscript to establish no-go results.

Given a (discrete) ensemble of quantum states $\mathcal{E}=\{\ket{\psi}\}$, we define its corresponding mixture
\begin{equation}
\Psi_{\mathcal{E}}^{(M)}:=\frac{1}{|\mathcal{E}|}\sum_{\ket{\psi}\in\mathcal{E}}\ket{\psi}\bra{\psi}^{\otimes M}\,.
\end{equation}
This definition can be readily extended to continuous ensembles of quantum states. In particular, we will focus on the ensemble of all quantum pure states, which mixture is defined as
\begin{equation}
\Psi_{\text{Haar}}^{(M)}=\int_{\text{Haar}}d\psi\, \ket{\psi}\bra{\psi}^{\otimes M}
\end{equation}
where $\int_{\text{Haar}}$ is the integral with respect to the Haar measure on the set of pure quantum states (see also Ref.~\cite{jiPseudorandomQuantumStates2018} for technical details).

Two ensembles $\mathcal{E}_1$ or $\mathcal{E}_2$ are said to be \emph{statistically indistinguishable} if any distinguisher, provided with polynomially many copies of a quantum state vector $\ket{\psi}$ uniformly drawn from one of the two ensembles, cannot discern which ensemble the state originates from (with high probability). Equivalently, any algorithm capable of distinguishing the ensembles $\mathcal{E}_1$ and $\mathcal{E}_2$ (with high probability) must have a sample complexity $\omega(\poly(n))$.  It follows from the Holevo-Helstrom theorem \cite[Theorem 3.4]{watrousTheoryQuantumInformation2018} that a sufficient condition for statistical indistinguishability of $\mathcal{E}_1$ and $\mathcal{E}_2$ is closeness in trace distance of their corresponding mixtures $\Psi_{\mathcal{E}_1}^{(M)}$, $\Psi_{\mathcal{E}_2}^{(M)}$. 

\begin{definition}[Statistical indistinguishability]
The ensemble $\mathcal{E}_{1}$ is statistically indistinguishable from $\mathcal{E}_{2}$ if, for every $M=O(\poly(n))$, the following holds
\begin{equation}
\norm{\Psi_{\mathcal{E}_1}^{(M)}-\Phi_{\mathcal{E}_2}^{(M)}}_1=\mathrm{negl}(n)\,, 
\end{equation}
where $\mathrm{negl}(n):=o(\operatorname{poly}^{-1}(n))$. 
\end{definition}

A strictly weaker notion is that of \emph{computational indistignuishability}. Two ensembles $\mathcal{E}_1,\mathcal{E}_2$ are computationally indistinguishable if any distinguishing algorithm, provided with polynomially many copies of a quantum state vector $\ket{\psi}$ uniformly drawn from one of the two ensembles, and bounded to run in \emph{polynomial time}, cannot discern which ensemble the state originates from (with high probability). Formally, we have the following.

\begin{definition}[Computational indistinguishability]
    The ensemble $\mathcal{E}_1$ is computationally indistinguishable from $\mathcal{E}_1$ iffor any $M=O(\poly(n))$ and any polynomial-time quantum algorithm $\mathcal{A}$, the following holds
\begin{equation}
\left|\Pr_{\ket{\psi}\in\mathcal{E}_1}\left[\mathcal{A}(\ket{\psi})=1\right]-\Pr_{\ket{\psi}\in\mathcal{E}_2}\left[\mathcal{A}(\ket{\psi})=1\right]\right|=\mathrm{negl}(n),\label{eq:compind}
\end{equation}
where $\mathrm{negl}(n):=o(\operatorname{poly}^{-1}(n))$.  
\end{definition}

Computational indistinguishability is strictly weaker than statistical indistinguishability. That is, statistical indistinguisability implies computational indistinguishability, but not the other way around.

Before concluding the section, let us provide some useful lemmas that will be used throughout the manuscript.
\begin{lemma}[Statistical indistinguishability]\label{lem:statlemma}
Let $\mathcal{E}$ be an ensemble of quantum states that is statistically indistinguishable from the set of Haar random states. Then for every unitary $V$, the set $\mathcal{E}_{V}:\{V\ket{\psi}\,|\, \ket{\psi}\in\mathcal{E}\}$ is statistically indistinguishable from $\text{Haar}$.
\end{lemma}
\begin{proof}
The proof just uses the unitarily invariance of the trace norm. Indeed,
\begin{eqnarray}
\norm{\Psi_{\mathcal{E}_V}-\Psi_{\text{Haar}}}_{1}&=& \norm{V^{\otimes M}\Psi_{\mathcal{E}}^{(M)}V^{\dag\otimes M}-\Psi_{\text{Haar}}^{(M)}}_{1}\nonumber\\
&=&\norm{\Psi_{\mathcal{E}}^{(M)}-V^{\dag\otimes M}\Psi_{\text{Haar}}^{(M)}V^{\otimes M}}_{1}\nonumber\\
&=&\norm{\Psi_{\mathcal{E}}^{(M)}-\Psi_{\text{Haar}}^{(M)}}_{1}=\mathrm{negl}(n)
\end{eqnarray}
where we have used the fact that $V^{\dag\otimes M}\Psi_{\text{Haar}}^{(M)}V^{\otimes M}=\Psi_{\text{Haar}}^{(M)}$ for the left/right invariance of the Haar measure.
\end{proof}
\begin{lemma}\label{lem:statlemma2}
Given two ensembles $\mathcal{E}_1$ and $\mathcal{E}_2$ that are statistically indistinguishable from an ensemble $\mathcal{E}_3$. Then $\mathcal{E}_1$ is statistically indistinguishable from $\mathcal{E}_2$.
\end{lemma}
\begin{proof}
We can use the triangular inequality and write
\begin{eqnarray}
\norm{\Psi_{\mathcal{E}_1}^{(M)}-\Psi_{\mathcal{E}_2}^{(M)}}_1&\le& \norm{\Psi_{\mathcal{E}_1}^{(M)}-\Psi_{\mathcal{E}_3}^{(M)}}_1+\norm{\Psi_{\mathcal{E}_3}^{(M)}-\Psi_{\mathcal{E}_2}^{(M)}}_1
\nonumber
\\
&=&\mathrm{negl}(n).
\end{eqnarray}
\qedhere
\end{proof}

\begin{lemma}\label{lem:complemma}
Let $\mathcal{E}$ be an ensemble of quantum states that is computationally indistinguishable from the set of Haar random states. Then for every poly-size circuit $V$, the set $\mathcal{E}_{V}:\{V\ket{\psi}\,|\, \ket{\psi}\in\mathcal{E}\}$ is computationally indistinguishable from $\text{Haar}$.
\end{lemma}
\begin{proof}
For the sake of contradiction, let us assume that there is a polynomial-time quantum algorithm $\mathcal{A}$ able to distinguish the ensemble $\mathcal{E}_V$ and Haar. Let us show that using $\mathcal{A}$, we can distinguish $\mathcal{E}$ from Haar in polynomial time. Let $\ket{\psi}$ drawn either from $\mathcal{E}$ or from Haar. Let us apply a polynomial-size circuit $V$ to the unknown state vector $\ket{\psi}$. Note that, if $\ket{\psi}\in\mathcal{E}$, then $V\ket{\psi}\in\mathcal{E}_V$. Conversely, if $\ket{\psi}\in \text{Haar}$, then $V\ket{\psi}\in \text{Haar}$ for the left invariance of the Haar measure. Therefore, we can use the algorithm $\mathcal{A}$ that is able to \textit{efficiently} distinguish $\mathcal{E}_V$ from Haar to effectively distinguish $\mathcal{E}$ from Haar. This is a contradiction. Therefore, we conclude that $\mathcal{E}_V$ is computationally indistinguishable from Haar. 
\end{proof}

\begin{lemma}\label{lem:complemma2}
Given two ensembles $\mathcal{E}_1$ and $\mathcal{E}_2$ of quantum states that are computationally indistinguishable from an ensemble $\mathcal{E}_3$. Then $\mathcal{E}_1$ is computationally indistinguishable from $\mathcal{E}_2$. 
\begin{proof}
Using the definition of computational indistinguishability in Eq.~\eqref{eq:compind}, we can write
\begin{align}
&\left|\Pr_{\ket{\psi}\in\mathcal{E}_1}\left[\mathcal{A}(\ket{\psi})
=1\right]-\Pr_{\ket{\psi}\in\mathcal{E}_2}\left[\mathcal{A}(\ket{\psi})=1\right]\right|\\
\nonumber
&\le \left|\Pr_{\ket{\psi}\in\mathcal{E}_1}\left[\mathcal{A}(\ket{\psi})=1\right]-\Pr_{\ket{\psi}\in\mathcal{E}_3}\left[\mathcal{A}(\ket{\psi})=1\right]\right|\\
&+
 \left|\Pr_{\ket{\psi}\in\mathcal{E}_3}\left[\mathcal{A}(\ket{\psi})=1\right]-\Pr_{\ket{\psi}\in\mathcal{E}_2}\left[\mathcal{A}(\ket{\psi})=1\right]\right|=\mathrm{negl}(n)\nonumber\qedhere
\end{align}
\end{proof}
for any polynomial-time quantum algorithm $\mathcal{A}$. 
\end{lemma}

\subsection{Subset phase states} \label{sec:subset-phase-states}

In the following section, we introduce two discrete ensembles of states, \emph{random subset phase states} and \emph{pseudorandom subset phase states} that are, respectively, statistically and computationally indistinguishable from the ensemble of Haar random states.
 
\begin{definition}[Random subset phase states ~\cite{aaronsonQuantumPseudoentanglement2024}]
    We denote the ensemble of \emph{random subset phase states} $\mathcal{E}_{f,S}$ as the set of states 
\begin{equation}\label{eq:subset_phase_states}
   \ket{\psi_{f,S}}=\frac{1}{\sqrt{|S|}}\sum_{x\in S}(-1)^{f(x)}\ket{x} 
\end{equation}
for some random subset of bit-strings $S\subset\{0,1\}^{n}$ and $f\,:\{0,1\}^{n}\mapsto \{0,1\}$ a random Boolean function.
\end{definition}

The ensemble of random subset phase states is statistically indistinguishable from Haar random states for $|S|=\omega(\poly (n))$ ~\cite{aaronsonQuantumPseudoentanglement2024}. 
Further, it has been shown that the ensemble of subset phase states features low magic and low entanglement, in contrast to the ensemble of Haar random states, see Ref.~\cite{guPseudomagicQuantumStates2024}. 
\begin{lemma}[Restatement of Theorem 1 in 
Ref.\ \cite{guPseudomagicQuantumStates2024} and Theorem 2.7 in 
Ref.\ \cite{aaronsonQuantumPseudoentanglement2024}]
\label{lem:subset-phase-states-properties}
    For all $\abs{S}$ such that $\omega(\log(n)) \leq \log|S| \leq n$, there exists an ensemble of random subset phase states $\mathcal{E}_{f,S}$ such that
    \begin{enumerate}
        \item for all $\alpha=\Theta(1)$, $M_{\alpha}(\psi_{f,S})=O(\log|S|)$ and $M_{\alpha}(\psi_{f,S})=\omega(\log n)$,
        \item for all $\alpha$, $\mathcal{S}_{\alpha}\left(\psi_{f,S}\right)=O(\log|S|) $ and $\mathcal{S}_{\alpha}(\psi_{f,S})=\omega(\log n)$ over every cut $(A:B)$ \footnote{The bound $\mathcal{S}_{\alpha}(\psi_{f,S})=\omega(\log n)$ applies only to cuts where the subsystem size is also at least $\omega(\log n)$.},
    \end{enumerate}
    with high probability over $\ket{\psi_{f,S}}\in\mathcal{E}_{f,S}$.
\end{lemma}

A limitation of subset phase states lies in their construction from random Boolean functions and random permutations, rendering their implementation inefficient. To address this limitation, we consider the ensemble of pseudorandom subset phase states. These differentiate themselves from random subset phase states only in the fact that $S$ is taken as a pseudorandom subset and $f$ a pseudorandom Boolean function. As outlined in Ref.~\cite{aaronsonQuantumPseudoentanglement2024}, we can efficiently prepare pseudorandom subset phase states using pseudorandom permutations and pseudorandom Boolean functions, assuming the existence of \emph{one-way-functions} (OWF)—Boolean functions. 
We denote the ensemble pseudorandom subset phase states as $\tilde{\mathcal{E}}_{f,S}$. In Ref.~\cite{aaronsonQuantumPseudoentanglement2024}, it has been shown that the ensemble of pseudorandom subset phase states is computationally indistinguishable from random subset phase states for any $|S|=\omega(\poly(n))$ and thus is also computationally indistinguishable from Haar random states.

\section{Pauli sampling}
\label{sec:pauli-sampling}

In this section, we discuss the core subroutine at the heart our approach to distributed quantum inner product estimation. We refer to this as (approximate) \textit{Pauli sampling} and begin with a formal definition. 

\begin{definition}[Approximate Pauli sampling]
    Given $\Delta>0$ and access to the unknown quantum state $\rho$, we define approximate Pauli sampling as the task of sampling from a distribution $q$ such that
    \begin{equation}
      \left\Vert q - p_{\rho}\right\Vert _{\mathrm{TV}}\leq\Delta .
    \end{equation}
    Here, $p_{\rho}$ is the Pauli distribution corresponding to $\rho$, as defined in \Cref{eq:definition_pauli_distribution}.
\end{definition}

We start our discussion with an observation regarding the feasibility of Pauli sampling for arbitrary states using single-copy algorithms. By single-copy algorithms, we mean algorithms without quantum memory which use their access to copies of an unknown state $\rho$ by processing each copy individually (c.f. Ref.~\cite{chenExponentialSeparationsLearning2022}). The following lemma is a direct consequence of the limitations of such single-copy algorithms which were demonstrated by Ref.~\cite{chenExponentialSeparationsLearning2022}:
\begin{lemma}[Exponential sample complexity of Pauli sampling for general states for single-copy algorithms]
\label{lem:single-copy-lower-bound}
    Let $\Delta>0$ and let $\rho$ be an unknown $n$-qubit (possibly mixed) quantum state. Then, any single-copy algorithm for approximate Pauli sampling up to error $\Delta<1/4$ requires $\Omega(2^{n/2})$ many copies of $\rho$.
\end{lemma}
\begin{proof}
    Consider the task of purity testing discussed in Ref.\ \cite[Section 5.2]{chenExponentialSeparationsLearning2022}. Purity testing is the task of deciding if an unknown state $\rho$ is either a pure state or the maximally mixed state. Ref.~\cite{chenExponentialSeparationsLearning2022} demonstrates a sample complexity lower bound of $\Omega\left(2^{n/2}\right)$ for this distinguishing task that applies to any single-copy algorithm. Here, we note that this lower bound must also apply to the task of approximate Pauli sampling. To see this, consider the Pauli distribution of the maximally mixed state $\rho=I/2^n$,
    \begin{equation}
        p_{I/2^n}(x) = \begin{cases}
            1 & x = 0^{2n} \\
            0 & \text{else} \,.
        \end{cases} 
    \end{equation}
    Now, assume there is an algorithm for $\Delta$-approximate Pauli sampling from an unknown state $\rho$ which is either the maximally mixed state or a pure state. In case $\rho$ is maximally mixed, this alleged algorithm would sample $x=0^{2n}$ with probability at least $q(x=0^{2n})>1-\Delta$. On the other hand, for a pure state, we have $p_{\rho}(x=0^{2n})=2^{-n}$, i.e., sampling the outcome $x=0^{2n}$ is exponentially unlikely under $p_\rho$ and so $q(x=0^{2n})< 1/2^n + \Delta$. Hence, the approximate Pauli sampling algorithm facilitates deciding between these cases with high probability and thus inherits the lower bound for purity testing.
\end{proof}
\Cref{lem:single-copy-lower-bound} shows that, in general, Pauli sampling requires multi-copy algorithms such as Bell sampling which is a 2-copy algorithm. However, note that \Cref{lem:single-copy-lower-bound} crucially uses the fact that we require an algorithm that works for both pure and mixed states. If we assume from the start, that the input state $\rho$ is pure, then the limitation of \Cref{lem:single-copy-lower-bound} does not necessarily apply. This is an important caveat in the context of our work, as we mostly work only with pure states.

Next, let us remark on existing approaches to Pauli sampling. In particular, for certain classes of pure states, Pauli sampling has been shown to admit an efficient solution. In particular, if $\rho$ is a real pure state, i.e., a state with real coefficients in the computational basis, then we can directly Pauli sample using Bell sampling by associating each outcome $x\in \01^{2n}$ to its corresponding Pauli word \cite{montanaroLearningStabilizerStates2017}. Examples of such states are the ground states of stoquastic Hamiltonians as can be seen by invoking the Perron-Frobenius theorem \cite{bravyiComplexityStoquasticLocal2008}.  In fact, most natural 
many-body Hamiltonians considered in the literature have this feature
\cite{sachdevQuantumPhaseTransitions2011}.

The other class of states for which we can Pauli sample directly and efficiently are pure stabilizer states. 
It was shown in Refs.\ \cite{montanaroLearningStabilizerStates2017,grossSchurWeylDualityClifford2021} that for these states, \emph{Bell difference sampling} corresponds to Pauli sampling. Bell difference sampling is the sampling procedure consisting of independently performing Bell sampling on $\rho \otimes \rho$ and adding the two outcomes (mod 2), again associating the outcome to its corresponding Pauli operator.

Finally, it is sometimes possible to perform Pauli sampling based on an efficient classical representation of the state \cite{lamiNonstabilizernessPerfectPauli2023}. Therefore, one alternative approach to Pauli sampling relies on learning such a classical representation of the unknown state first. This strategy can be applied to stabilizer states \cite{grewalImprovedStabilizerEstimation2024a} as well as states which admit an efficient MPS representation \cite{cramerEfficientQuantumState2010a}.

\subsection{Hardness of approximate sampling even for bounded magic and entanglement} \label{sec:hardness-pauli}
In this section, we provide a rigorous proof of \cref{ithm:main-result-Pauli-sampling-no-go}, which we restate below for convenience. Our proof follows via a reduction between Pauli sampling and a property testing problem. In particular, we show that Pauli sampling allows us to test imaginarity of an unknown state. The \emph{imaginarity} of a state vector $\ket{\psi}$ is defined as $I(\ket{\psi}):=1-|\langle\psi|\psi^{*}\rangle|^2$ where $\ket{\psi^{*}}$ denotes the conjugate state with respect to the computational basis. Imaginarity testing is the task of deciding whether an unknown state $\psi$ has $I(\ket{\psi})=1$ or far from it. It has previously been shown that testing imaginarity of an unknown state $\rho$ requires exponentially many copies of $\rho$ \cite{haugPseudorandomUnitariesAre2023}. We defer the proofs of the auxiliary Lemmas in this section to \Cref{sec:proofsoflemmas}.

Subset phase states as defined in \cref{eq:subset_phase_states} are real states with $I(\ket{\psi})=0$. In \cref{lem:imaginary-ensemble}, by conjugating these real states with randomly drawn local Clifford unitaries, we construct a new (pseudo)-random ensemble of \emph{imaginary} states and show that they are statistically indistinguishable from subset phase states. In \cref{lem:imaginarityfrompaulisampling}, we show how access to samples from the Pauli distribution can be used to estimate the imaginarity of an unknown state. Putting these ingredients together, we obtain a lower bound on the sample complexity of Pauli sampling.

\begin{lemma}[Pseudo-random states with high imaginarity]\label{lem:imaginary-ensemble}
Let $U=\bigotimes_{i=1}^{n}C_{i} $ for $C_{i}$ being random single-qubit Clifford unitaries. Then define the ensemble $\mathcal{E}_{U}=\{U\ket{\psi_{f,S}}\,|\, \ket{\psi_{f,S}}\in \mathcal{E}_{f,S}\}$ where $\mathcal{E}_{f,S}$ is an ensemble of random subset phase states. Denote $\ket{\psi^{*}}$ the conjugate state in the computational basis. We have the following list of results:

\begin{enumerate}
    \item \textbf{Imaginarity:}
    $\Pr_{\ket{\psi}\sim \mathcal{E}_{U}}[I(\ket{\psi}) > 1/100 ]= 1-\operatorname{negl}(n)$.
    \item \textbf{Statistical indistinguishability:} For $|S|=\omega(\poly(n))$, the two ensembles $\mathcal{E}_{U}$ and $\mathcal{E}_{f,S}$ are statistically indistinguishable. 
\end{enumerate}

\end{lemma}

    Note that an analogous result (with computational rather than statistical indistinguishability) holds between ensembles $\tilde{\mathcal{E}}_U$ and  $\tilde{\mathcal{E}}_{f,S}$ where $\tilde{\mathcal{E}}_U=\{U\ket{\psi_{f,S}}\,|\, \ket{\psi_{f,S}}\in\tilde{\mathcal{E}}_{f,S}\}$ and $\tilde{\mathcal{E}}_{f,S}$ are pseudorandom subset phase states.

\begin{lemma}[Pauli sampling implies estimating imaginarity]\label{lem:imaginarityfrompaulisampling}
Let $\rho=\ket{\psi}\bra{\psi}$ be a pure quantum state and $p_{\rho}$ its associated Pauli distribution. Then, given black box access to an algorithm for sampling from a distribution $q$ such that $\left\Vert p_{\rho} - q\right\Vert_\mathrm{TV}\leq \Delta$, for $0 \leq \Delta < 1$, there exists an efficient algorithm to estimate the imaginarity $I(\ket{\psi})=1-|\langle\psi|\psi^{*}\rangle|^2$ within additive error $\epsilon > \Delta$ and failure probability $\delta$ using $\frac{2}{(\epsilon-\Delta)^2}\ln\frac{2}{\delta}$ samples from $q$. 
\end{lemma}

Using the preceding lemmata, we are now ready to prove \cref{ithm:main-result-Pauli-sampling-no-go}, that we restate below.
\begin{theorem}[Formal version of \Cref{ithm:main-result-Pauli-sampling-no-go}]\label{th:hardnesspauli}
Let $\ket{\psi}$ be an unknown $n$-qubit pure state and $p_{\psi}$ the Pauli distribution associated to $\ket{\psi}$. Then, there is no algorithm that can sample from a distribution $q$ such that $\left\Vert p_{\rho} - q\right\Vert_\mathrm{TV}\leq 1/3$ using $O(\poly(n))$ copies of $\ket{\psi}$.
In fact, this is the case even if one is promised that the state has bounded magic and entanglement, such that $M_\alpha(\psi) < g(n)$ (with $\alpha=O(1)$) and $S_{\alpha}(\psi) < f(n)$ (over all cuts $(A:B)$) where $f(n), g(n) = \omega(\log n)$.

\begin{proof}

By \Cref{lem:subset-phase-states-properties}, for any two functions $f(n), g(n)=\omega(\log n)$, there is an ensemble of subset phase states $\mathcal{E}_{f,S}$ with $|S|=\omega(\poly(n))$ that satisfies $M_{\alpha}(\ket{\psi_{f,S}})=O(g(n))$ and $S_{\alpha}(\ket{\psi_{f,S}})=O(f(n))$ over all cuts $(A:B)$. We can further construct an ensemble $\mathcal{E}_{U}=\{U\ket{\psi_{f,S}}\,|\, \ket{\psi_{f,S}}\in\mathcal{E}_{f,S}\}$ with $U$ being a local random Clifford circuit.

Note that local Clifford operations on some state $\ket{\psi}$ preserve its entanglement and magic. 
Moreover, by virtue of \cref{lem:imaginary-ensemble}, $\mathcal{E}_U$ is statistically indistinguishable from $\mathcal{E}_{f,S}$ and it contains states obeying $|\langle\psi|\psi^{*}\rangle|^2<1/100$ with high probability over the random $U$, while $|\langle\psi|\psi^{*}\rangle|^2=1$ for all states in $\mathcal{E}_{f,S}$. 

By \cref{lem:imaginarityfrompaulisampling}, we know that $O(1)$ samples from $q$ suffice to obtain estimate of $|\langle\psi|\psi^{*}\rangle|^2$ up to precision $\epsilon=1/3+1/100$ with high probability. Therefore, assuming we could sample from $q$, we would also be able to distinguish between the two ensembles $\mathcal{E}_{U}$ and $\mathcal{E}_{f,S}$ with high probability, using only  $O(\poly(n))$ copies of $\ket{\psi}$, depending on whether the estimate of $|\langle\psi|\psi^{*}\rangle|^2$ is in $[0,1/3+1/200]$ or in $[2/3-1/100,1]$. This contradicts the statistical indistinguishability of $\mathcal{E}_{U}$ and $\mathcal{E}_{f,S}$ as proven in \cref{lem:imaginary-ensemble} and thus concludes the proof.
\end{proof}
\end{theorem}

In \cref{th:hardnesspauli}, we demonstrated that there is no sample-efficient algorithm capable of 
sampling, even approximately, from the Pauli distribution of an unknown state $\ket{\psi}$ with bounded magic and entanglement.
In what follows, we restrict the class of states those states that are efficiently preparable by a quantum circuit of size scaling at most as $\mathrm{poly}(n)$. For this class of states we can show that no computational efficient algorithm exists that samples from the Pauli distribution modulo a cryptographic assumption.  
In particular, under the assumption on the existence of One-Way-Functions (OWFs) we can construct \emph{pseudorandom} subset phase states efficiently (see \Cref{sec:subset-phase-states}).
\begin{theorem}
Let $\ket{\psi}$ be an unknown, yet efficiently preparable $n$-qubit pure state and $p_{\psi}$ the Pauli distribution associated to $\ket{\psi}$. Moreover, let  $\ket{\psi}$ be such that $M_\alpha(\psi) < g(n)$ (with $\alpha=O(1)$) and $S_{\alpha}(\psi) < f(n)$ (over all cuts $(A:B)$) where $f(n), g(n) = \omega(\log n)$. Then, there is no polynomial-time quantum algorithm that can sample from a distribution $q(x)$ such that $\left\Vert p_{\rho} - q\right\Vert_\mathrm{TV}\leq 1/3$ that uses $O(\poly(n))$ copies of $\ket{\psi}$, assuming the existence of OWFs.
\end{theorem}
\begin{proof}
The proof follows identically to the one of \cref{th:hardnesspauli}. However, it uses the computational indistinguishability of the two ensembles of states $\tilde{\mathcal{E}}_{U}$ and $\tilde{\mathcal{E}}_{f,S}$, as opposed to the (stronger) statistical indistinguishability of $\mathcal{E}_{U}$ and $\mathcal{E}_{f,S}$.
\end{proof}

\subsection{Sampling algorithm}
\label{ssec:ancestral-sampling-algorithm}

We have seen that for "complex states", i.e., states with high entanglement and high magic, Pauli sampling cannot be performed efficiently. Now, we turn our attention to the complementary regime and will provide an efficient algorithm for Pauli sampling for states with bounded entanglement and magic.

\paragraph{Estimating marginals:}Our starting point for the description of the algorithm is the insight that Bell measurement data can be used to estimate marginals of the Pauli distribution $p_\rho$. In particular, in \Cref{sec:bell_basis_bell_sampling}, we have introduced the concept of Bell sampling as the measurement routine where we perform Bell measurements across two copies of a state $\rho$.
Now, recall that using the decomposition of the SWAP operator in \Cref{eq:swap_pauli_identity}, we can express the marginals as (c.f. \cref{eq:marginals_with_swap})
\begin{align}\label{eq:marginal_to_estimate}
p_{\rho}\left(x_{1},\dots,x_{k}\right) =p_{\rho}\left(x_{1:k}\right)=\frac{1}{2^{k}\tr\left(\rho^{2}\right)}\langle P_{x_{1:k}}^{\otimes2}\otimes\mathbb{S}^{\otimes n-k}\rangle_{\rho\otimes\rho}.
\end{align}
Here, we use the shorthand notation $x_{1:k}$ to denote $x_1,\dots,x_k$.
Since the Bell states are eigenstates of the SWAP operator $\mathbb{S}$ as well as of $P^{\otimes{2}}$ for any single-qubit Pauli $P$ with
eigenvalues $\pm1$, we can estimate any expectation value of the form $\langle P_{x_{1:k}}^{\otimes2}\otimes\mathbb{S}^{\otimes n-k}\rangle_{\rho\otimes\rho}$
to within $\epsilon$ precision using $O\left(1/\epsilon^{2}\right)$
Bell sampling outcomes. Furthermore, since the purity is given by $\trace\left(\rho^2 \right) = \langle \mathbb{S}^{\otimes n}\rangle_{\rho\otimes\rho}$, it can also be estimated from the Bell measurement data. Hence, we can obtain an estimate for the RHS of \Cref{eq:marginal_to_estimate} by estimating $\langle P_{x_{1:k}}^{\otimes2}\otimes\mathbb{S}^{\otimes n-k}\rangle_{\rho\otimes\rho}$ and $\trace\left(\rho^2 \right)$ from the Bell data. For more details on the estimation, we refer the reader to \Cref{app:estimation}.
Importantly, the same measurement data can be used to estimate all marginals. That is, we can collect the data once and then classically estimate $p_{\rho}\left(x_{1:k}\right)$ for any given string $x_{1:k}$ and $1\leq k\leq n$ in post-processing. In the following lemma, we record the sample complexity for simultaneously estimating all such marginals. We note that a similar observation about simultaneous estimation of all $\langle P_{x}^{\otimes2}\rangle_{\rho\otimes\rho}$ for all $x\in \{0,1\}^{2n}$ has made been already in Ref.\ \cite[Appendix E]{huangInformationtheoreticBoundsQuantum2021a}. Here, we extend this observation to observables also containing $\mathbb{S}$ factors.

\begin{lemma}[Estimating all even marginals]\label{lem:estimating_marginals}
Let $\epsilon>0,\delta>0$ and let $\rho$ be a pure state. Then, 
$N=O\left(n\log\left(1/\delta\right)/\epsilon^{2}\right)$  pairs
of copies $\rho\otimes\rho$ suffice to produce, with probability $1-\delta$, estimates $\pi_{\rho}(x_{1:k})$
such that 
\begin{equation}
\left|\pi_{\rho}(x_{1:k})-p_{\rho}\left(x_{1:k}\right)\right|\leq\frac{\epsilon}{2^{k}}
\end{equation}
for all $x\in \01^{2n}$ and all $1\leq k\leq n$.
\end{lemma}
The proof of this lemma is presented in \Cref{app:estimation}. There, we also present the slight generalization of this lemma to mixed states leading to an overall sample complexity of $N=O\left(n\log\left(1/\delta\right)/(\epsilon^2 \tr(\rho^2)^2 )\right)$. Note that the purity $\tr(\rho^2)$ enters into the sample complexity because the size of the marginals scales with the purity (c.f. \cref{eq:marginal_to_estimate}
).

Let us remark on a subtlety concerning \Cref{lem:estimating_marginals}. Recall that $x=(v_1,w_1, \dots ,v_n,w_n)\in \{0,1\}^{2n}$ such that $x_i=(v_i,w_i)$. Hence, the guarantees given in \cref{lem:estimating_marginals} only apply to \emph{even} marginals of $p_\rho$, i.e., those marginals that correspond to a Pauli string $P_{x_{1:k}}$ of length $k$. It turns out, however, that the algorithm we will use for sampling from the Pauli distribution requires also odd marginals. These correspond to bit-strings of length $2k-1$. Using the decomposition 
\begin{align}
    p_\rho(v_1,w_1, \dots ,v_{k}) &=p _\rho(v_1,w_1, \dots ,v_{k},w_k=0) \nonumber \\ 
    &+p_\rho(v_1,w_1, \dots ,v_{k},w_k=1)
\end{align}
we can compute estimates for the odd marginals from estimates of the even ones. We thus obtain the following result which applies to both even and odd marginals:
\begin{corollary}[Estimating even and odd marginals]\label{cor:estimating_all_marginals}
Let $\epsilon>0,\delta>0$ and let $\rho$ be a pure state. Then, $N=O\left(n\log\left(1/\delta\right)/\epsilon^{2}\right)$ pairs
of copies $\rho\otimes\rho$ suffice to produce, for all $1\leq k \leq n$ with probability $1-\delta$, 
\begin{itemize}
    \item  estimates $\pi_{\rho}(v_1,w_1, \dots ,v_k,w_k)$ such that \begin{equation}
\left|\pi_{\rho}(v_1,w_1, \dots ,v_k,w_k)-p_{\rho}(v_1,w_1, \dots ,v_k,w_k)\right|\leq\frac{\epsilon}{2^{k}}
\end{equation}
    \item estimates $\pi_{\rho}(v_1,w_1, \dots ,v_k)$ such that \begin{equation}\label{eq:additive_marginals}
\left|\pi_{\rho}(v_1,w_1, \dots ,v_k)-p_{\rho}(v_1,w_1, \dots ,v_k)\right|\leq\frac{2\epsilon}{2^{k}}
\end{equation}
\end{itemize}
for all $x=(v_1,w_1, \dots ,v_n,w_n)\in \{0,1\}^{2n}$ and all $1\leq k\leq n$.
\end{corollary}

\paragraph{Ancestral sampling:}The algorithm we propose for Pauli sampling is based on the so-called ancestral sampling algorithm and, in particular, a variant thereof put forward in  Ref.\ \cite{bremnerAchievingQuantumSupremacy2017a}. We will begin by reviewing the standard algorithm and the adaptation thereof: Consider a distribution $p$ over $\01^n$. The ancestral sampling algorithm makes use of the fact that the joint distribution $p\left(x\right)$ can be factorized in terms of conditionals as follows,
\begin{equation}
p\left(x\right)=p\left(x_{1}\right)p\left(x_{2}|x_{1}\right)p\left(x_{3}|x_{1},x_{2}\right)\dots p\left(x_{n}|x_{1},\dots,x_{n}\right) \,.
\end{equation}
The sampling algorithm proceeds by sampling a bit-string $x$ in a bit-by-bit fashion. That is, it samples $x_{1}$ according to $p\left(x_{1}\right)$,
then samples $x_{2}$ according to $p\left(x_{2}|x_{1}\right)$, where
$x_{1}$ was fixed by the previous step, and so on and so forth, until
$x$ is completely sampled. Furthermore, we can express the conditional probabilities in terms of marginals 
\begin{equation}
    p\left(x_{k}|x_{1},\dots,x_{k-1}\right)=\frac{p\left(x_{1},\dots,x_{k}\right)}{p\left(x_{1},\dots,x_{k-1}\right)} \,.
\end{equation}
Hence, the ancestral algorithm can be run whenever we have access to marginals of the target distribution $p$ that we want to sample from.

\subsection{Adapted ancestral sampling}
In previous works featuring the ancestral sampling algorithm, access to these marginals $p(x_1,\dots,x_k)$ is by some explicit computation. For instance, in the classical simulation literature, the marginals are computed from say the circuit description of the quantum circuit to be simulated (see, e.g., Refs.\  \cite{bremnerAchievingQuantumSupremacy2017a,cliffordClassicalComplexityBoson2018,bravyiHowSimulateQuantum2022}). In contrast, in the context of our work, these marginals are estimated from measurement data as outlined above. Let $\pi(x_{1:k})$ denote such an estimate for $p(x_{1:k})$. Then, in particular, the approximation guarantees of \cref{lem:estimating_marginals} and \Cref{cor:estimating_all_marginals} apply to $\pi(x_{1:k})$. 

\begin{figure}[t]
     \centering
    \includegraphics[trim={0.1cm 0.1cm 0.1cm 0.1cm},clip,width=0.5\textwidth]{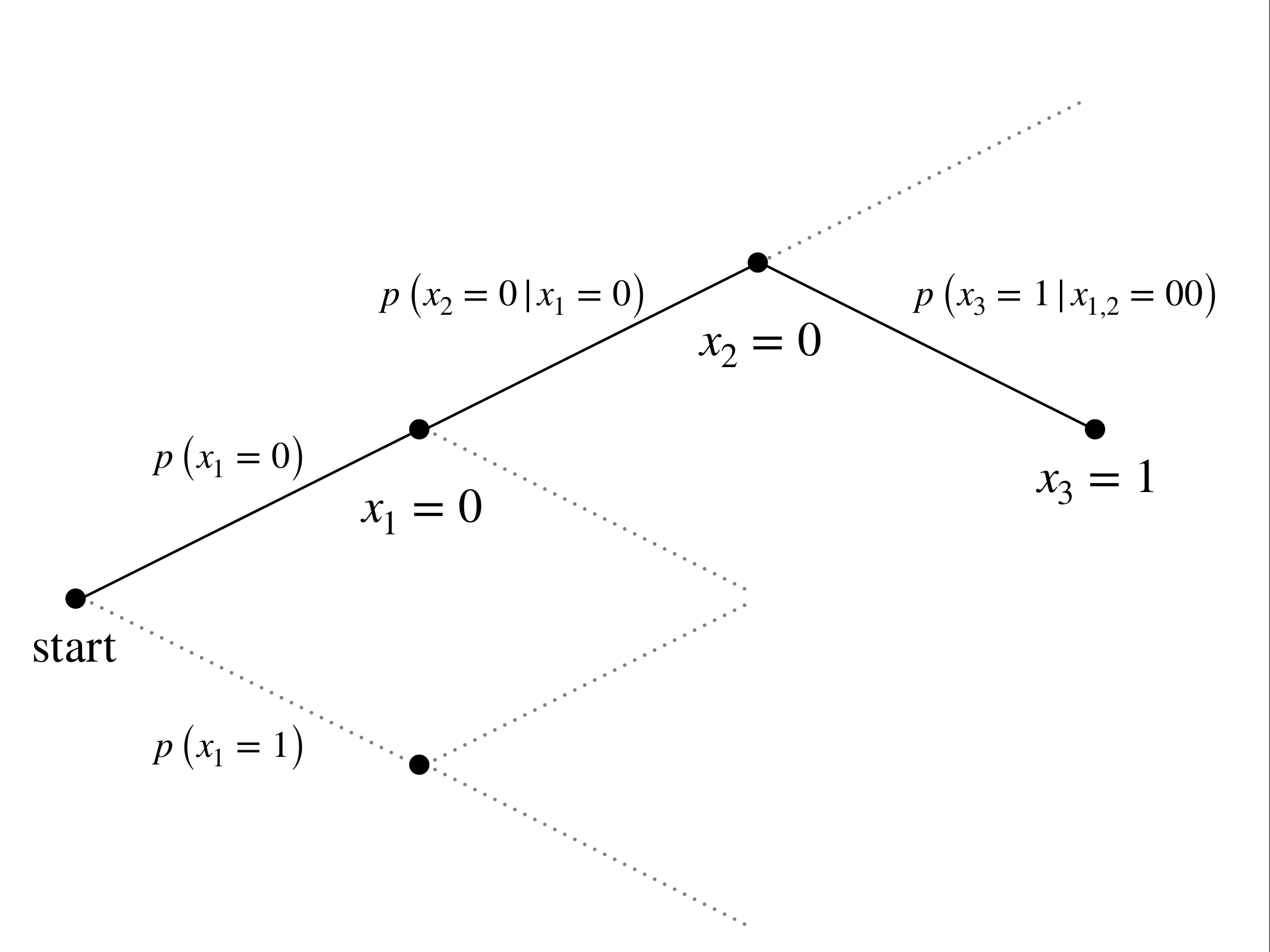}
    \caption{Schematic example run of the ancestral sampling algorithm. The algorithm can be visualized as a walk through a binary tree with its output corresponding to the entire path taken. The $k^{th}$ layer of the tree corresponds to the $k^{th}$ bit and each node corresponds to either of the (binary) values assigned to the corresponding bit. For example in the second layer, the upper node corresponds to $x_2=0$ and the lower to $x_2=1$. The algorithm proceeds by traversing the tree from left to right, according to the conditional probabilities $p(x_k|x_1,\dots, x_{k-1})$. In the figure, we see in dark color the example path giving rise to the sample $x=0001$ while the dotted lines show (parts) of other potential paths that could have been taken leading to different outputs.}
    \label{fig:ancestral_sampling_tree}
 \end{figure}
 
Note that, in general, additive error estimates $\pi(x_{1:k})$ of the marginal probabilities can take negative values. This would be an issue when trying to run the standard ancestral sampling algorithm based on such possibly negative estimates. To deal with this issue, we turn to a variant of the ancestral sampling algorithm put forward in Ref.\ \cite{bremnerAchievingQuantumSupremacy2017a} which we refer to as the \emph{adapted ancestral sampling algorithm}. Their adapted algorithm can be run also based on possibly negative approximations to the marginals. 

To allow such negative inputs, the algorithm differs in one aspect from the standard ancestral algorithm discussed above: Recall that in the $k^{th}$ step of the ancestral algorithm, we set the $k^{th}$ bit to $0/1$ according to the probabilities $p(x_{1:k-1}, x_k =0)$ and $p(x_{1:k-1}, x_k =1)$, both of which are non-negative. With the adapted algorithm, taking the approximations $\pi(x_{1:k})$ as input, there are now two possible cases:
\begin{enumerate}
    \item Either both estimates are non-negative,  ${\pi(x_{1:k-1},x_k=0)}, {\pi(x_{1:k-1},x_k=1)\geq 0}$, 
    \item or one of the two estimates is negative, so  $\pi(x_{1:k-1},x_k=0)<0$ or $\pi(x_{1:k-1},x_k=1)<0$.
\end{enumerate}
In the first case, the algorithm functions in the same way as the standard ancestral sampling algorithm described above, namely it sets $x_k=0$ with probability given by the ratio 
\begin{equation}
    \mathfrak{p} = \frac{\pi(x_{1:k-1},x_k=0)}{\left(\pi(x_{1:k-1},x_k=0) + \pi(x_{1:k-1},x_k=1) \right)}\,, 
\end{equation}
and it set $x_k = 1$ with probability $1-\mathfrak{p}$.
In the second case, however, if $\pi(x_{1:k-1},x_k=0)<0$ the algorithm sets $x_k=1$ (and respectively if $\pi(x_{1:k-1},x_k=1)<0$, $x_k=0$). In the pictorial representation of \cref{fig:ancestral_sampling_tree}, this corresponds to deterministically taking the step along the path corresponding to the positive estimate of the marginal. 

Below, we will provide a set of sufficient conditions for approximate Pauli sampling. We will first state these conditions with respect to an arbitrary distribution $p$ over $\01^n$ as we believe that our performance guarantees can be of independent interest for other settings where one wants to sample a distribution $p$ given access to additive error estimates of its marginals. We then proceed to reconnect to Pauli sampling by considering Pauli distributions $p_\rho$ We, then give sufficient conditions on properties of the state $\rho$ in terms of its magic and entanglement to give a characterisation of efficiently sample-able states.  

To state our performance guarantee, we need one additional definition:
\begin{definition}[Corresponding marginals and $\mathfrak{f}(\gamma)$]\label{def:set-s-gamma}
     Let $p$ be a distribution over $\01^n$ and let $\gamma>0$. Then, for a fixed outcome $x\in \01^{n}$, we say that the marginals $p(x_{1:k})$ for $k\in[n]$ are the marginals \emph{corresponding} to $x$. Now, consider the following set: 
    \begin{equation}
        S_\gamma = \left\{x\in \01^n : \forall k\in \left[n \right], \, p(x_{1:k})\geq \frac{\gamma}{2^{k}}\right\}
        .
    \end{equation}
    In words, an outcome $x$ is in $S_{\gamma}$, if all its corresponding marginals are lower bounded as $p(x_{1:k})\geq \frac{\gamma}{2^{k}}$.
    We define 
    \begin{equation}
        \mathfrak{f}(\gamma):=1 - p(S_\gamma) \,.
    \end{equation} 
\end{definition}

To exemplify this definition, take for example the outcome $x=001$ then $p(x_1=0),p(x_1=0,x_2=0), p(x_1=0,x_2=0,x_3=1)$ are its corresponding marginals.

We are now in the position to state the main theorem of this section.

\begin{theorem}[Performance guarantee of adapted ancestral sampling algorithm]\label{thm:sampling_conditions_marginals}
Let $p$ be a distribution over $\01^n$ and $\gamma>0$. For all $\eps < \gamma /2$, given black-box access to estimates of the marginals such that
\begin{equation}\label{eq:condition_on_marginals_thm_sampling}
    | \pi(x_{1:k})-p(x_{1:k}) |\leq \frac{\epsilon}{2^{k}} \,,
\end{equation}
the adapted ancestral algorithm samples from a distribution $q$ which satisfies
\begin{equation}\label{eq:tv-distance-error}
    \left \Vert q-p \right \Vert_{\mathrm{TV}}\leq  \mathfrak{f}(\gamma) + \exp(\frac{4 \epsilon n}{\gamma}) -1 \, . 
\end{equation}
\end{theorem}
Note that the overall TV distance can be made arbitrarily small for any given distribution $p$ by choosing $\gamma$ sufficiently small and further choosing $\eps = O\left(\frac{\gamma}{n}\right)$. In particular, this choice allows one to suppress the second error term on the RHS of \Cref{eq:tv-distance-error} arbitrarily. For this choice of $\epsilon$, we will also apply the theorem in the subsequent section.

A detailed proof of \Cref{thm:sampling_conditions_marginals} is presented in \Cref{sec:Appendix_pauli_sampling_algorithm}. The key insight underlying the proof is that the additive errors on the estimates $\pi(x_{1:k})$ of the marginals correspond to multiplicative errors if the marginals are sufficiently large. This lower bound on the size of the marginals is the purpose of \Cref{def:set-s-gamma}. On the other hand, the outcomes $x\not\in S_{\gamma}$ whose corresponding marginals are not sufficiently large contribute to the sampling error as measured by TV distance. These outcomes are captured by the error term $\mathfrak{f}(\gamma)$.

\subsection{Efficiently sampleable states} \label{ssec:efficiently-samplable-states}

In \cref{thm:sampling_conditions_marginals}, we have established sufficient conditions for the approximate sampling from a distribution when given estimates to its marginals. In this section, we want to establish sufficient conditions on properties of a pure state $\rho$ that would result in efficient approximate Pauli sampling. In particular we will show that as long as the state $\rho$ has bounded entanglement and magic, we can approximately Pauli sample from $p_\rho$ efficiently using the adapted ancestral sampling algorithm discussed in the previous section. 

First, let us explain our approximate Pauli sampling algorithm. It proceeds along the following steps:
\begin{enumerate}
    \item \textbf{Data acquisition:} Use $2N$ copies of the unknown state $\rho$ and perform Bell sampling on pairs of copies $\rho^{\otimes 2}$ obtaining a list of $N$ outcomes, namely $\{ y_1,\dots, y_{N}\in \01^{2n} \}$.
    \item \textbf{Classical post-processing:} Run the adapted ancestral sampling algorithm on the estimates $\pi_\rho(x_{i:k})$ of the marginals $p_{\rho}(x_{i:k})$ to produce samples $x \in \01^{2n}$. As discussed in \Cref{cor:estimating_all_marginals}, all required marginals can be estimated from the Bell measurement data obtained in the data acquisition phase.
\end{enumerate}

The ancestral sampling algorithm samples a bit-string $x\in \01^{2n}$ in a bit-by-bit fashion. Alternatively, since each pair of bits $x_i=(v_i,w_i)$ corresponds to a single-qubit Pauli $P_{x_i}$, the algorithm can be viewed as sampling a full $n$-qubit Pauli string $P_x$ by drawing single-qubit Paulis qubit-by-qubit. In each pass of the ancestral algorithm, we have to estimate the marginal probabilities $p_{\rho}(x_1), p_{\rho}(x_1,x_2)$, etc., which correspond to bi-partitions $(1|2, \dots ,n), (1,2|3, \dots ,n)$, etc. of the system of $n$ qubits. It becomes clear then that the algorithm presupposes a certain qubit ordering, i.e., an assignment of which qubit is sampled first, which second, and so on. This choice of ordering can be crucial for the efficiency of the algorithm in our case. To see that, note that every ordering gives rise to a different sequence of bi-partitions. 
By \Cref{lem:bounding_marginals_with_entanglement}, we know that the size of the marginals crucially depends on the entanglement across these bi-partitions. Hence, by \cref{thm:sampling_conditions_marginals}, a path giving rise to bi-partitions with little entanglement across is necessary in order for the sampling algorithm to be efficient. 

To make this observation rigorous, we give the following definition:

\begin{definition}[Qubit ordering]
    We identify a qubit ordering with a permutation $\pi \in S_n$ where $S_n$ is the symmetric group. It acts on state vectors via 
    \begin{equation}
        \pi \ket{a_1, \dots, a_n} = \ket{a_{\pi(1)}, \dots, a_{\pi(n)}} \,.
    \end{equation}
    Hence, given a state $\rho$ with respect to the canonical ordering corresponding to the identity permutation, the state with respect to a different ordering is given by $\pi \rho \pi^{\dagger}$.  
\end{definition}

 Next, we define a measure of (multi-partite) entanglement that is ordering-specific and captures entanglement across the sequence of bi-partitions $(1|2, \dots ,n), (1,2|3, \dots ,n), \dots, (1,\dots,n-1|n)$: 
\begin{definition}[Path entanglement]\label{definition:path_entanglement}
Let $\rho$ be an $n$-qubit pure state and let $\pi\in S_n$ specify a qubit ordering. Then we define the \emph{path entanglement} of $\rho$ with respect to this ordering $\pi$ as
\begin{equation}
    E_{0}^{\pi}(\rho)=\max \left( \mathcal{S}_0(\rho_{\pi(1)}), \mathcal{S}_{0}(\rho_{\pi(1),\pi(2)}), \dots ,\mathcal{S}_{0}(\rho_{\pi(1),\dots,\pi(n-1)}) \right).
\end{equation}
Here, $\rho_{\pi(1)}, \rho_{\pi(1),\pi(2)}, \dots$ denote reduced states.
\end{definition}
In words, this is the maximum Rényi-0 entanglement entropy across all bi-partitions encountered by the ancestral sampling algorithm when a certain qubit-ordering is fixed. 

To convey the importance of the choice of ordering, we present a few illustrative examples of states:
\begin{enumerate}
    \item \textbf{Product states}: in the case of product states (e.g., $\ket{0}^{\otimes n}$), the Pauli distribution $p_{\rho}$ takes a product form such that $p_{\rho}(x) = p_{\rho}(x_1) \ldots p_{\rho}(x_n)$, which makes the choice of ordering irrelevant. Accordingly, the entanglement entropy is $0$ along all bi-partitions.
    \item \textbf{Bell pairs}: an interesting case where the choice of ordering can render our algorithm efficient or inefficient is that of $n/2$ Bell pairs, i.e., $\rho = \ket{\Phi_{n}^{+}} \bra{\Phi_{n}^{+}}$
    (see \cref{eq:2n-Bell-state}). Here, the Bell pairs are between qubits $(1,n/2+1), (2, n/2+2),\dots,(n/2,n).$ In this case, the canonical ordering, corresponding to sampling qubit 1, then 2, then 3 and so on, leads to large path entanglement and hence requires an exponential number of Bell sampling data. However, choosing the ordering $\pi$ to be matching up with the pairing of the Bell pairs, such that one samples first qubit 1, then $n/2 +1$, then 2, then $n/2 + 2$, and so on, leads to a path entanglement $E_0^{\pi}(\rho)=2$ and hence to an efficient algorithm.
    \item \textbf{2D cluster states}: another interesting case is that of 2D cluster states \cite{briegelPersistentEntanglementArrays2001}
    \begin{equation}
        \ket{G} = \prod_{(i,j)\in E} \text{CZ}_{i,j} \bigotimes_{k \in V} \ket{+}_k,
    \end{equation}
    where $V,E$ are the vertices and edges of a $\sqrt{n}\times\sqrt{n}$ square lattice $G$. For these states, all choices of orderings lead to a path entanglement of the order $E_0^{\pi}(\rho) = O(\sqrt{n})$, which renders our algorithm inefficient. One way to see this is by considering the bi-partition at the half-way point of the ordering, i.e., $(\pi(1),\dots,\pi(n/2) : \pi(n/2+1),\dots,\pi(n)) = (A:B)$. Regardless of how the $n/2$ qubits in $A$ are chosen, the entanglement entropy corresponding to this bi-partition is exactly the number of edges in $E$ at the boundary between $A$ and $B$ in the lattice $G$. Moreover, it is easy to see that any choice of ordering will lead to a boundary of size $O(\sqrt{n})$ at this half-way point.
\end{enumerate}

Now, we finally state our performance guarantee for the approximate Pauli sampling algorithm described above.

\begin{theorem}[Performance guarantee for the approximate Pauli sampling algorithm]\label{theorem:main_theorem_sampling_bounded_magic_entanglement}
    Let $1>\Delta>0, \delta>0$. Let $\rho$ be a pure, $n$-qubit state with magic $M_0(\rho)$ (or $M_1(\rho)$) and let there be a qubit ordering $\pi\in S_n$ with path entanglement $E_0^{\pi}(\rho)$. Then, there exists an algorithm sampling from a distribution $q$ such that
    \begin{equation}
        \left \Vert q-p_{\rho}\right \Vert_{\mathrm{TV}} \leq \Delta \, ,
    \end{equation}
    with probability at least $1-\delta$ provided that
    \begin{equation}\label{eq:sample-complexity-M0}
     N=  O\left(\frac{n^3 2^{2E_0^{\pi}(\rho)} 2^{2M_0(\rho)} \log(1/\delta)}{\Delta^4 }\right)\,,
    \end{equation}
    or alternatively,
     \begin{equation}\label{eq:sample-complexity-M1}
     N= O\left(\frac{n^3 2^{2E_0^{\pi}(\rho)} 2^{4M_1(\rho)/\Delta} \log(1/\delta)}{\Delta^4 }\right) \,.
    \end{equation}

Bell samples of the state $\rho^{\otimes2}$ are taken. Producing a single sample takes  $O(Nn^2)$ time in classical post-processing on the obtained Bell measurement data. The success probability of $1-\delta$ is with respect to the randomness in the outcomes of the Bell measurements.
\end{theorem}

\begin{proof}[Proof of \Cref{theorem:main_theorem_sampling_bounded_magic_entanglement}]
    The general idea going into proving the theorem is that states with bounded magic give rise to Pauli distributions supported on ``few'' elements whose respective probabilities are relatively large. Concretely, we obtain sufficient conditions on the state $\rho$ such that the conditions of \cref{thm:sampling_conditions_marginals} are satisfied, namely we show that:
    \begin{itemize}
        \item An upper bound on the magic of the state translates into a lower bound on the overall probability weight distributed among Paulis with high expectation values $\alpha_{\rho}(x)^2$. Put differently, it translates into an upper bound on the CDF associated to the state, $F_{\rho}(\tau)$ (see \Cref{lem:bounding_probabilities_with_M0,lem:bounding-probabilities-via-M1}).
        
        \item An upper bound on the path entanglement leads to a lower bound on the marginals of the Pauli distribution (see \cref{lem:bounding_marginals_with_entanglement}).   
    \end{itemize}

    Let $R:=2^{E_0^{\pi}(\rho)}$, it is the maximal Schmidt rank across the set of bi-partitions corresponding to the assumed qubit ordering $\pi$. Now, consider the set 
    \begin{equation}
        S=\{x\in \01^{2n}: \tr\left(\rho P_x \right)^2 > \gamma R \} \,.
    \end{equation}
    By \Cref{lem:bounding_marginals_with_entanglement}, for all $x\in \01^n$ in $S$, all their corresponding marginals satisfy, $p_{\rho}(x_{1:k})\geq \gamma /2^k$.  Hence, $S \subseteq S_{\gamma}$ where $S_{\gamma}$ has been defined in \Cref{def:set-s-gamma} and also $\mathfrak{f}(\gamma)=1-p_\rho(S_{\gamma})$. This implies $p_\rho(S) \leq p_\rho(S_{\gamma})$ and therefore $\mathfrak{f}(\gamma)\leq 1 - p_\rho(S)$.

    By \Cref{lem:bounding_probabilities_with_M0}, we have that
    \begin{eqnarray}
       1 - p_\rho(S) = F_\rho(\gamma R) \leq 2^{M_0(\rho)} \gamma R \\ 
       \Rightarrow \quad  \mathfrak{f}(\gamma) \leq 1 - p_\rho(S) \leq 2^{M_0(\rho)}\gamma R.\nonumber
    \end{eqnarray}
     Similarly, by \Cref{lem:bounding-probabilities-via-M1}, we find that
    \begin{eqnarray}
       1-p_\rho (S) = F_\rho(\gamma R) \leq \frac{M_1(\rho)}{\log (\frac{1}{\gamma R})} \\ 
       \Rightarrow \quad \quad \mathfrak{f}(\gamma) \leq 1 - p_\rho(S) \leq  \frac{M_1(\rho)}{\log (\frac{1}{\gamma R}) .}
       \nonumber
    \end{eqnarray}
    
    From \Cref{thm:sampling_conditions_marginals}, we have that the adapted ancestral sampling algorithm samples from a distribution $q$ within TV distance
    \begin{equation}
         \left \Vert q-p_{\rho}\right \Vert_{\mathrm{TV}}  \, \leq \mathfrak{f}(\gamma) + \exp(\frac{4 \epsilon n}{\gamma}) -1 \,.
    \end{equation}
    So, to obtain the bound in terms of $M_0(\rho)$ in \Cref{eq:sample-complexity-M0}, we choose
    \begin{equation}\label{eq:bound-M0}
        \gamma = \frac{\Delta }{2\cdot 2^{M_0(\rho)}\,R} \quad , \ \eps \leq \frac{ \gamma \, \Delta }{4\cdot 4 n} \,,
    \end{equation}
    or alternatively, to obtain the bound in terms of $M_1(\rho)$ in \Cref{eq:sample-complexity-M1}, we choose
    \begin{equation}\label{eq:bound-M1}
        \gamma = \frac{1}{2^{2M_1(\rho)/\Delta}R} 
     \quad , \ \eps \leq \frac{ \gamma \, \Delta }{4\cdot 4 n} \,.
    \end{equation}
    These choices guarantee that $ \mathfrak{f}(\gamma) \leq \frac{\Delta}{2}$ and $\exp(\frac{4 \epsilon n}{\gamma}) \leq \exp(\frac{\Delta}{4}) \leq 1 + \frac{\Delta}{2}$ (using $\Delta < 1$ and $\exp(x) \leq 1 + 2x$ for $x \in [0,1]$), and hence $\left \Vert q-p_{\rho}\right \Vert_{\mathrm{TV}} \leq \Delta$.

    Finally, by \Cref{cor:estimating_all_marginals}, the number of copies $N$ required to obtain $\epsilon/2^{k}$ additive estimates for all marginals with probability $1-\delta$, is given  as $N=O\left(n\log\left(1/\delta\right)/\epsilon^{2}\right)$.
    Hence, 
    \begin{equation}
        N = O\left(\frac{n^3\log(1/\delta)}{\Delta^2\gamma^2}\right),
    \end{equation}
    which, combined with \cref{eq:bound-M0,eq:bound-M1}, gives the stated complexities for $N$. 

    For the time complexity bound, note that, to produce a single sample $x$ via the algorithm, $O(n)$ marginals have to be estimated from the Bell measurement data. Following the proof of \cref{lem:estimating_marginals} in \Cref{app:estimation}, in particular \Cref{eq:marginal-estimator}, we see that each estimated marginal is written as a sum of $N$ terms, each a product of $n$ numbers, 
and can be computed in time $O(nN)$. Hence, the total time is of the order $O(n^2 N)$.
\end{proof}

Note that the sample complexity of the algorithm scales exponentially with both the magic and the entanglement. Nonetheless, we can efficiently sample from the Pauli distribution of $\rho$ as long as it is a state with magic and entanglement bounded by $O(\log(n))$.
\begin{corollary}\label{corollar:sampleable_states}
    Let $\mathcal{C}$ be the class of $n$-qubit pure state such that for every $\rho \in \mathcal{C} $ we have $M_0(\rho)=O(\log n)$ and $E_0^{\pi}(\rho)=O\left( \log (n)\right)$ for all qubit orderings $\pi \in S_n$. Then there exists an approximate Pauli sampling algorithm, consuming $\poly (n)$ Bell measurements,  with success probability at least $1-\delta$, for $\mathcal{C}$.
\end{corollary}

Note that a stronger statement is possible if we are willing to drop computational efficiency. Namely, it suffices that there exists \emph{some} ordering such that $E_0^\pi(\rho)=O\left(\log (n) \right)$. The reason we cannot provide guarantees on the computational complexity in this case is because it might be computationally inefficient to find such an ordering even if it exists. 

\subsection{Beyond pure states}
 
In the previous section we have established sample complexity upper bounds for Pauli sampling of pure states in terms of their magic and entanglement. In this section, we argue that our algorithm extends naturally to mixed states. However, the characterization in terms of magic and entanglement gets more difficult.\newline
First, we emphasize that the key ingredient to our bound derived in the previous section, \cref{thm:sampling_conditions_marginals}, applies to any distribution $p$. In particular, \Cref{thm:sampling_conditions_marginals} considers as input the access to estimates $\pi(x_{1:k})$ of the marginals $p(x_{1:k})$ of the distribution $p$ that are within an additive error of at most $\epsilon/2^k$. As we show in \Cref{lem:marginals_mixed}, in case of Pauli distributions $p_\rho$, we can obtain such estimates for both pure and mixed states, using Bell measurements, with an overhead in the sample complexity that scales inversely with the purity of the state $\rho$. Hence, \Cref{thm:sampling_conditions_marginals} can be readily applied to Pauli distributions of mixed states as well.\newline 
However, on a high level, \Cref{thm:sampling_conditions_marginals} captures the finding that the complexity of our algorithm depends on the size of the marginals of the distribution in question. For pure states we were able to connect the size of the marginals of the Pauli distribution $p_\rho$ to the entanglement and magic of $\rho$ through \cref{lem:bounding_marginals_with_entanglement} and \cref{lem:bounding_probabilities_with_M0,lem:bounding-probabilities-via-M1}, respectively. Here, we made use of the fact that there are pure state entanglement and magic measures given directly in terms of the Pauli distribution $p_\rho$. For mixed states, however, it is not clear how to quantify magic and entanglement in terms of the Pauli distribution $p_\rho$ anymore (see e.g. \cite{plenioIntroductionEntanglementMeasures2007} for a review on mixed-state entanglement measures and \cite{liuManybodyQuantumMagic2022c, leoneStabilizerEntropiesAre2024} for mixed state measures of magic). Hence, we also lack generalizations of the above-mentioned lemmas to mixed states. We leave this direction to future work.\newline
While in the mixed state case the connection of the Pauli distribution to magic and entanglement is not so clear, we emphasize that we can still analyze the complexity of our approximate Pauli sampling algorithm in terms of the Pauli distribution $p_\rho$ of a mixed state $\rho$ directly. This will be sufficient for many practical purposes, for instance, when trying to understand the robustness of our algorithm to experimental imperfections. To exemplify, we demonstrate how to analyze the complexity of our Pauli sampling algorithm when applied to copies of a mixed state $\rho$ which we assume to be the result of a Pauli noise channel $\Lambda$ acting on an ideal pure state $\sigma$. That is, $\rho=\Lambda(\sigma)$ and we are interested in sampling from $p_\rho$. In this case, we can easily relate $p_\rho$ to $p_\sigma$ and hence obtain a performance guarantee in terms of the pure input state $\sigma$ and the noise parameter of the Pauli channel.

\begin{theorem}\label{thm:robustness_to_Pauli_noise}
    Let $\sigma$ be a pure state and let $\Lambda$ be a Pauli-channel such that
     \begin{equation}
     \rho:=\Lambda(\sigma)= (1-\xi)\sigma+\sum_{y\in \01^{2n}\setminus\{ 0^{2n}\}} \xi_y P_y \sigma P_y,   
     \end{equation}
     where for all $y$, we have that $\xi_y\geq0$, and $\sum_y \xi_y = \xi$ and $0\leq\xi\leq1$. Then, there exists an algorithm for sampling, with probability at least $1-\delta$, from a distribution $q_{\rho}$ such that
     \begin{equation}
         \Vert p_{\rho}-q_{\rho} \Vert_{TV}\leq \Delta,
     \end{equation}
      for any $\Delta>4\xi$, using
     \begin{equation}
         N = N_{\sigma}/(1-2\xi)^4
     \end{equation}
     Bell samples of the state $\rho^{\otimes 2}$. Here, $N_\sigma$ denotes the number of Bell samples necessary to achieve the same TV distance $\Delta$ if  the algorithm had been applied to the ideal pure state $\sigma$ instead of $\rho$. 
 \end{theorem}

Note that $N_{\sigma}$ can be read off from \Cref{theorem:main_theorem_sampling_bounded_magic_entanglement}. We can obtain \cref{thm:robustness_to_Pauli_noise} since we can directly relate the Pauli distributions $p_\rho$ and $p_\sigma$ via $p_{\rho}(x)\geq (1-\xi)^2 p_{\sigma}(x)$ which holds for all $x$.
The full proof of \Cref{thm:robustness_to_Pauli_noise} is deferred to \Cref{ssec:robustness-Pauli-noise}.

\section{Distributed inner product estimation}
In this section, we focus on the task of distributed inner product estimation (IP): estimating the overlap $\tr(\rho\sigma)$ between states $\rho$ and $\sigma$, with the states $\rho,\sigma$ being distributed between two parties that can only use local quantum operations and classical communication. 

One strategy for this task relies on either of the two parties (say Alice) learning a full classical representation of her unknown state $\rho$ and sending it to the other (Bob). He could then perform direct fidelity estimation \cite{flammiaDirectFidelityEstimation2011b} in order to estimate $\trace(\rho \sigma)$. Alternatively, he could also learn a classical representation of his state $\sigma$ and then compute the overlap classically based on the classical representations of $\rho$ and $\sigma$. Both approaches only result in efficient protocols when applied to certain classes of states such as stabilizer states, $t$-doped stabilizer states and states that admit efficient MPS representations. Other approaches based on randomized measurements have been proposed \cite{elbenCrossPlatformVerificationIntermediate2020}, and 
in
Ref.\ \cite{anshuDistributedQuantumInner2022}, general lower bounds on the sample complexity of distributed inner product estimation have been proven.

Following the analysis of 
Ref.\ \cite{anshuDistributedQuantumInner2022} we begin by providing a lower bound on the sample complexity of IP even for restricted classes of states, namely those with $\omega( \log (n))$ entanglement and magic. We further provide two protocols for IP based on coordinated Pauli measurements on the states $\rho,\sigma$ and provide sufficient conditions for the protocols to be efficient.

\subsection{Hardness of IP for states with large magic and entanglement}

We are mainly interested in the \emph{inner product} (IP) estimation problem defined in \Cref{def:IP}. However, for establishing lower bounds, it will be convenient to work with a decision-version of this problem. Again, following the 
formulation of Ref.\ \cite{anshuDistributedQuantumInner2022}, we define the \emph{decisional inner product estimate problem}, abbreviated as DIPE.

\begin{definition}[Decisional inner product estimation, DIPE]
\label{def:DIPE}
Alice and Bob are each given $k$ copies of a pure state in $\mathbb{C}^d$. They are promised that one of the following two cases hold:
\begin{enumerate}
    \item Alice and Bob both have $\ket{\phi}^{\otimes k}$, where $\ket{\phi}$ is a uniformly random state vectors from an ensemble $\E$.
    \item Alice has $\ket{\phi}^{\otimes k}$ and Bob has $\ket{\psi}^{\otimes k}$, where $\ket{\phi}$ and $\ket{\psi}$ are independent uniformly random state vectors in $\E$.
\end{enumerate}
Their goal is to decide which case they are in with success probability at least $2/3$, using an interactive protocol that involves local quantum operations and classical communication.
\end{definition}

Anshu \emph{et al.} \cite{anshuDistributedQuantumInner2022} considered an instance of DIPE where the ensemble $\E$ is taken to be all quantum states in $\mathbb{C}^d$. For this general instance, they have shown a $k=\Omega(\sqrt d)$ sample complexity lower bound:

\begin{theorem}[Lower bound on DIPE, Theorem 4 in Ref.\ \cite{anshuDistributedQuantumInner2022}]\label{thm:DIPE-lower-bound}
    $k=\Omega(\sqrt{d})$ copies are necessary for Alice and Bob to solve DIPE, when they are allowed arbitrary interactive protocols (or arbitrary LOCC operations).
\end{theorem}

Then, via a chain of reductions, they arrive at a sample complexity lower bound $k=\Omega(\sqrt{d}/\epsilon)$ to solve IP up to additive error $\epsilon$, assuming that the input state vectors $\ket{\psi}, \ket{\phi}$ are unrestricted in $\mathbb{C}^d$.

We generalize these lower bounds, both for DIPE and IP, to the restricted case where the states $\ket{\psi}, \ket{\phi}$ have bounded magic and entanglement.

We first consider an instance of DIPE where the ensemble $\E$ is taken to be the subset phase states $\E_{f,S} = \left\{ \ket{\psi_{f,S}}=\frac{1}{\sqrt{|S|}}\sum_{x\in S}(-1)^{f(x)}\ket{x} \right\}_{f,S}$ defined in \Cref{sec:subset-phase-states}. We call this instance DIPE*. Showing hardness of DIPE* is rather straightforward, as it follows from the hardness of DIPE for the Haar measure and the statistical (and computational) indistinguishability of the ensemble $\E_{f,S}$ from the Haar measure:

\begin{theorem}[Lower bound on DIPE*]\label{thm:DIPE*-lower-bound}
    $k=\omega(\text{poly}(n))$ copies are necessary for Alice and Bob to solve DIPE*, when they are allowed arbitrary interactive protocols (or arbitrary LOCC operations).
\end{theorem}
\begin{widetext}
\begin{proof}
We prove this statement by contradiction. Assume there exists an interactive protocol to solve DIPE* using $k=O(\text{poly}(n))$ copies. Call this protocol $\A$. Then
\begin{align}
   & \Big|\Pr_{\substack{\ket{\psi}\sim\E_{f,S}\\ \ket{\phi}\sim\E_{f,S}}} [\A(\ket{\psi}^{\otimes k},\ket{\phi}^{\otimes k})=1]
    - \Pr_{\substack{\ket{\psi}\sim\text{Haar}\\ \ket{\phi}\sim\text{Haar}}} [\A(\ket{\psi}^{\otimes k},\ket{\phi}^{\otimes k})=1]\Big| \\
    &\leq \Big|\Pr_{\substack{\ket{\psi}\sim\E_{f,S}\\ \ket{\phi}\sim\E_{f,S}}} [\A(\ket{\psi}^{\otimes k},\ket{\phi}^{\otimes k})=1] - \Pr_{\substack{\ket{\psi}\sim\text{Haar}\\ \ket{\phi}\sim\E_{f,S}}} [\A(\ket{\psi}^{\otimes k},\ket{\phi}^{\otimes k})=1]\Big| \nonumber \\
    &\quad + \Big|\Pr_{\substack{\ket{\psi}\sim\text{Haar}\\ \ket{\phi}\sim\E_{f,S}}} [\A(\ket{\psi}^{\otimes k},\ket{\phi}^{\otimes k})=1] - \Pr_{\substack{\ket{\psi}\sim\text{Haar}\\ \ket{\phi}\sim\text{Haar}}} [\A(\ket{\psi}^{\otimes k},\ket{\phi}^{\otimes k})=1]\Big| \nonumber \\
    &\leq 2\,\text{negl}(n) \nonumber
\end{align}
using the triangular inequality and the computational indistinguishability of $\E_{f,S}$ from the Haar measure (a consequence of their statistical indistinguishability, see \cref{ssec:statistical-computational-indistinguishability}).
This would imply that $\A$ would also solve DIPE for Haar random inputs using $k=O(\text{poly}(n))$ copies, therefore contradicting \Cref{thm:DIPE-lower-bound}.
\end{proof}
\end{widetext}

From here, we extend this hardness result to IP*, the analogue instance of IP where input states are restricted to have bounded magic and entanglement. This is rather straightforward in the regime of additive error $\epsilon \in \left(0,\frac{1}{2}-O\left(\frac{1}{\text{poly(n)}}\right)\right)$:

\begin{lemma}\label{lem:IP*-lower-bound}
    Suppose Alice has input $\ket{\phi}^{\otimes k}$ and Bob has input $\ket{\psi}^{\otimes k}$, for arbitrary unknown pure state vectors $\ket{\phi},\ket{\psi}$ with bounded magic and entanglement, i.e., $M_\alpha(\psi) < f(n)$ (with $\alpha=O(1)$) and $S_{\alpha}(\psi) < g(n)$ (over all cuts $(A:B)$) where $f(n), g(n) = \omega(\log n)$. Then $k=\omega(\poly(n))$ copies are necessary for them to estimate $\left|\braket{\phi}{\psi}\right|^2$ up to additive error $\epsilon\in\left(0,\frac{1}{2}-O\left(\frac{1}{\text{poly(n)}}\right)\right)$ with success probability $2/3$, when they are allowed arbitrary interactive protocols (or arbitrary LOCC operations).
\end{lemma}

\begin{proof}
    This comes from the observation that DIPE* is a special case of IP*. To see this, notice that the overlap between $\ket{\psi}$ and $\ket{\phi}$ in \Cref{def:DIPE} is either $\abs{\braket{\phi}{\psi}}^2 = 1$ in case $1$, while in case $2$:
\begin{equation}\label{eq:bounded-overlap}
    \Ex_{\ket{\psi},\ket{\phi}\sim\E_{f,S}}[\abs{\braket{\phi}{\psi}}^2] = \text{negl}(n).
\end{equation}
This last equation comes from the fact that, for Haar random states,
we have
\begin{equation}
\Ex_{\ket{\psi},\ket{\phi}\sim\text{Haar}}[\abs{\braket{\phi}{\psi}}^2] = \frac{1}{d},
\end{equation}
and by considering, e.g., the SWAP test between state vector $\ket{\psi}$ and $\ket{\phi}$ as an algorithm $\A$, we have
\begin{align}
&\Big|\Pr_{\ket{\psi},\ket{\phi}\sim\E_{f,S}} [\A(\ket{\psi},\ket{\phi})=1] - \Pr_{\ket{\psi},\ket{\phi}\sim\text{Haar}} [\A(\ket{\psi},\ket{\phi})=1]\Big| \nonumber\\
&= \frac{1}{2}\Big| \Ex_{\ket{\psi},\ket{\phi}\sim\E_{f,S}}[\abs{\braket{\phi}{\psi}}^2] - \Ex_{\ket{\psi},\ket{\phi}\sim\text{Haar}}[\abs{\braket{\phi}{\psi}}^2] \Big| \nonumber\\
    &\leq \text{negl}(n).
\end{align}
Therefore, by applying IP* with accuracy say $\epsilon=0.1$, one can distinguishing between case $1$ and $2$ with high probability, and therefore solve DIPE*. Similarly, applying IP* with an arbitrary accuracy in $\epsilon \in \left(0,\frac{1}{2}-O\left(\frac{1}{\text{poly(n)}}\right)\right)$ would still allow to differentiate an inner product of $1$ from $\text{negl}(n)$, as both estimate would be separated from $\frac{1}{2}$. The $k=\omega(\poly(n))$ lower bound then comes from \Cref{thm:DIPE*-lower-bound}.
\end{proof}

From here, similarly to Anshu \emph{et al.}, we can extend this lower bound to the case 
\begin{equation}
\epsilon\in\left(\frac{1}{2}-O\left(\frac{1}{\text{poly(n)}}\right),1\right), 
\end{equation}
by using the same chain of reductions from DIPE to IP. This, however, involves considering input states state vectors of the form
\begin{align}\label{def:tilted-state}
    \ket{\psi_{\epsilon,\theta}} &= \sqrt{1-\epsilon}e^{i\theta}\ket{0}\ket{0^{\otimes n}} + \sqrt{\epsilon}\ket{1}\ket{\psi},
    \\
    \ket{\phi_{\epsilon,\theta'}} &= \sqrt{1-\epsilon}e^{i\theta'}\ket{0}\ket{0^{\otimes n}} + \sqrt{\epsilon}\ket{1}\ket{\phi} ,
\end{align}
for $\theta,\theta'$ independent uniformly random phases in $[0,2\pi]$, and $\ket{\psi},\ket{\phi} \in \E_{f,S}$ as opposed to being Haar random in Ref.\ \cite{anshuDistributedQuantumInner2022}. So, in order to extend \Cref{lem:IP*-lower-bound}, we need to show that these states also have bounded magic and entanglement. We establish this in the following lemma, proven in Appendix \ref{appdx:proof-bounded-magic-entanglement}.

\begin{lemma}\label{lem:bounded-magic-entanglement}
    Let a phase state vector $\ket{\psi}\in \mathcal{E}_{f,S}$ that satisfies $S_{\alpha}(\ket{\psi})=O(\log|S|)$ over all cuts $(A:B)$ and $M_{\alpha}(\ket{\psi})=O(\log|S|)$, for $\alpha \geq 0$. Then, for $\epsilon \in (0,1), \theta \in [0,2\pi]$, its corresponding tilted state $\ket{\psi_{\epsilon,\theta}}$ defined in \Cref{def:tilted-state} also satisfies $S_{\alpha}(\ket{\psi_{\epsilon,\theta}})=O(\log|S|)$ over all cuts $(A:B)$ and $M_{\alpha}(\ket{\psi_{\epsilon,\theta}})=O(\log|S|)$, for $\alpha \geq 0$.
\end{lemma}

With this, we arrive at our general lower bound statement for IP*:
\begin{theorem}[Formal version of \Cref{ithm:main-result-IP-no-go}]\label{thm:IP*-lower-bound}
    Suppose Alice has input $\ket{\phi}^{\otimes k}$ and Bob has input $\ket{\psi}^{\otimes k}$, for two arbitrary unknown state vectors $\ket{\phi},\ket{\psi}$ with bounded magic and entanglement, i.e., $M_\alpha(\psi) < f(n)$ (with $\alpha=O(1)$) and $S_{\alpha}(\psi) < g(n)$ (over all cuts $(A:B)$) where $f(n), g(n) = \omega(\log n)$. Then $k=\omega(\poly(n))$ copies are necessary for them to estimate $\left|\braket{\phi}{\psi}\right|^2$ up to additive error $\epsilon\in\left(0,1\right)$ with success probability $2/3$, when they are allowed arbitrary interactive protocols (or arbitrary LOCC operations).
\end{theorem}
\begin{proof}
    The same chain of reductions of Ref.~\cite[Section 5.3]{anshuDistributedQuantumInner2022} from DIPE to IP apply from DIPE* to IP*.\break This is because the only property of the states $\ket{\psi},\ket{\phi} \in \E$ they use is that $\abs{\braket{\phi}{\psi}}\leq\frac{1}{200}$ holds with high probability for $d$ larger than some constant. We also have this property for 
    $\E=\E_{f,S}$ following \Cref{eq:bounded-overlap}. The bounded magic and entanglement of the states used in the reduction follows from \Cref{lem:bounded-magic-entanglement}.
\end{proof}

We can restrict the class of states we consider such that, in addition to having bounded magic and entanglement, they are furthermore efficiently preparable. Similarly to our lower bounds on the sample complexity of approximate Pauli sampling (\cref{sec:hardness-pauli}), we obtain a lower bound on the computational complexity by utilizing pseudorandom states.

\begin{theorem}
Suppose Alice has input $\ket{\phi}^{\otimes k}$ and Bob has input $\ket{\psi}^{\otimes k}$, for two efficiently preparable unknown state vectors $\ket{\phi},\ket{\psi}$ with bounded magic and entanglement, i.e., $M_\alpha(\psi) < f(n)$ (with $\alpha=O(1)$) and $S_{\alpha}(\psi) < g(n)$ (over all cuts $(A:B)$) where $f(n), g(n) = \omega(\log n)$. Then $k=\omega(\poly(n))$ samples necessary for them to estimate $\left|\braket{\phi}{\psi}\right|^2$ up to additive error $\epsilon\in\left(0,1\right)$ with success probability $2/3$, when they are allowed arbitrary polynomial-time interactive protocols (or arbitrary polynomial-time LOCC operations), assuming the existence of OWFs.
\end{theorem}
\begin{proof}
    The proof follows the same steps as the proof of \cref{thm:IP*-lower-bound} with the difference that instead of considering the ensemble of random subset phase states $\mathcal{E}_{f,S}$, we consider the ensemble of pseudorandom phase states $\tilde{\mathcal{E}}_{f,S}$ (see \Cref{sec:subset-phase-states}). More precisely, we construct similar titled states to those in \cref{def:tilted-state} but for $\ket{\psi},\ket{\phi} \in \tilde{\E}_{f,S}$, for which \cref{lem:bounded-magic-entanglement} can be straightforwardly generalized, and the same proof of \cref{thm:IP*-lower-bound} can be applied.
\end{proof}
\subsection{Algorithm for IP based on Pauli sampling}
\label{subsec:algorithm_ip}
In this section, we describe our approach to distributed inner product estimation based on Pauli sampling. Specifically, we will present two variants of a protocol based on coordinated Pauli measurements. Throughout this section, we will use the shorthand notation $\alpha_{\rho}(x) = \tr(\rho P_x)$. Essentially, our approach is motivated by the expansion of $\tr(\rho \sigma)$ in the Pauli basis as
\begin{eqnarray}
\trace\left(\rho\sigma\right)&=&\frac{1}{2^{n}}\sum_{x\in \01^{2n}}\trace\left(\rho P_x\right)\trace\left(\sigma P_x\right) 
\nonumber\\
&=& \frac{1}{2^{n}}\sum_{x\in \01^{2n}} \alpha_{\rho}(x) \alpha_{\sigma}(x)
\end{eqnarray}
Assuming that $\rho,\sigma$ are pure states such that $\trace\left(\rho^2\right)=\trace\left(\sigma^2\right)=1$, this can be further rewritten as 
\begin{eqnarray}\label{eq:asymmetric-expansion}
\trace\left(\rho\sigma\right)&=&\sum_{x\in \{0,1\}^{2n}} \frac{\trace\left(\rho P_x\right)^2}{2^{n}} \frac{\trace\left(\sigma P_x\right)}{\trace\left(\rho P_x\right)} \\
&=& \sum_{x\in \{0,1\}^{2n}} p_{\rho}(x) \frac{\alpha_{\sigma}(x)}{\alpha_{\rho}(x)}\nonumber.
\end{eqnarray}

This expression suggests a protocol for Monte Carlo estimation of $\trace\left(\rho\sigma\right)$ through sampling $x$ from the Pauli distribution $p_{\rho}$ and averaging  $\frac{\alpha_{\sigma}(x)}{\alpha_{\rho}(x)}$.  We note that this is the same idea as in the popular direct fidelity estimation protocol \cite{dasilvaPracticalCharacterizationQuantum2011,flammiaDirectFidelityEstimation2011b}.
As we discuss below in \Cref{ssec:asymmetric-protocol}, this estimation protocol can also be adapted to the distributed setting of this work and we will henceforth refer to this protocol as the \textit{asymmetric} protocol. A downside of this protocol is that the estimator 
$\frac{\alpha_{\sigma}(x)}{\alpha_{\rho}(x)}$ is essentially unbounded. Thus, the implementation requires dealing with the event where a sampled $x$ is such that $|\alpha_{\rho}(x)|$ is very small. Otherwise, the sample complexity of the protocol would explode leading to an inefficient protocol. In fact, this is also the case for the DFE protocol mentioned above.
We propose to address such ``bad events" with very small $|\alpha_{\rho}(x)|$ by applying a filtering function which rescales the estimator (see \Cref{ssec:asymmetric-protocol} for more details). However, rescaling the estimator, in turn, can lead to a large bias in the final estimate.

Here, as our main contribution, we propose a different protocol for the distributed estimation of $\tr(\rho \sigma)$ which we will call the \emph{symmetric} protocol. Crucially, this newly proposed protocol does not require filtering or post-selecting bad events because it uses a bounded estimator. 

The symmetric protocol is based on expressing $\tr(\rho \sigma)$ in terms of the mixture distribution 
$\pmix =\tfrac{1}{2}\left(p_\rho + p_\sigma\right)$.
In particular, one can check that for pure $\rho,\sigma$
\begin{equation}
f := \frac{1}{2}\left(1+ \trace\left(\rho\sigma\right) \right)=  
\sum_{x\in \01^{2n}} \pmix(x) g(x)\, ,
\label{eq:symmetric-protocol-expansion}
\end{equation}
where 
\begin{equation}
    g(x) = G\left(\alpha_{\rho}(x),\alpha_{\sigma}(x)
\right) =
\frac{1}{2} \frac{(\alpha_{\rho}(x)+ \alpha_{\sigma}(x))^2}{ \alpha_{\rho}(x)^2 +\alpha_{\sigma}(x)^2}\,.
\end{equation}
This expression suggests that we can estimate $f$ and hence $\tr(\rho \sigma)$ by sampling $x$ from $\pmix =\tfrac{1}{2}\left(p_\rho + p_\sigma\right)$ and averaging $g(x)$.
 In the following, we give an informal overview of how to implement this idea in the distributed setting of IP. For concreteness, we assume Alice having access to copies of $\rho$ and Bob having access to copies of $\sigma$. 
\begin{enumerate}
    \item In a first step, Alice and Bob both perform (approximate) Pauli sampling according to $p_\rho$ and $p_\sigma$, respectively. This allows them to approximately sample from the mixture $\pmix$ and obtain a list $L=\{{x_1}, \dots, x_{N_1}\}$ of bit-strings corresponding to $N_1$ many Pauli operators which they communicate with each other.
    \item In the next step, Alice and Bob will both measure $N_2$ times, each of the Paulis in $L$ on copies of their respective state. From this measurement data, Alice computes estimates $\hat{\alpha}_{\rho}(x_i)$ for $\alpha_{\rho}(x_i)$ and Bob computes estimates $\hat{\alpha}_{\sigma}(x_i)$ for $\alpha_{\sigma}(x_i)$. Bob then sends his list of estimates $\{\hat{\alpha}_{\sigma}(x_1),\dots \hat{\alpha}_{\sigma}(x_{N_1})\}$ to Alice. 
    \item In a third step, Alice uses their combined data to obtain an estimate of $\tr(\rho \sigma)$. She does so by computing $\hat{g}(x_i)=G \left(
\hat{\alpha}_{\rho}(x_i),
\hat{\alpha}_{\sigma}(x_i)
\right)$. She repeats this for all Paulis in the list and averages the estimates to obtain the final estimate of $f$ and hence $\tr(\rho \sigma)$ via \Cref{eq:symmetric-protocol-expansion}.
\end{enumerate}

We note that step 2 of the above protocol, which involves the estimation of expectation values of many Pauli operators, can be improved by making use of more clever measurement strategies, such as grouping strategies \cite{crawfordEfficientQuantumMeasurement2021a, wuOverlappedGroupingMeasurement2023} or locally biased classical shadows \cite{hadfieldMeasurementsQuantumHamiltonians2022}.

We call this protocol symmetric because, in the first step, both Alice and Bob perform Pauli sampling according to their respective state in order to sample from the mixture $\pmix$. Additionally, in the third step, the quantity to be averaged, $G\left(\alpha_{\rho}(x),\alpha_{\sigma}(x)\right)$, is symmetric in $\alpha_{\rho}(x),\alpha_{\sigma}(x)$. Importantly, the function $G\left(\alpha_{\rho}(x),\alpha_{\sigma}(x)\right)$ is bounded between 0 and 1 and hence makes for a better-behaved estimator in comparison to $\frac{\alpha_{\sigma}(x)}{\alpha_{\rho}(x)}$.

In the following, we will analyze both the symmetric and the asymmetric protocols for IP in detail and state their performance guarantees. As mentioned at the beginning of \Cref{sec:pauli-sampling}, depending on the class of states, different approaches to Pauli sampling are more suitable than others. To account for this, in the following, we will explain and analyze our protocols assuming black-box access to an algorithm performing approximate Pauli sampling. 

On a high-level, our findings are that both the asymmetric and the symmetric protocol run efficiently for input states with low magic. However, we find that that symmetric protocol is efficient for a larger class of states including those with $M_1(\rho)=O(\log (n))$ (c.f. \cref{cor:main-result-IP-low-magic-symmetric}) whereas the asymmetric protocol has the stricter requirement that $M_0(\rho)=O(\log (n))$ (c.f. \cref{cor:main-result-IP-low-magic-asymmetric}). This is an important distinction as $M_0(\rho)$ is not a robust measure of magic as we have remarked at the end of \Cref{ssec:preliminaries-measures-magic-entanglement}. Importantly, this distinction also implies that the symmetric protocol can deal with states beyond those generated from Clifford circuits doped with few $T$-gates. Furthermore, due to the bounded estimator, the symmetric protocol exhibits a better scaling of the error with the number of copies required. 

On the flip side, the asymmetric protocol also has its advantages: It is arguably much less demanding to implement since only one party needs to perform (approximate) Pauli sampling which typically requires 2-copy measurements like Bell measurements. Moreover, due to the asymmetric nature of the protocol, only one of the input states $\rho,\sigma$ needs to have bounded magic in order for the protocol to run efficiently, namely the state of the party performing Pauli sampling. The other state is essentially unconstrained.

\subsection{The symmetric protocol}\label{ssec:sym}
\begin{algorithm}[H]
    \centering
    \caption{Symmetric protocol}\label{alg:symmetric-protocol}
    \raggedright\textbf{Input:} $k$ copies of unknown pure states $\rho,\sigma$\\
    \textbf{Output:} an estimate of $f=\frac{1}{2}\left(1 + \tr(\rho \sigma) \right)$
    \begin{algorithmic}[1]    
        \State  Alice and Bob sample $N_1$ times from the mixture distribution $\tilde{p}_{\mathrm{mix}}$ and obtain $L=\{x_1,\dots,x_{N_1}\}$
         \ForAll{$x \in L$}
            \State Alice measures $N_2$ times $P_x$ on  
            $\rho$ and obtains estimate $\hat{\alpha}_{\rho}(x)$
            \State Bob  measures $N_2$ times $P_x$ on  
            $\sigma$ and obtains estimate $\hat{\alpha}_{\sigma}(x)$
        \EndFor
        
        \State \textbf{Return} $\frac{1}{N_1}\sum_{i=1}^{N_1} G \left(
        \hat{\alpha}_{\rho}(x_i),
        \hat{\alpha}_{\sigma}(x_i)
        \right)$
    \end{algorithmic}
\end{algorithm}

In the symmetric protocol, both Alice and Bob perform approximate Pauli sampling in the first step. In particular, they can sample from the mixture $\tilde{p}_{\mathrm{mix}} = \frac{1}{2}(\tilde{p}_{\rho} + \tilde{p}_{\sigma})$ by sampling from $\tilde{p}_{\rho}$ and $\tilde{p}_{\sigma}$, respectively. Here, $\tilde{p}$ denotes an approximate version of $p$. The full estimation protocol is given in \Cref{alg:symmetric-protocol}. Note that the final estimator reads
\begin{equation}
    f(N_1, N_2) := \frac{1}{N_1}\sum_{i=1}^{N_1} \hat{g}(x_i) =   \frac{1}{N_1}\sum_{i=1}^{N_1} G \left(
\hat{\alpha}_{\rho}(x_i),
\hat{\alpha}_{\sigma}(x_i)
\right)
\label{eq:estimator-f}
\end{equation}
where the $x_i$ are drawn i.i.d from $\tilde{p}_{\mathrm{mix}}$ and $\hat{\alpha}_{\rho}(x_i)$ is an estimate of $\alpha_{\rho}(x) = \tr(\rho P_x)$ obtained from $N_2$ many Pauli measurements.
In the protocol, Alice and Bob each use a total of $N_1\cdot N_2$ many copies of their respective state.

The estimator in \Cref{eq:estimator-f}, although bounded, is a not an unbiased estimator because the function $G:\left[-1,1\right]^2 \setminus \{0,0\}\to \left[0,1\right]$ given by
\begin{equation}
    G(u,v) = \frac{1}{2} \frac{(u + v)^2}{u^2 + v^2}
\end{equation}
is non-linear. This is why we have to take into account properties of $\tilde{p}_{\mathrm{mix}}$ when bounding the error in the estimate $f(N_1,N_2)$ as we briefly describe below (see Appendix~\ref{app:proof-sym} for more details).

We note that the function $G$ varies rapidly for arguments $(u,v)$ around the origin i.e when $\left\Vert (u,v) \right\Vert_2$ is small. Hence, $G \left(
\hat{\alpha}_{\rho}(x_i),
\hat{\alpha}_{\sigma}(x_i)
\right)$ will be close to $G\left(\alpha_{\rho}(x_i),\alpha_{\sigma}(x_i)\right)$, only if $x_i$ is such that at least one of $|\alpha_{\rho}(x_i)|,|\alpha_{\sigma}(x_i)|$ is reasonably large. In \cref{subsec:relating-pauli-distribution-to-entanglement-magic}, we have shown, on a high level, that the more magical a state $\rho$ is, the more weight can accumulate in the tail of the distribution $p_\rho$. This tail probability is captured by the behavior of the \emph{cumulative distribution function} (CDF) $F_\rho$ introduced in \Cref{def:cdf}. Thus, the smaller the magic of the states, the smaller the probability that we will sample a Pauli $P_{x_i}$ such that both $|\alpha_{\rho}(x_i)|,|\alpha_{\sigma}(x_i)|$ are small and therefore the smaller the error we make (for a fixed amount of samples taken). This is the intuitive reason for why $F_\rho$ and $F_\sigma$ appear in the error contribution of our estimate.

We now state our performance guarantees for the symmetric protocol in terms of the CDFs $F_\rho$ and $F_\sigma$. The proof of the following theorem is given in \Cref{app:proof-ip}.

\begin{theorem}[Performance guarantee of the symmetric protocol]
    \label{thm:error-symmetric}
    Let $\epsilon_1>0,\epsilon_2>0$ and $\delta>0$. Let $\tilde p_{\rm mix}$ be a distribution such that $\left\Vert p_{\rm mix}-\tilde p_{\rm mix}\right\Vert_{\rm TV}<\Delta$. Let $f(N_1,N_2)$ be our estimate for $f = \frac{1}{2}\left(1+ \trace\left(\rho\sigma\right) \right)$ as defined in \Cref{eq:estimator-f}. Then,
    \begin{equation}
    \big|f(N_1,N_2)-f\big|\le2\epsilon_1+2\sqrt{\epsilon_2}+\frac{F_\rho(\epsilon_2)}2+\frac{F_\sigma(\epsilon_2)}2+2\Delta \,
    \end{equation}
    with probability at least $1-\delta$, provided that 
    \begin{align}
        N_1 &\ge (2\epsilon_1^2)^{-1}\ln(8/\delta) \,,\\
        N_2&\ge(2/\epsilon_2^{2})\ln(8N_1/\delta)\,.
    \end{align}
    Here, $F_\rho$ and $F_\sigma$ denote the CDFs previously introduced in \Cref{def:cdf}.
\end{theorem}

While the functions $F_\rho, F_\sigma$ precisely capture the performance of the symmetric protocol, they are also, in general, as complex as their underlying distributions $p_\rho, p_\sigma$. To provide a more coarse-grained and informative statement, we now formulate a Corollary to \Cref{thm:error-symmetric} in terms of the stabilizer entropies which associate a single number to a state.

\begin{corollary}[Performance guarantee of the symmetric protocol in terms of $M_0$]
\label{cor:main-result-IP-low-magic-symmetric-M0}
Let $\rho,\sigma$ be $n$-qubit pure states and let $\epsilon>0$. Let $\pmix =\tfrac{1}{2}\left(p_\rho + p_\sigma\right)$ and $\tilde p_{\rm mix}$ be a distribution such that $\left\Vert p_{\rm mix}-\tilde p_{\rm mix}\right\Vert_{\rm TV}<\Delta$. Let $M_0 = \max(M_0(\rho), M_0(\sigma))$. Let $f(N_1,N_2)$ be our estimate for $f = \frac{1}{2}\left(1+ \trace\left(\rho\sigma\right) \right)$ as defined in \Cref{eq:estimator-f}. Then, 
\begin{equation}
    \big|f(N_1,N_2)-f\big|\leq 5\epsilon+2\Delta
\end{equation}
with probability at least $1-\delta$, provided that
\begin{align}
    N_1 &\geq \frac{1}{2\epsilon^2}\log (8/\delta)\, ,\\
    N_2 &\geq  \frac{2}{\min(\epsilon^4,\epsilon^2 2^{-2M_0})}\log (8N_1/\delta)\, .
\end{align}
\end{corollary}

\begin{proof}
Let us write $M_0 = \max(M_0(\rho), M_0(\rho))$.
Consider \Cref{thm:error-symmetric} and recall from \Cref{lem:bounding_probabilities_with_M0} that 
$F_{\rho}(\epsilon_2) \leq 2^{M_0 (\rho)} \epsilon_2$ and $F_{\sigma}(\epsilon_2) \leq 2^{M_0 (\sigma)} \epsilon_2$. So that, $\frac{F_\rho(\epsilon_2)}2+\frac{F_\sigma(\epsilon_2)}2\leq2^{M_0}\epsilon_2$ and \Cref{thm:error-symmetric} implies that
\begin{align}
\big|f(N_1,N_2)-\tr(\rho\sigma)\big|\le2\epsilon_1+2\sqrt{\epsilon_2}+2^{M_0}\epsilon_2+2\Delta
\end{align}
with probability at least $1-\delta$ provided that $N_1\ge(2\epsilon_1^2)^{-1}\log(8/\delta)$ and $N_2\ge(2/\epsilon_2^2)\log(8N_1/\delta)$. Then we take $\epsilon_1=\epsilon$ and $\epsilon_2=\min(\epsilon^2,\epsilon/2^{M_0})$ to obtain
\begin{equation}
\big|f(N_1,N_2)-\tr(\rho\sigma)\big|\le5\epsilon+2\Delta
\end{equation}
with probability at least $1-\delta$ provided that $N_1\ge(2\epsilon^2)^{-1}\log(8/\delta)$ and $N_2\ge(2/\min(\epsilon^4,\epsilon^2 2^{-2 M_0})\log(8N_1/\delta)$.
\end{proof}

\begin{corollary}[Performance guarantee of the symmetric protocol in terms of $M_1$ - formal version of \Cref{ithm:main-result-IP}]
\label{cor:main-result-IP-low-magic-symmetric}
Let $\rho,\sigma$ be $n$-qubit pure states and let $\epsilon>0$. Let $\pmix =\tfrac{1}{2}\left(p_\rho + p_\sigma\right)$ and $\tilde p_{\rm mix}$ be a distribution such that $\left\Vert p_{\rm mix}-\tilde p_{\rm mix}\right\Vert_{\rm TV}<\Delta$. Let $M_1 = \max(M_1(\rho), M_1(\sigma))$. Let $f(N_1,N_2)$ be our estimate for $f = \frac{1}{2}\left(1+ \trace\left(\rho\sigma\right) \right)$ as defined in \Cref{eq:estimator-f}. Then, 
\begin{equation}
    \big|f(N_1,N_2)-f\big|\leq 5\epsilon+2\Delta
\end{equation}
with probability at least $1-\delta$, provided that
\begin{align}
    N_1 &\geq \frac{1}{2\epsilon^2}\log (8/\delta)\, ,\\
    N_2 &\geq  \frac{2}{\min(\epsilon^4,\epsilon^2 2^{-2M_1/\epsilon})}\log (8N_1/\delta)\, .
\end{align}
\end{corollary}

\begin{proof}
    Let us write $M_1 = \max(M_1(\rho),M_1(\sigma))$. In this case, we have from \Cref{lem:bounding-probabilities-via-M1} that $F_{\rho}(\epsilon_2) \leq M_1(\rho)/\log_2(1/\epsilon_2)$ and $F_{\sigma}(\epsilon_2) \leq M_1(\rho)/\log_2(1/\epsilon_2)$. So that, we have that 
$\frac{F_\rho(\epsilon_2)}2+\frac{F_\sigma(\epsilon_2)}2\leq M_1/\log_2(1/\epsilon_2)$ and \Cref{thm:error-symmetric} implies that
\begin{align}
\big|f(N_1,N_2)-\tr(\rho\sigma)\big|\le2\epsilon_1+2\sqrt{\epsilon_2}+\frac{M_1}{\log_2(1/\epsilon_2)}+2\Delta
\end{align}
with probability at least $1-\delta$ provided that $N_1\ge(2\epsilon_1^2)^{-1}\log(8/\delta)$ and $N_2\ge(2/\epsilon_2^2)\log(8N_1/\delta)$. Now choose $\epsilon_1 = \epsilon$ and $\epsilon_2= \min(\eps^2, 2^{-M_1/\epsilon})$ to obtain
\begin{equation}
\big|f(N_1,N_2)-\tr(\rho\sigma)\big|\le5\epsilon+2\Delta
\end{equation}
with probability at least $1-\delta$ provided that $N_1\ge(2\epsilon^2)^{-1}\log(8/\delta)$ and $N_2\ge(2/\min(\epsilon^4,\epsilon^2 2^{-2 M_1/\epsilon})\log(8N_1/\delta)$.
\end{proof}

\begin{remark}
    The conditions of low magic $M_0(\psi)=O(\log n)$ and/or $M_1(\psi)=O(\log n)$ for $\psi=\rho,\sigma$ are not necessary. For instance, given a desired error $\tau$, by \Cref{thm:error-symmetric} the following condition is clearly also sufficient: $F_\psi(\epsilon)\le\tau$ for a sufficiently small $\epsilon=1/{\rm poly}(n)$. This however, does not imply that the magics $M_0(\psi)$ and/or $M_1(\psi)$ are $O(\log n)$. In Appendix~\ref{app:m1} we give the example of a pure state $\psi$ with $M_0(\psi)$ larger than $n-1$, $M_1(\psi)$ larger than $\sqrt{n}$ and $F(\tau^2)\le 2\tau$ for $\tau=1/\sqrt{n}$.
\end{remark}

\subsection{The asymmetric protocol}

\begin{algorithm}[H]
    \centering
    \caption{Asymmetric protocol}\label{alg:asymmetric-protocol}
     \raggedright\textbf{Input:} $k$ copies of unknown pure states $\rho,\sigma$.\\
    \textbf{Output:} An estimate of $f=\trace(\rho \sigma)$.
    \begin{algorithmic}[1]
        \State  Alice samples $N_1$ many bit-strings $x$ from $\tilde{p}_{\rho}$ and obtains $L=\{x_1,\dots,x_{N_1}\}$.
        \ForAll{$x \in L$}
            \State Alice measures $N_\rho$ times $P_x$ on $\rho$ and obtains an estimate $\hat{\alpha}_{\rho}(x)$.
            \State Bob measures $N_\sigma$ times $P_x$ on $\sigma$ and obtains an estimate $\hat{\alpha}_{\sigma}(x)$.
            \State Bob computes $\frac{\hat{\alpha}_{\sigma}(x)}{s_\lambda(\hat{\alpha}_{\rho}(x))}$.
        \EndFor
        \State \textbf{Return}   $\frac{1}{N_1}\sum_{i=1}^{N_1}\frac{\hat{\alpha}_{\sigma}(x_i)}{s_\lambda(\hat{\alpha}_{\rho}(x_i))}$.
    \end{algorithmic}
\end{algorithm}
\label{ssec:asymmetric-protocol}
In this section, we will analyse the behaviour of the asymmetric protocol as described earlier. A complete description of this protocol is given in \Cref{alg:asymmetric-protocol}. The asymmetric protocol is closer to the \emph{direct fidelity estimation} \cite{flammiaDirectFidelityEstimation2011b} protocol as only one party is required to sample Paulis. As we will see, the main drawback of the asymmetric protocol is that the naive protocol uses an estimator that is a priori unbounded. This has indeed already been noted and discussed in the setting of direct fidelity estimation. One strategy to overcome this issue has been suggested in Refs.\ \cite{dasilvaPracticalCharacterizationQuantum2011,flammiaDirectFidelityEstimation2011b} and consists of post-selecting the sampled strings $x$ on having $|\alpha_{\rho}(x)|>\lambda$ for some threshold $\lambda$ of our choice. Here, we take a slightly different approach where we apply a ``filtering'' function to the estimates $\hat{\alpha}_{\rho}(x_i)$, ensuring that the overall estimate remains bounded. More precisely, in the estimation protocol given in \Cref{alg:asymmetric-protocol}, we use the estimator $f(N_1,N_\sigma,N_\rho)$ that reads
\begin{equation}\label{eq:f-n1-n2-asym}
    f(N_1, N_\sigma, N_\rho) :=\frac1{N_1}\sum_{i=1}^{N_1}\frac{\hat\alpha_\sigma(x_i)}{s_\lambda(\hat\alpha_\rho(x_i))}
\end{equation}
with the function $s_\lambda$ defined as
    \begin{equation}
    s_\lambda(z)=\begin{cases}
        z,&|z|>\lambda,\\
        \text{sign}(z)\lambda,&|z|\le\lambda,
    \end{cases}
    \end{equation}
and where the $x_i$ are drawn i.i.d.~from $\tilde p_\rho$, $\hat{\alpha}_{\rho}(x_i)$ is an estimate of $\alpha_{\rho}(x) = \tr(\rho P_x)$ obtained from $N_\rho$ many Pauli measurements, and $\hat{\alpha}_{\sigma}(x_i)$ is an estimate of $\alpha_{\sigma}(x) = \tr(\sigma P_x)$ obtained from $N_\sigma$ many Pauli measurements. In this protocol, Alice and Bob use a total of $N_1 \cdot N_\rho$ and $N_1 \cdot N_\sigma$ many copies of their states, respectively. For comparison with the symmetric protocol, one should compare $N_2$ with $\max(N_\sigma,N_\rho)$, although it is interesting to notice that, due to the asymmetry of this protocol, we generally have $N_\rho > N_\sigma$.

We now state the performance guarantee for the asymmetric protocol. Importantly, in contrast to the symmetric protocol, the error depends directly on $M_0(\rho)$.

\begin{theorem}[Performance guarantee of the asymmetric protocol]\label{th:asymm}
    Let $\epsilon_\rho>0$, $\epsilon_\sigma>0$, $\delta>0$, $\gamma>0$ and $\lambda>0$. Let $\tilde p_\rho$ be a distribution such that $||p_\rho-\tilde p_\rho||_{\rm TV}<\Delta$. We write $f(N_1,N_\sigma,N_\rho)$ for our estimate of $\tr(\rho\sigma)$ using $N_1$ samples from $\tilde p_\rho$, and $N_\rho, N_\sigma$ measurements per sample for $\rho$ and $\sigma$ respectively. One has
    \begin{equation}
    \Big|f(N_1,N_\sigma,N_\rho)-\tr(\rho\sigma)\Big|\le\frac1{\lambda}\left(\epsilon_\sigma+\frac{\epsilon_\rho}{\lambda}\right)+\gamma+\frac{\Delta}{\lambda}+\frac{F_\rho(\lambda^2)}{\lambda}+2^{M_0(\rho)}\lambda
    \end{equation}
    with probability at least $1-\delta$, provided that \footnote{\label{note}A median-of-means estimator could be used to reduce the complexity of $N_1$ to $\frac{2(1+\Delta/\lambda^2)}{\gamma^2}\ln(\frac{3}{\delta})$ (or $O\left(\frac{\log(\delta^{-1})}{\epsilon^2}\right)$), but for ease of exposition, we opt for a simpler empirical mean estimator.}
    \begin{align}
        N_1&\geq\frac{3(1+\Delta/\lambda^2)}{\gamma^2\delta},\\
        N_\sigma&\geq\frac{2}{\epsilon_\sigma^2}\ln\left(\frac{6N_1}{\delta}\right),\\
        N_\rho&\geq\frac{2}{\epsilon_\rho^2}\ln\left(\frac{6N_1}{\delta}\right),
    \end{align}
    where $F_\rho(\cdot)$ denotes the CDF of $\rho$ previously introduced, and $M_0(\rho)$ is its stabilizer entropy for $\alpha=0$. 
\end{theorem}

Analogously to the symmetric protocol, we have the following corollary.

\begin{corollary}[Performance guarantee of the asymmetric protocol in terms of $M_0$]
    \label{cor:main-result-IP-low-magic-asymmetric}
    Let $\epsilon>0$, $\delta>0$. Assuming that the state $\rho$ has bounded magic scaling as $M_0(\rho)\leq c\log(n)$ and that we can sample Paulis from a distribution $\tilde p_\rho$ such that $||p_\rho-\tilde p_\rho||_{\rm TV}<O(\epsilon^2/n^{2c})$, then our asymmetric protocol returns an estimate $f(N_1,N_\sigma,N_\rho)$ of $\tr(\rho\sigma)$ using $N_1$ (approximate) Pauli samples and $N_\rho, N_\sigma$ measurements per sample for $\rho$ and $\sigma$ respectively, that satisfies
    \begin{equation}
    \Big|f(N_1,N_\sigma,N_\rho)-\tr(\rho\sigma)\Big|\le \epsilon
    \end{equation}
    with probability at least $1-\delta$, for\cref{note}
    \begin{align}
        N_1&=O\left(\frac{1}{\epsilon^2\delta}\right),\\
        N_\sigma&=O\left(\frac{n^{4c}\log(N_1\delta^{-1})}{\epsilon^4}\right),\\
        N_\rho&=O\left(\frac{n^{8c}\log(N_1\delta^{-1})}{\epsilon^6}\right).
    \end{align}
\end{corollary}
We defer the proof of Theorem~\ref{th:asymm} to Appendix~\ref{app:proof-asym}. The proof follows a similar approach as the one for the symmetric protocol, with one main difference, namely that several error terms of the form $\sim1/\lambda$ which can be attributed to the fact that the range of the estimator is $[-1/\lambda,1/\lambda]$. 

\section*{Acknowledgements}
We sincerely thank Matthias Caro and Tommaso Guaita for providing incredibly helpful feedback on the manuscript and for discussions. We thank David Gross, Gregory White and Alexander Nietner for clarifying discussions on MPS tomography schemes. We further thank Barbara Kraus and Janek Denzler for insightful discussions. This work has been supported by the BMBF (DAQC, MUNIQC-Atoms), the Munich Quantum Valley (K-4 and K-8), the BMWK (EniQmA), the Quantum Flagship (PasQuans2, Millenion), QuantERA (HQCC), the Cluster of Excellence MATH+, the DFG (CRC183), the Einstein Foundation (Einstein Research Unit on Quantum Devices), and the ERC (DebuQC).

\bibliographystyle{myapsrev4-1}
\bibliography{literature_bibtex}

\newpage
\begin{widetext}

\section{Estimation from Bell measurement data}
\label{app:estimation}
In this section, for the sake of the work being self-contained, we explain how to estimate the marginals of the Pauli distribution from Bell measurement data obtained via Bell sampling. A similar exposition can be found in 
Ref.\ \cite[Appendix E]{huangInformationtheoreticBoundsQuantum2021a}.

Recall that a single round of Bell sampling on two copies $\rho^{\otimes 2}$ results in a bit-string $y=(y_1,\dots,y_n)$ of length $2n$ corresponding to a Bell state vector $\ket{P_y}$.
For convenience, we define the $4\times4$ matrix $\boldsymbol{M}$ with entries
$M_{a,b}=\trace\left(P_{a}^{\otimes2}\ket{P_{b}}\bra{P_{b}}\right)$
where $a,b$ are $2$-bit strings labelling the single-qubit Paulis
and $2$-qubit Bell basis
\begin{equation}
\boldsymbol{M}=\left(\begin{array}{c|cccc}
 & \Phi^{+} & \Phi^{-} & \Psi^{+} & \Psi^{-}\\
\hline I & +1 & +1 & +1 & +1\\
Z & +1 & +1 & -1 & -1\\
X & +1 & -1 & +1 & -1\\
Y & -1 & +1 & +1 & -1
\end{array}\right)=\left(\begin{array}{c|cccc}
 & 00 & 01 & 10 & 11\\
\hline 00 & +1 & +1 & +1 & +1\\
01 & +1 & +1 & -1 & -1\\
10 & +1 & -1 & +1 & -1\\
11 & -1 & +1 & +1 & -1
\end{array}\right).
\end{equation}
Furthermore, we define the vector $\boldsymbol{s}$ with entries $s_{a}=\trace\left(\mathbb{S}\ket{P_{a}}\bra{P_{a}}\right)$
so that
\begin{equation}
\boldsymbol{s}=\begin{pmatrix}+1 & +1 & +1 & -1\end{pmatrix}.
\end{equation}

We now prove the following lemma asserting that we can estimate all the marginals of $p_\rho$ from few Bell measurement data.

\begin{lemma}[\Cref{lem:estimating_marginals} restated]\label{lem:marginals_mixed}
Let $\epsilon>0,\delta>0$ and let $\rho$ be a quantum state on $n$ qubits. Then, 
$N=O\left(n\log\left(1/\delta\right)/(\epsilon\tr(\rho^2))^2\right)$  pairs
of copies $\rho\otimes\rho$ suffice to produce, with probability $1-\delta$, estimates $\pi_{\rho}(x_{1:k})$
such that 
\begin{equation}
\left|\pi_{\rho}(x_{1:k})-p_{\rho}\left(x_{1:k}\right)\right|\leq\frac{\epsilon}{2^{k}}
\end{equation}
for all $x\in \01^{2n}$ and all $1\leq k\leq n$.
\end{lemma}

\begin{proof}
Let $\{y^{(1)}, \dots, y^{(N)}\}$ be the $N$ bit-strings of length $2n$ that were sampled independently via Bell sampling on $\rho^{\otimes 2}$. In this proof, it is useful to introduce the following quantities for the sake of clarity. We introduce
\begin{align}
    Q_k(x_{1:k})&=\langle P_{x_{1:k}}^{\otimes 2}\otimes S^{\otimes {n-k}}\rangle_{\rho\otimes\rho},\\
    \hat Q_k(x_{1:k})&=\frac1N\sum_{i=1}^N\left(\prod_{j=1}^{k} M_{x_j,y^{(i)}_j}\prod_{l=k+1}^n s_{y^{(i)}_l})\right)\,, \label{eq:marginal-estimator}
\end{align}
i.e., $Q_k(x_{1:k})$ is the ideal expectation value $\langle P_{x_{1:k}}^{\otimes 2}\otimes S^{\otimes {n-k}}\rangle_{\rho\otimes\rho}$ and $\hat Q_k(x_{1:k})$ is our estimate of $Q_k(x_{1:k})$ computed from $N$ Bell measurements outcomes obtained from copies of $\rho\otimes\rho$. This is an empirical estimate because all the operators $P_{x_{1:k}}^{\otimes 2}\otimes S^{\otimes {n-k}}$ acting on two copies commute and a common eigenbasis is precisely the Bell basis with eigenvalues $\prod_{j=1}^{k} M_{x_j,y^{(i)}_j}\prod_{l=k+1}^n s_{y^{(i)}_l}$. Note that the previous expressions also work for $k=0$ so that $Q_0:=\tr\rho^2$ is the purity and $\hat Q_0$ is our estimate for the purity. Both are again obtained from Bell measurements on $N$ copies of $\rho\otimes\rho$ . Explicitly,
\begin{align}
    Q_0&=\langle S^{\otimes {n}}\rangle_{\rho\otimes\rho},\\
    \hat Q_0&=\frac1N\sum_{i=1}^N\left(\prod_{l=1}^n s_{y^{(i)}_l}\right)\,.
\end{align}

Now, any single $Q_k(x_{1:k})$ can be estimated to precision $\epsilon_1$ with probability at least $1-\delta$ using $N=O(\log(1/\delta)/\epsilon_1^2)$ many copies of $\rho^{\otimes 2}$. Hence, by a union bound over all $x\in \{0,1\}^{2n}$, using  $N=O(n\log(1/\delta)/\epsilon_1^2)$ copies suffices to estimate the quantities $Q_k(x_{1:k})$ for all $x\in \{0,1\}^{2n}$ and $1\leq k\leq n$ simultaneously up to an additive
error smaller than $\epsilon_1$ with probability larger than $1-\delta$.
Now, it remains to study how the error $\epsilon_1$ in $\hat Q_k(x_{1:k})$
propagates in the estimates $\pi(x_{1:k})$ of the marginals $p(x_{1:k})$. Here, recall that the marginals and our estimates are given by
\begin{align}
    p(x_{1:k})&=\frac{Q_k(x_{1:k})}{2^k Q_0}\,,\\
    \pi(x_{1:k})&=\frac{\hat Q_k(x_{1:k})}{2^k\hat Q_0}\,.
\end{align}

Now, we shall use the following simple
  inequalities,
 \begin{equation}
       \frac{Q_k(x_{1:k})-\epsilon_1}{Q_0+\epsilon_1}\le
    2^k\pi(x_{1:k})\le \frac{Q_k(x_{1:k})+\epsilon_1}{Q_0-\epsilon_1}\,,
 \end{equation}
and letting $\epsilon_1=\epsilon Q_0/4$ where $0<\epsilon<1$ is final desired precision. The proof will be completed by showing
  \begin{equation}
       \frac{Q_k(x_{1:k})}{Q_0}-\epsilon\le\frac{Q_k(x_{1:k})-\epsilon_1}{Q_0+\epsilon_1}\le
    2^k\pi(x_{1:k})\le \frac{Q_k(x_{1:k})+\epsilon_1}{Q_0-\epsilon_1}\le\frac{Q_k(x_{1:k})}{Q_0}+\epsilon
  \end{equation}
  when $\epsilon_1=\epsilon Q_0/4$.
  Putting
  everything together, the upper bound reads
\begin{equation}
        \frac{Q_k+\epsilon_1}{Q_0-\epsilon_1}=\frac{Q_k-\epsilon_1+2\epsilon_1}{Q_0-\epsilon_1}=\frac{Q_k-\epsilon_1}{Q_0-\epsilon_1}+\frac{2\epsilon_1}{Q_0-\epsilon_1}\le\frac{Q_k}{Q_0}+\frac{2\epsilon_1}{Q_0}\cdot\frac1{1-\epsilon_1/Q_0}
        \leq \frac{Q_k}{Q_0}+\frac{2\epsilon_1}{Q_0}\cdot \left(1+\frac{2\epsilon_1}{Q_0}\right) \leq \frac{Q_k}{Q_0} +\epsilon \,  .
\end{equation}

For the lower bound, we have analogously,
\begin{equation}
     \frac{Q_k-\epsilon_1}{Q_0+\epsilon_1}=\frac{Q_k+\epsilon_1-2\epsilon_1}{Q_0+\epsilon_1}=\frac{Q_k+\epsilon_1}{Q_0+\epsilon_1}-\frac{2\epsilon_1}{Q_0+\epsilon_1}\ge\frac{Q_k}{Q_0}-\frac{2\epsilon_1}{Q_0}\cdot\frac1{1+\epsilon_1/Q_0}\ge\frac{Q_k}{Q_0}-\epsilon\,.
\end{equation}
  Recalling that $2^k p(x_{1:k})=Q_k(x_{1:k})/Q_0$,
  we have shown that $2^k\pi(x_{1:k})$ approximates
  $2^k p(x_{1:k})$ within additive error $\epsilon$ with $N=O(n\log(1/\delta)/\epsilon_1^2)$ copies, where $\epsilon_1=\epsilon Q_0/4$. In other words, a sufficient number of copies is $N=O(n\log(1/\delta)/(\epsilon Q_0)^2)$, as claimed.
\end{proof}

\section{Proofs of \cref{sec:hardness-pauli}: Indistinguishable ensembles with imaginarity gap}\label{sec:proofsoflemmas}
In this appendix, we prove Lemma~\ref{lem:imaginary-ensemble} and Lemma~\ref{lem:imaginarityfrompaulisampling} of the main text.
In the following lemma, we give a construction of two statistically indistinguishable ensembles such that states from the one ensemble are real while states from the other have high imaginarity $I(\ket{\psi})$.

\begin{lemma}[\Cref{lem:imaginary-ensemble} restated]
Let $U=\bigotimes_{i=1}^{n}C_{i} $ for $C_{i}$ being random single-qubit Clifford unitaries. Then define the ensemble $\mathcal{E}_{U}=\{U\ket{\psi_{f,S}}\,|\, \ket{\psi_{f,S}}\in \mathcal{E}_{f,S}\}$ where $\mathcal{E}_{f,S}$ is an ensemble of random subset phase states. Denote $\ket{\psi^{*}}$ the conjugate state in the computational basis. We have the following list of results:

\begin{enumerate}
    \item \textbf{Imaginarity:}
    $\Pr_{\ket{\psi}\sim \mathcal{E}_{U}}[I(\ket{\psi}) > 1/100 ]= 1-\operatorname{negl}(n)$.

    \item \textbf{Statistical indistinguishability:} For $|S|=\omega(\poly(n))$, the two ensembles $\mathcal{E}_{U}$ and $\mathcal{E}_{f,S}$ are statistically indistinguishable. 

\end{enumerate}

\end{lemma}

\begin{proof} 
To prove the first point, let $\psi=\ket{\psi}\bra{\psi}$, notice that $\psi^{*}=\psi^{T}$ where $T$ denotes the transposition. Then, using the replica trick, we can rewrite the overlap as
\begin{equation}
|\langle\psi|\psi^{*}\rangle|^2=\tr(\psi\psi^{*})=\tr(\mathbb{S}\psi\otimes\psi^{T})=\tr(\mathbb{S}^{T_2}\psi^{\otimes 2})  
\end{equation}
where $\mathbb{S}$ is the swap operator in $\mathcal{H}^{\otimes 2}$, and we use the invariance of the trace under partial transpose $T_{2}$, i.e., the transposition on the second copy of $\psi$ only. Notice that
\begin{equation}
\mathbb{S}^{T_2}=2^n\ket{\Omega}\bra{\Omega}^{\otimes n}
\end{equation}
where $\ket{\Omega}=\frac{1}{\sqrt{2}}(\ket{0,0}+\ket{1,1})$
as defined above. Let us average over the choice of $U$ using standard Haar measure techniques. We find that for any pure state $\ket\psi$, it holds that,
\begin{equation}
\mathbb{E}_{U}[|\langle\psi|\psi^{*}\rangle|^2]=\left(\prod_{i=1}^{n}\mathbb{E}_{C_i}\right)\tr[\bigotimes_{i=1}^{n}(C_{i}^{\otimes 2}\ket{\Omega}\bra{\Omega}C_{i}^{\dag \otimes 2})\psi^{\otimes 2}] =\frac{1}{3^n}\tr[\bigotimes_{i=1}^{n}(I_i+\mathbb{S}_i)\psi^{\otimes 2}]\le \left(\frac{2}{3}\right)^n\,.
\end{equation}
The last inequality can be derived by upper bounding each term by $1$. Using Markov inequality, we have that, for any $\ket{\psi}$, it holds that,

\begin{equation}
\underset{U}{\operatorname{Pr}}\left[|\langle\psi|\psi^{*}\rangle|^2<\frac{1}{100}\right]\ge 1-100\left(\frac{2}{3}\right)^n\,.
\end{equation}

Now, notice that we can write $\psi=U\psi_{f,S}U^{\dag}$ for every $\ket{\psi}\in\mathcal{E}_{U}$, where $\psi_{f,S}=\ket{\psi_{f,S}}\bra{\psi_{f,S}}$. Since, $\psi_{f,S}\sim \mathcal{E}_{f,s}$ and $U$ are drawn independently, we find

\begin{equation}
    \Pr_{\ket{\psi}\sim \mathcal{E}_{U}} [I(\ket{\psi}) > 1/100 ] = \sum_{f,S} \Pr_{\ket{\psi_{f,S}}\sim \mathcal{E}_{f,s}} [\psi_{f,S}]\Pr_{U}[I(U{\psi_{f,S}}U^{\dagger}) > 1/100 ] \ge 1-100\left(\frac{2}{3}\right)^n\,.
\end{equation}
This shows that while the ensemble $\mathcal{E}_{f,s}$ is strictly real, the Clifford-conjugated version $\mathcal{E}_{U}$ is highly imaginary.
 
To show the second point, i.e., that $\mathcal{E}_U$ is statistically indistinguishable from $\mathcal{E}_{f,S}$, we can use \cref{lem:statlemma} and \cref{lem:statlemma2} by setting $|S|=\omega(\poly(n))$.
Using \cref{lem:complemma} and \cref{lem:complemma2}, we conclude that $\tilde{\mathcal{E}}_{U}$ is computationally indistinguishable from $\mathcal{E}_{f,S}$ for $|S|=\omega(\poly(n))$.
\end{proof}

In the following lemma, we show that given samples from a distribution close to the Pauli distribution, allows us to estimate the imaginarity of the state using a number of samples that scales inverse polynomially with the estimation error.

\begin{lemma}[\Cref{lem:imaginarityfrompaulisampling} restated]
Let $\rho=\ket{\psi}\bra{\psi}$ be a pure quantum state and $p_{\rho}$ its associated Pauli distribution. Then, given black box access to an algorithm for sampling from a distribution $q$ such that $\left\Vert p_{\rho} - q\right\Vert_\mathrm{TV}\leq \Delta$, for $0 \leq \Delta < 1$, there exists an efficient algorithm to estimate the imaginarity $I(\ket{\psi})=1-|\langle\psi|\psi^{*}\rangle|^2$ within additive error $\epsilon > \Delta$ and failure probability $\delta$ using $\frac{2}{(\epsilon-\Delta)^2}\ln\frac{2}{\delta}$ samples from $q$. 
\end{lemma}

\begin{proof}
    Let us first expand $I(\psi)$ in the Pauli basis
\begin{equation}
I(\psi)=\frac{1}{d}\sum_{x}\tr(P_x\psi)\tr(P_x\psi^{*})=\sum_{x}\frac{\tr(P_x\psi^{*})}{\tr(P_x\psi)}p_{\psi}(x).
\end{equation}
Note that the components $\tr(P_x\psi^{*})$ can be easily computed by $\tr(P_x\psi)$ by 
\begin{equation}
\tr(P_x\psi^{*})=\tr(P^{T}_x\psi)=(-1)^{\sum_{i=1}^{n}s^{x}_{i}t^{x}_{i}}\tr(P_x\psi)
\end{equation}
where we have used $(i)$ the invariance of the trace under trasposition and $(ii)$ the fact that all Paulis are symmetric, except $Y$, for which $Y^{T}=-Y$. Indeed, in the above formula,  $s^{P}_{i}$ and $t^{P}_{i}$ are binary variables corresponding to the Gottesman encoding of Pauli operators and ${\sum_{i=1}^{n}s_{i}^{P}t_{i}^{P}}$ is counting the number of $Y$s in the expression of $P$. Therefore, we have
\begin{equation}
\tr(\psi\psi^{*})=\sum_{x}(-1)^{\sum_{i}s_{i}^{x}t_{i}^{x}}p_{\psi}(x)=\langle (-1)^{\sum_{i}s_{i}^{x}t_{i}^{x}}\rangle_{x\sim p_{\psi}(x)}.
\end{equation}
In other words, we can sample from $p_\psi(x)$ and then sum up signs $(-1)^{\sum_{i}s_{i}^{x}t_{i}^{x}}$ depending on the Pauli $P_x$ sampled. Let us say that we sample $k$ times from the distribution $p_{\psi}$ obtaining the random variables $v_i\in\{\pm 1\}$ for each sample $i=1,\ldots, k$. An unbiased estimator $\tilde{I}$ for the quantity $I(\psi)$, is
\begin{equation}\label{eq:estimator}
\tilde{I}=\frac{1}{k}\sum_{i=1}^{k}v_i
\end{equation}
and, using Hoeffding's inequality and the fact that $|s_i|=1$ for all $i$, we can write
\begin{equation}\label{eq:Hoeffding}
\operatorname{Pr}[|I-\tilde{I}|\ge \epsilon]=\operatorname{Pr}[|\Ex[\tilde{I}]-\tilde{I}|\ge \epsilon]\le 2e^{-k\epsilon^{2}/2},
\end{equation}
which in turn says that $k=2\epsilon^{-2}\ln(2/\delta)$ are sufficient to estimate $I$ within an error $\epsilon$ and failure probability $\delta$.

However, we are not sampling from $p_\psi(x)$ directly, but rather from a distribution $q(x)$ such that $TV(p_\psi,q) = \sum_x \abs{p_\psi(x)-q(x)} \leq \epsilon'$. Therefore, a similar estimator $\hat{I}$ to that in \cref{eq:estimator} constructed out of samples from $q(x)$ satisfies:
\begin{align}
\nonumber
    \abs{\Ex[\tilde{I}]-\Ex[\hat{I}]} &= \Big| \sum_x p_\psi(x) (-1)^{\sum_{i}s_{i}^{x}t_{i}^{x}} - \sum_x q(x) (-1)^{\sum_{i}s_{i}^{x}t_{i}^{x}}\Big|\\
    &\leq \sum_x \abs{p_\psi(x)-q(x)} \leq \Delta \label{eq:diff-exp-values}
\end{align}
where the first inequality follows from the triangular inequality. Similarly to \cref{eq:Hoeffding}, it also satisfies:
\begin{equation}
    \operatorname{Pr}[|\Ex[\hat{I}]-\hat{I}|\ge \epsilon-\Delta]\le 2e^{-k(\epsilon-\Delta)^{2}/2},
\end{equation}
which means that $k=2(\epsilon-\Delta)^{-2}\ln(2/\delta)$ samples are sufficient to guarantee that $|\Ex[\hat{I}]-\hat{I}| \leq \epsilon-\Delta$ with failure probability $\delta$, and therefore:
\begin{equation}
    |I-\hat{I}| = |I-\Ex[\hat{I}] + \Ex[\hat{I}]-\hat{I}| \le |I-\Ex[\hat{I}]| + |\Ex[\hat{I}]-\hat{I}| \le \epsilon,
\end{equation}
by applying the triangular inequality and \cref{eq:diff-exp-values}.
\end{proof}

\section{Bounding magic and entanglement of tilted subset phase states}\label{appdx:proof-bounded-magic-entanglement}

In this appendix, we prove \Cref{lem:bounded-magic-entanglement}, restated below for convenience.

\begin{lemma}[\Cref{lem:bounded-magic-entanglement} restated]
    Let a phase state $\ket{\psi}\in \mathcal{E}_{f,S}$ that satisfies $S_{\alpha}(\ket{\psi})=O(\log|S|)$ over all cuts $(A:B)$ and $M_{\alpha}(\ket{\psi})=O(\log|S|)$, for $\alpha \geq 0$. Then, for $\epsilon \in (0,1), \theta \in [0,2\pi]$, its corresponding tilted state $\ket{\psi_{\epsilon,\theta}}$ defined in \Cref{def:tilted-state} also satisfies $S_{\alpha}(\ket{\psi_{\epsilon,\theta}})=O(\log|S|)$ over all cuts $(A:B)$ and $M_{\alpha}(\ket{\psi_{\epsilon,\theta}})=O(\log|S|)$, for $\alpha \geq 0$.
\end{lemma}
\begin{proof}
    Given a state $\ket{\phi}\in\mathbb{C}^{\otimes 2^n}$, call $Q(\ket{\phi})=\{P\in\mathcal{P}_n | \bra{\phi}P\ket{\phi} \neq 0\}$. Then $M_0(\ket{\phi}) = \log\left(\frac{\abs{Q(\ket{\phi})}}{2^n}\right)$.
    
    In the case of the subset phase state $\ket{\psi}$ in the statement of the lemma, we have
    \begin{equation}\label{eq:bound-magic-phase-state}
        M_0(\ket{\psi}) = \log\left(\frac{\abs{Q(\ket{\psi})}}{2^n}\right) \leq O(\log\abs{S}).
    \end{equation}
    From this, we aim to bound $M_0(\ket{\psi_{\epsilon,\theta}})$ for the tilted state
    \begin{equation}
        \ket{\psi_{\epsilon,\theta}} = \sqrt{1-\epsilon}e^{i\theta}\ket{0}\ket{0^{\otimes n}} + \sqrt{\epsilon}\ket{1}\ket{\psi}.
    \end{equation}
    We do so by upper bounding the size of $Q(\ket{\psi_{\epsilon,\theta}})$.
    Consider $P\in\mathcal{P}_{n+1}$ and divided it into $P=P_1 \otimes P_n$. We have
    \begin{align}
        \bra{\psi_{\epsilon,\theta}} P_1 \otimes P_n \ket{\psi_{\epsilon,\theta}} &= \epsilon \bra{1} P_1 \ket{1} \bra{\psi}P_n\ket{\psi} + (1-\epsilon) \bra{0} P_1 \ket{0}\bra{0^{\otimes n}}P_n\ket{0^{\otimes n}}\label{eq:first-line}\\
        &\quad\quad + \sqrt{\epsilon(1-\epsilon)}[\bra{0} P_1 \ket{1} \bra{0^{\otimes n}}P_n\ket{\psi} + \text{c.c.}] \label{eq:second-line}.
    \end{align}
    Consider first the case where $P_1 \in \{I,Z\}$. Then $\bra{0} P_1 \ket{1} = 0$, which means that the terms in the second line (\Cref{eq:second-line}) vanish. In the first line, the terms do not vanish iif $P_n \in Q(\ket{\psi})$ or $\bra{0^{\otimes n}}P_n\ket{0^{\otimes n}} \neq 0$ (i.e., $P_n \in \{I,Z\}^{\otimes n}$). This implies that there are at most $2(Q(\ket{\psi}) + 2^n)$ terms in $Q(\ket{\psi_{\epsilon,\theta}})$ such that $P_1 \in \{I,Z\}$.
    As for the case where $P_1 \in \{X,Y\}$, we have $\bra{0} P_1 \ket{0} = \bra{1} P_1 \ket{1} = 0$, which means that the terms in the first line (\Cref{eq:first-line}) vanish. Now note that $\bra{0^{\otimes n}} P_n \ket{\psi}$ is always of the form $\braket{x}{\psi}$ (up to a potential phase $i$) for a certain $x\in\{0,1\}^n$ that depends on $P_n$, and that each $x$ has $2^n$ Paulis $P_n$ that lead to it. Given that the subset phase state vector $\ket{\psi}$ is a superposition of $\abs{S}$ computational basis states, then this means that there are at most $2\abs{S}2^n$ terms in $Q(\ket{\psi_{\epsilon,\theta}})$ such that $P_1 \in \{X,Y\}$.
    Overall, this leads to
    \begin{equation}
        \abs{Q(\ket{\psi_{\epsilon,\theta}})} \leq 2 (Q(\ket{\psi}) + 2^n + \abs{S}2^n),
    \end{equation}
    which, from \Cref{eq:bound-magic-phase-state}, results in $M_0(\ket{\psi_{\epsilon,\theta}})\leq O(\log\abs{S})$. Since $M_\alpha(\ket{\phi}) \leq M_0(\ket{\phi})$, for all $\ket{\phi}$ and $\alpha \geq 0$, then $M_{\alpha}(\ket{\psi_{\epsilon,\theta}})\leq O(\log|S|)$.

    As discussed around \Cref{eq:Shmidt-decomposition}, the Rényi-0 entanglement entropy over a cut $(A:B)$ is related to the Schmidt rank $r$ of a quantum state over that cut as
    \begin{equation}
        S_{0}(\ket{\psi})= \log(r).
    \end{equation}
    Looking at the Schmidt decomposition of the subset phase state $\ket{\psi}$ in the statement of the lemma
    \begin{equation}
        \ket{\psi}= \sum_{i=1}^{r}\sqrt{\lambda_{i}}\ket{i_{A}}\ket{i_{B}},
    \end{equation}
    we can derive the Schmidt decomposition of the tilted state vector 
    \begin{equation}
        \ket{\psi_{\epsilon,\theta}} = \sum_{i=1}^{r}\sqrt{\epsilon\lambda_{i}}\ket{1}\ket{i_{A}}\ket{i_{B}} + \sqrt{1-\epsilon}e^{i\theta}\ket{0}\ket{0^{\otimes n}},
    \end{equation}
    irrespective of the cut $(A:B)$, since the tilted state is a superposition of two orthogonal subspaces. This means that the Schmidt rank of $\ket{\psi_{\epsilon,\theta}}$ is simply $r+1$, and therefore $S_0(\ket{\psi_{\epsilon,\theta}})\leq O(\log\abs{S})$. Since $S_\alpha(\ket{\phi}) \leq S_0(\ket{\phi})$, for all $\ket{\phi}$ and $\alpha \geq 0$, then $S_{\alpha}(\ket{\psi_{\epsilon,\theta}})\leq O(\log|S|)$.
\end{proof}
\section{Pauli sampling via ancestral sampling}\label{sec:Appendix_pauli_sampling_algorithm}
In the following section, we present the proof of \Cref{thm:sampling_conditions_marginals} which restate here for convenience:
\begin{theorem}[\Cref{thm:sampling_conditions_marginals} restated]\label{thm:sampling_conditions-marginals-restated}
Let $p$ be a distribution over $\01^n$ and $\gamma>0$. For all $\eps < \gamma /2$, given black-box access to estimates of the marginals such that
\begin{equation}
    | \pi(x_{1:k})-p(x_{1:k}) |\leq \frac{\epsilon}{2^{k}} \,,
\end{equation}
the adapted ancestral algorithm samples from a distribution $q$ which satisfies
\begin{equation}
    \left \Vert q-p \right \Vert_{\mathrm{TV}}\leq  \mathfrak{f}(\gamma) + \left|\exp(\frac{4 \epsilon n}{\gamma}) -1\right| \, .
\end{equation}
Here, $\mathfrak{f}(\gamma)$ is defined in \Cref{def:set-s-gamma}.
\end{theorem}

To prove this theorem, we will need two additional lemmata.
The first lemma gives a bound on the total variation distance between two distributions whose individual probabilities are multiplicatively close on a large subset $S\subseteq \01^n$ of the sample space.
\begin{lemma}[Bound on TV distance via multiplicative error approximation on a heavy subset]
\label{lem:multiplicative-error-on-heavy-subset}
Let $\tau\geq0$ and let $p,q$ be distributions
over $\left\{ 0,1\right\} ^{n}$. Assume that there is a
subset $S\subseteq\left\{ 0,1\right\} ^{n}$ such that 
\begin{equation}
p(S)=\sum_{x\in S}p\left(x \right)\geq1-\tau \,,
\end{equation}
and for all $x\in S$, we have $
\left|p\left(x\right)-q\left(x\right)\right|\leq\Delta p\left(x\right)\,$.
Then,  
\begin{equation}
\left\Vert q-p\right\Vert _{\mathrm{TV}}\leq \tau + \Delta
\end{equation}
\end{lemma}

\begin{proof}
We start from
\begin{align}
\left\Vert q-p\right\Vert _{\mathrm{1}} & =\sum_{x\in S}\left|p\left(x\right)-q\left(x\right)\right|+\sum_{x\not\in S}\left|p\left(x\right)-q\left(x\right)\right|\\
 & \leq\Delta\,p\left(S\right)+\sum_{x\not\in S}\left|p\left(x\right)-q\left(x\right)\right|.
\end{align}
As for the second term, we have
\begin{align}
\sum_{x\not\in S}\left|p\left(x\right)-q\left(x\right)\right| & =\sum_{\substack{x\not\in S\\
p\left(x\right)\geq q\left(x\right)
}
}p\left(x\right)-q\left(x\right)+\sum_{\substack{x\not\in S\\
p\left(x\right)<q\left(x\right)
}
}q\left(x\right)-p\left(x\right)\\
 & \leq\sum_{\substack{x\not\in S\\
p\left(x\right)\geq q\left(x\right)
}
}p\left(x\right)+\sum_{\substack{x\not\in S\\
p\left(x\right)<q\left(x\right)
}
}q\left(x\right)\\
 & \leq1-p\left(S\right)+\sum_{x\not\in S}q\left(x\right)
\end{align}
and
\begin{equation}
\sum_{x\not\in S}q\left(x\right)=1-\sum_{x\in S}q\left(x\right)\leq1-\left(1-\Delta\right)\sum_{x\in S}p\left(x\right)=1-\left(1-\Delta\right)p\left(S\right)
\end{equation}
so that
\begin{equation}
\sum_{x\not\in S}\left|p\left(x\right)-q\left(x\right)\right|\leq2-2p\left(S\right)+\Delta p\left(S\right)
\end{equation}
and overall
\begin{equation}
\left\Vert q-p\right\Vert _{\mathrm{1}}\leq2\left[1-p\left(S\right)+\Delta\,p\left(S\right)\right]\leq2\left[\tau+\Delta\,\left(1-\tau\right)\right]\leq 2\left[\tau+\Delta \right].
\end{equation}
\end{proof}

The second lemma shows that the adapted ancestral sampling algorithm can sample from a distribution $q$ which approximates $p$ multiplicatively on individual outcomes $x\in\01^n$ such that $|p(x)-q(x)|<\Delta p(x)$. The condition for this is that the additive error on the estimates of the marginals, which the algorithm is given access to, is sufficiently small compared to the size of the marginals. 
\begin{lemma}\label{lem:multiplicative-approximation}
    Let $\gamma>0, \epsilon<\gamma/2$ and let $x \in \mathbb \{0,1\}^n$ be an outcome such that for all corresponding  marginals it holds that
\begin{equation}\label{eq:bound-marginals-restated}
p\left(x_{1},\dots,x_{k}\right)\geq\frac{\gamma}{2^{k}} \,.
\end{equation}
Assume that for all $k\in [n]$, we are given access to additive error approximations $\pi(x_{1:k})$, i.e., access to all the corresponding marginals, such that
\begin{equation}\label{eq:error_probabilities_additive-restated}
\left|\pi\left(x_{1:k}\right)-p\left(x_{1:k}\right)\right|\leq\frac{\epsilon}{2^{k}} \,.
\end{equation}
 Then, the adapted ancestral algorithm samples from a distribution $q$ such that
\begin{equation}
\left|p\left(x\right)-q\left(x\right)\right|\leq \left|\exp(\frac{4 \epsilon n}{\gamma}) -1\right| p\left(x\right) \, ,
\end{equation}
in time $t=O(n)$.
\end{lemma}
\begin{proof}

Consider the $k$-th step of the sampling procedure so
that $x_{1},\dots,x_{k-1}$ are already determined. 
There are two possible scenarios, either both marginals $a=p(x_1, \dots ,x_k=0), \,b=p(x_1, \dots ,x_k=1)$ are positive or at most one of them is negative. We will treat each scenario independently.

\textbf{Case 1:} $a>0,b>0$, such that $q\left(x_{k}=0|x_{1:k-1}\right)=\frac{a}{a+b}=\frac{q\left(x_{1:k-1},x_k=0\right)}{q\left(x_{1:k-1}\right)}$.
Using the closeness of the estimates $\pi(x_{1:k})$ from \Cref{eq:error_probabilities_additive-restated}, we have that 
\begin{equation}
\frac{p\left(x_{1:k-1},x_k=0\right)-\frac{\epsilon}{2^{k}}}{p\left(x_{1:k-1}\right)+\frac{\epsilon}{2^{k-1}}}\leq\frac{q\left(x_{1:k-1},x_{k}=0\right)}{q\left(x_{1:k-1}\right)}\leq\frac{p\left(x_{1:k-1},x_{k}=0\right)+\frac{\epsilon}{2^{k}}}{p\left(x_{1:k-1}\right)-\frac{\epsilon}{2^{k-1}}}
\end{equation}
and similarly for $\frac{q\left(x_{1:k-1},x_{k}=1\right)}{q\left(x_{1:k-1}\right)}$. We note that due to the assumption on the size of the marginals from \Cref{eq:bound-marginals-restated}, the additive errors in the numerator and denominator are of the same order of magnitude as the marginal. Thus, the additive errors can be turned into a multiplicative errors as follow,
\begin{equation}
\frac{\left(1-\frac{\epsilon}{\gamma}\right)}{\left(1+\frac{\epsilon}{\gamma}\right)}\frac{p\left(x_{1:k-1},q_{k}=0\right)}{p\left(x_{1:k-1}\right)}\leq\frac{q\left(x_{1:k-1},q_{k}=0\right)}{q\left(x_{1:k-1}\right)}\leq\frac{\left(1+\frac{\epsilon}{\gamma}\right)}{\left(1-\frac{\epsilon}{\gamma}\right)}\frac{p\left(x_{1:k-1},q_{k}=0\right)}{p\left(x_{1:k-1}\right)}
\end{equation}
Hence, using our assumption that $\frac{\epsilon}{\gamma}<1/2$, we obtain 
\begin{equation}
\left|q\left(x_{k}=0|x_{1:k-1}\right)-p\left(x_{k}=0|x_{1:k-1}\right)\right|\leq4\frac{\epsilon}{\gamma}p\left(x_{k}=0|x_{1:k-1}\right) 
=\tilde{\epsilon} p\left(x_{k}=0|x_{1:k-1}\right) \,.
\end{equation}
where we set $\tilde{\epsilon}=4 \epsilon / \gamma$.

In the second scenario, one of the two marginals $a,b$ is negative. By the properties of the ancestral sampling algorithm it is clear that if the outcome of interest has $x_k=0$ it has to be the case that $a>0$ and equivalently if $x_k=1$ it has to be the case that $b>0$. Without loss of generality we will only treat the first case here.  

\textbf{Case 2:} $a>0,b<0$, such that $q\left(x_{k}=0|x_{1:k-1}\right)=1$ and $q\left(x_{k}=0|x_{1:k-1}\right)=0$.
Note that
\begin{equation}
\left|p\left(x_{k}=0|x_{1:k-1}\right)-q\left(q_{k}=0|x_{1:k-1}\right)\right|=\left|p\left(x_{k}=1|x_{1:k-1}\right)-\underbrace{q\left(x_{k}=1|x_{1:k-1}\right)}_{=0}\right|=p\left(x_{k}=1|x_{1:k-1}\right)
\end{equation}
and
\begin{equation}
p\left(x_{k}=1|x_{1:k-1}\right)=\frac{p\left(x_{1:k-1},x_{k}=1\right)}{p\left(x_{1:k-1}\right)}\leq\frac{\frac{\epsilon}{2^{k}}}{\frac{\gamma}{2^{k-1}}}=\frac{1}{2}\frac{\epsilon}{\gamma}\leq\frac{\epsilon}{\gamma}
\end{equation}
where for the first inequality, we have again used the bound on the marginals from \Cref{eq:bound-marginals-restated} and the fact that $\pi(x_{1:k-1},x_k)\leq0$ implies that $p(x_{1:k-1},x_k)\leq\frac{\epsilon}{2^k}$.
Hence, 
\begin{equation}
p\left(x_{k}=0|x_{1:k-1}\right)\geq1-\frac{\epsilon}{\gamma}
\end{equation}
and again, since we assumed $\frac{\epsilon}{\gamma}<1/2$, we obtain
\begin{equation}
\left|p\left(x_{k}=0|x_{1:k-1}\right)-q\left(x_{k}=0|x_{1:k-1}\right)\right|\leq\frac{\frac{\epsilon}{\gamma}}{1-\frac{\epsilon}{\gamma}}p\left(x_{k}=0|x_{1:k-1}\right) \leq \tilde{\epsilon}  p\left(x_{k}=0|x_{1:k-1}\right) \,.
\end{equation}
We conclude that in both cases, we can show multiplicative closeness of the conditionals.
Now, if we have multiplicative closeness for all $k \in [n]$ conditionals 
\begin{equation}
    \left(1-\tilde{\epsilon}\right)
    p\left(x_{k}|x_{1},\dots,x_{k-1}\right)
    \leq q\left(x_{k}|x_{1},\dots,x_{k-1}\right)
    \leq\left(1+\tilde{\epsilon}\right)
    p\left(x_{k}|x_{1},\dots,x_{k-1}\right)
\end{equation}
then, it follows that the also the probabilities are close
\begin{align}
    \left|q\left(x\right)-p\left(x\right)\right|
    &=	\left|q\left(x_{n}|x_{1},\dots,x_{n-1}\right)\cdots q\left(x_{1}\right)-p\left(x_{n}|x_{1},\dots,x_{n-1}\right)\cdots p\left(x_{1}\right)\right| \\
    \nonumber
    &\leq |(1+\tilde{\epsilon})^n -1 | \,p(x) \\
    \nonumber
    &\leq |\exp(\tilde{\epsilon} \,n) -1| \, p(x) \,\\
    \nonumber
    &= |\exp(4 \epsilon n /\gamma) -1| \, p(x)\,.
    \nonumber
\end{align}

For the time complexity bound, we note that the algorithm needs to flip $n$ biased coins to output an $n$-bit string $x$.
\end{proof}

We end this section by providing the proof of \Cref{thm:sampling_conditions-marginals-restated}.

\begin{proof}[Proof of \Cref{thm:sampling_conditions-marginals-restated}]
The proof now follows immediately by combining \Cref{lem:multiplicative-error-on-heavy-subset} with \Cref{lem:multiplicative-approximation}. In particular, in \Cref{lem:multiplicative-error-on-heavy-subset} take the set $S$ to be $S_{\gamma}$ as defined in \Cref{def:set-s-gamma}. Then, by \Cref{lem:multiplicative-approximation}, for all $x\in S_{\gamma}$, the adapted ancestral sampling algorithm distribution $q$ approximates $p$ via
 \begin{equation}
       |p(x)-q(x)|\leq \left|\exp(\frac{4 \epsilon n}{\gamma}) -1\right| p\left(x\right) \,.
   \end{equation}
So, the TV distance bounded due to \Cref{lem:multiplicative-error-on-heavy-subset} is
\begin{equation}
    \left \Vert q-p \right \Vert_{\mathrm{TV}}\leq (1-p(S_\gamma)) + \left|\exp(\frac{4 \epsilon n}{\gamma}) -1\right| = \mathfrak{f}(\gamma) + \left|\exp(\frac{4 \epsilon n}{\gamma}) -1\right|\,.
\end{equation}    
\end{proof}

\subsection{Robustness to Pauli-noise}
\label{ssec:robustness-Pauli-noise}

The following is a restatement of \cref{thm:robustness_to_Pauli_noise} from the main text.
 \begin{theorem}
     Let $\sigma$ be a pure state and let $\Lambda$ be a Pauli-channel such that
     \begin{equation}
     \rho:=\Lambda(\sigma)= (1-\xi)\sigma+\sum_{y\in \01^{2n}\setminus\{ 0^{2n}\}} \xi_y P_y \sigma P_y,   
     \end{equation}
     where for all $y$, we have that $\xi_y\geq0$, and $\sum_y \xi_y = \xi$ and $0\leq\xi\leq1$. Then, there exists an algorithm for sampling, with probability at least $1-\delta$, from a distribution $q_{\rho}$ such that
     \begin{equation}
         \Vert p_{\rho}-q_{\rho} \Vert_{TV}\leq \Delta,
     \end{equation}
      for any $\Delta>4\xi$, using
     \begin{equation}
         N=O\left(\frac{n^3\log(1/\delta)2^{M_0(\sigma)}2^{4 E_0^{\pi}(\rho)}}{\Delta^4(1-2\xi)^4 } \right)
     \end{equation}
     Bell samples of the state $\rho^{\otimes 2}$.
 \end{theorem}
 \begin{proof}
    By the action of the channel $\Lambda$ we have that
\begin{align}
    \tr(  \rho P_x) &= (1-\xi) \tr( \sigma P_x) +\sum_y \xi_y  \tr(P_y \sigma P_y P_x) \, , \\
    &= \tr( \sigma P_x )\left[ (1-\xi) + \sum_y (-1)^{[x,y]} \xi_y \right] \, ,
\end{align}
where we used that $\tr(P_y \sigma P_y P_x) = \pm \tr( \sigma P_x) $ depending on whether $P_x, P_y$ commute or anti-commute. Their commutation relation is denoted via $[x,y] =0$ if they commute and $[x,y]=1$ if they anti-commute. It follows that for any $P_x$
    
     \begin{equation}
         |\tr( \rho P_x)|\geq (1-2\xi)|\tr( \sigma P_x) |.
     \end{equation}
    We can therefore obtain the following lower bound on the individual probabilities of the Pauli distribution $p_\rho$: For all $x\in \01^{2n}$
     \begin{equation}
         p_{\rho}(x)=\frac{\tr( \rho P_x )^2}{2^n \tr(\rho^2) }\geq (1-2\xi)^2\frac{\tr( P_x \sigma )^2}{2^n}\geq (1-2\xi)^2 p_{\sigma}(x) \, ,
     \end{equation}
     where we used that $\tr(\rho^2)\leq 1$.
     Further, we obtain for any marginal that
     \begin{equation}
         p_{\rho}(x_{1:k})=\sum_{x_{k+1:n}} p_{\rho}(x)\geq (1-2\xi)^2\sum_{x_{k+1:n}} p_{\sigma}(x)= (1-2\xi)^2 p_{\sigma}(x_{1:k}).
     \end{equation}
     It thus follows that 
     \begin{equation}
         \forall \ x\in S_{\gamma}(\sigma) \Rightarrow x\in S_{\gamma'}(\rho)
     \end{equation}
     with $\gamma'\leq (1-2\xi)^2\gamma$.
     Hence, from \cref{thm:sampling_conditions-marginals-restated} we see that 
     \begin{equation}
         \Vert q_{\rho}-p_{\rho} \Vert_{\mathrm{TV}} \leq
         \left[\exp(\frac{4 \epsilon n}{\gamma'}) -1\right]
         + \left[1 
         -\sum_{x\in S_{\gamma'}(\rho)} p_{\rho}(x) \right]\leq\exp\left( \frac{4\epsilon n}{\gamma'} \right) -(1-2\xi)^2\sum_{x\in S_{\gamma}(\sigma)} p_{\sigma}(x).   
\end{equation}
Now, using $1\geq\sum_{x\in S_{\gamma}(\sigma)} p_{\sigma}(x)\geq 1- 2^{M_0(\sigma)}\gamma R$ and choosing 
\begin{equation}
\epsilon \leq \frac{\Delta (1-2\xi)^2 \gamma}{4\times 4n } \ , \ \gamma= \frac{\Delta-4\xi}{2\times2^{M_0(\sigma)}R}     
\end{equation}
we get 
\begin{equation}
   \Vert q_{\rho}-p_{\rho} \Vert_{\mathrm{TV}} \leq \Delta 
\end{equation}
To obtain the sample complexity note that $\tr(\rho^2)\geq (1-\xi)^2$ and then make use of \cref{lem:marginals_mixed}.

\end{proof}
 
\section{Error analysis of the protocols}\label{app:proof-ip}
Here we show \Cref{thm:error-symmetric} and \Cref{th:asymm}. To this end, it will be useful to consider the following two basic lemmas that we just recall without proof.

\begin{lemma}\label{lem:lip}
    Given $0<r<1$, consider the compact set $C_r\subseteq[-1,1]^2$ defined as
    \begin{align}
    C_r=\big\{(x,y)\in[-1,1]^2:x^2+y^2\ge r^2\big\}\,.
    \end{align}
    The function $G:C_r\to{\mathbb R}$ defined as $G(x,y)=(1/2)(x+y)^2/(x^2+y^2)$ is Lipschitz continuous with Lipschitz constant $L(r)$ at most $1/r$.
\end{lemma}

\begin{lemma}\label{lem:lip2}
    Given $0<\lambda<1$, consider the compact set $C_\lambda\subseteq[-1,1]^2$ defined as
    \begin{align}
    C_\lambda=\big\{(x,y)\in[-1,1]^2:x^2\ge\lambda^2\big\}\,.
    \end{align}
    The function $H:C_\lambda\to{\mathbb R}$ defined as $H(x,y)=y/x$ is Lipschitz continuous with Lipschitz constant $L(\lambda)$ at most $\sqrt 2/\lambda$.
\end{lemma}

\subsection{The proof of \Cref{thm:error-symmetric}}\label{app:proof-sym}

Here we proof the following (equivalent) version of~\Cref{thm:error-symmetric},

\begin{theorem}[\Cref{thm:error-symmetric} restated]
    Let $\epsilon_1>0$ and $\epsilon_2>0$. Let $\tilde p_{\rm mix}$ be a distribution such that $||p_{\rm mix}-\tilde p_{\rm mix}||_{\rm TV}<\Delta$. We write $f(N_1,N_2)$ for our estimate of $\tr(\rho\sigma)$ with $N_1$ samples $\{x_1,\ldots,x_{N_1}\}$ from $\tilde p_{\rm mix}$ and $N_2$ measurements per sample. One has
    \begin{align}
    {\Pr}\left(\big|f(N_1,N_2)-\tr(\rho\sigma)\big|>2\epsilon_1+2\sqrt{\epsilon_2}+\frac{F_\rho(\epsilon_2)}2+\frac{F_\sigma(\epsilon_2)}2+2\Delta\right)<4\exp\left(-2\epsilon_1^2N_1\right)+4N_1\exp\left(-\epsilon_2^2N_2/2\right)\,,
    \end{align}
    where $F_\rho$ and $F_\sigma$ denote the CDFs previously introduced in Definition~\ref{def:cdf}.
\end{theorem}

\begin{proof}
The main tools needed are: (i) triangular inequality, (ii) Lipschitz continuity of the two-variable function $G(u,v)$ and (iii) Hoeffdings inequality together with the union bound. So that, first of all, we use triangular inequality to expand the error $|f(N_1,N_2)-f|$ as follows. Recall that $f=(1+\tr(\rho\sigma))/2$ and $f(N_1,N_2)$ is defined in Eq.~\eqref{eq:symmetric-protocol-expansion}:
\begin{multline*}
|f-f(N_1,N_2)|\le\Big|f-\sum_x\tilde p_{\rm mix}(x)G(\alpha_\rho(x),\alpha_\sigma(x))\Big|\\
+\Big|\sum_x\tilde p_{\rm mix}(x)G(\alpha_\rho(x),\alpha_\sigma(x))-\frac{1}{N_1}\sum_{i=1}^{N_1}G(\alpha_\rho(x_i),\alpha_\sigma(x_i))\Big|\\
+\frac{1}{N_1}\sum_{i=1}^{N_1}\left|\,G(\alpha_\rho(x_i),\alpha_\sigma(x_i))-G\left(\frac1{N_2}\sum_{j=1}^{N_2}m_\rho(i,j),\frac1{N_2}\sum_{j=1}^{N_2}m_\sigma(i,j)\right)\,\right|\,.
\end{multline*}
Now, observe that the function $G(u,v)$ is Lipschitz continuous when restricted to a domain that does not include the origin. This motivates the definition of the following set
\begin{equation}\label{eq:setdef}
I=\Big\{i:\{\alpha_\rho(x_i)^2>\epsilon_2\}\cup\{\alpha_\sigma(x_i)^2>\epsilon_2\}\Big\}\,.
\end{equation}
Now we split the sum in the third term of the RHS above into two parts: the contribution from $I$ and the contribution from its complement. One has
\begin{multline*}
\frac{1}{N_1}\sum_{i=1}^{N_1}\left|\,G(\alpha_\rho(x_i),\alpha_\sigma(x_i))-G\left(\frac1{N_2}\sum_{j=1}^{N_2}m_\rho(i,j),\frac1{N_2}\sum_{j=1}^{N_2}m_\sigma(i,j)\right)\,\right|\\
=\frac{1}{N_1}\sum_{i\in I}\left|\,G(\alpha_\rho(x_i),\alpha_\sigma(x_i))-G\left(\frac1{N_2}\sum_{j=1}^{N_2}m_\rho(i,j),\frac1{N_2}\sum_{j=1}^{N_2}m_\sigma(i,j)\right)\,\right|\\
+\frac{1}{N_1}\sum_{i\not\in I}\left|\,G(\alpha_\rho(x_i),\alpha_\sigma(x_i))-G\left(\frac1{N_2}\sum_{j=1}^{N_2}m_\rho(i,j),\frac1{N_2}\sum_{j=1}^{N_2}m_\sigma(i,j)\right)\,\right|\,.
\end{multline*}
When we consider the contribution to the sum of the terms $i\in I$, one can use the fact that $G(u,v)$ is Lipschitz continuous with Lipschitz constant $\sqrt{2/\epsilon_2}$ (see Lemma~\ref{lem:lip}). When we consider the contribution coming from the terms $i\not\in I$, we simply use the bound $|G(u,v)-G(u',v')|\le1$ which is always true. So one finally writes
\begin{multline}\label{eq:terms}
|f-f(N_1,N_2)|\le\Big|\tr(\rho\sigma)-\sum_x\tilde p_{\rm mix}(x)G(\alpha_\rho(x),\alpha_\sigma(x))\Big|\\
+\Big|\sum_x\tilde p_{\rm mix}(x)G(\alpha_\rho(x),\alpha_\sigma(x))-\frac{1}{N_1}\sum_{i=1}^{N_1}G(\alpha_\rho(x_i),\alpha_\sigma(x_i))\Big|\\
+\frac{1}{N_1}\sum_{i\in I}\left|\,G(\alpha_\rho(x_i),\alpha_\sigma(x_i))-G\left(\frac1{N_2}\sum_{j=1}^{N_2}m_\rho(i,j),\frac1{N_2}\sum_{j=1}^{N_2}m_\sigma(i,j)\right)\,\right|+\frac{1}{N_1}\sum_{i\not\in I}1\,.
\end{multline}

Now we analyze the four terms in the RHS of Eq.~\eqref{eq:terms} above separately in the following. First of all, the first term is easily bounded since $||p-\tilde p_{\rm mix}||_{\rm TV}\le\Delta$ and $|G(u,v)|\le1$. One has
\begin{align}
\Big|f-\sum_x\tilde p_{\rm mix}(x)G(\alpha_\rho(x),\alpha_\sigma(x))\Big|=\Big|\sum_xp_{\rm mix}(x)G(\alpha_\rho(x),\alpha_\sigma(x))-\sum_x\tilde p_{\rm mix}(x)G(\alpha_\rho(x),\alpha_\sigma(x))\Big|\le\Delta\,.
\end{align}

Now, one can use Hoeffdings inequality to bound the second and fourth terms in the RHS of Eq.~\eqref{eq:terms}. Indeed,
\begin{align}
\begin{aligned}
    {\Pr}\left(\Big|\sum_x\tilde p_{\rm mix}(x)G(\alpha_\rho(x),\alpha_\sigma(x))-\frac{1}{N_1}\sum_{i=1}^{N_1}G(\alpha_\rho(x_i),\alpha_\sigma(x_i))\Big|>\epsilon_1\right)&<2\exp(-2N_1\epsilon_1^2)\\
    {\Pr}\left(\frac{1}{N_1}\sum_{i\not\in I}1>\epsilon_1+\Delta+\frac{F_\rho(\epsilon_2)}2+\frac{F_\sigma(\epsilon_2)}2\right)&<2\exp(-2N_1\epsilon_1^2)\,,
\end{aligned}
\end{align}
where, in the second line, we have used that ${\mathbb E}[N_1^{-1}\sum_{i\not\in I}1]\le \Delta+F_\rho(\epsilon_2)/2+F_\sigma(\epsilon_2)/2$. To see this, note that
\begin{align}
\left|{\mathbb E}[N_1^{-1}\sum_{i\not\in I}1]-\sum_{x\not\in S}p_{\rm mix}(x)\right|\le\Delta
\end{align}
where $S=\{x:\{\alpha_\rho(x)^2>\epsilon_2\}\cup\{\alpha_\sigma(x)^2>\epsilon_2\}\}$ and
\begin{align}
\sum_{x\not\in S}p_{\rm mix}(x)=\frac12\sum_{x\not\in S}p_\rho(x)+\frac12\sum_{x\not\in S}p_\sigma(x)\le\sum_{\{x:\alpha_\rho(x)^2\le\epsilon_2\}}\frac{p_\rho(x)}2+\sum_{\{x:\alpha_\sigma(x)^2\le\epsilon_2\}}\frac{p_\sigma(x)}2=\frac12 F_\rho(\epsilon_2)+\frac12 F_\sigma(\epsilon_2)\,.
\end{align}
(Recall Definition~\ref{def:cdf}.) Finally, we consider the third term of the RHS of Eq.~\eqref{eq:terms}. Note first of all that, using Hoeffdings inequality again
\begin{align}
\begin{aligned}
    {\Pr}\left(\Big|\alpha_\rho(x_i)-\frac{1}{N_2}\sum_{j=1}^{N_2}m_\rho(i,j))\Big|>\epsilon_2\right)&<2\exp(-N_2\epsilon_2^2/2)\\
    {\Pr}\left(\Big|\alpha_\sigma(x_i)-\frac{1}{N_2}\sum_{j=1}^{N_2}m_\sigma(i,j))\Big|>\epsilon_2\right)&<2\exp(-N_2\epsilon_2^2/2)
\end{aligned}
\end{align}
for each index $i\in[N_1]$. So that, by the union bound
\begin{align}
 {\Pr}\left(\bigcup_{i=1}^{N_1}\Big\{\big|\alpha_\rho(x_i)-\frac{1}{N_2}\sum_{j=1}^{N_2}m_\rho(i,j))\big|>\epsilon_2\Big\}\cup\Big\{\big|\alpha_\sigma(x_i)-\frac{1}{N_2}\sum_{j=1}^{N_2}m_\sigma(i,j))\big|>\epsilon_2\Big\}\right)<4N_1\exp(-N_2\epsilon_2^2/2)\,.
\end{align}
Now, note that for each $i\in I$, we can assume that $|\alpha_\rho(x_i)|>\sqrt{\epsilon_2}$ without loss of generality (see the definition of the set $I$ in Eq.~\eqref{eq:setdef}). So that, with probability at least $1-4N_1\exp(-N_2\epsilon_2^2/2)$, we have $|N_2^{-1}\sum_{j=1}^{N_2} m_\rho(i,j)|>\sqrt{\epsilon_2}(1-\sqrt{\epsilon_2})$. In general, one has \footnote{We assume now that $\epsilon_2<1/16$ so that $\sqrt{\epsilon_2}<1/4$ and $1-\sqrt{\epsilon_2}>3/4>1/\sqrt2$}
\begin{align}
\max\left\{\frac1{N_2}\sum_{j=1}^{N_2} m_\rho(i,j),\frac1{N_2}\sum_{j=1}^{N_2} m_\sigma(i,j)\right\}>\sqrt{\frac{\epsilon_2}2}
\end{align}
for all $i\in I$ with probability at least $1-4N_1\exp(-N_2\epsilon_2^2/2)$. Thus
\begin{align}
\left(\frac1{N_2}\sum_{j=1}^{N_2} m_\rho(i,j)\right)^2+\left(\frac1{N_2}\sum_{j=1}^{N_2} m_\sigma(i,j)\right)^2>\frac{\epsilon_2}2
\end{align}
for all $i\in I$ with with probability at least $1-4N_1\exp(-N_2\epsilon_2^2/2)$. Now, according to Lemma~\ref{lem:lip}, the Lipschitz constant of $G(u,v)$ can be taken to be $\sqrt{2/\epsilon_2}$ when $u^2+v^2>\epsilon_2/2$. So that
\begin{multline}
\left|\,G(\alpha_\rho(x_i),\alpha_\sigma(x_i))-G\left(\frac1{N_2}\sum_{j=1}^{N_2}m_\rho(i,j),\frac1{N_2}\sum_{j=1}^{N_2}m_\sigma(i,j)\right)\,\right|\\
<\sqrt{\frac2{\epsilon_2}}\sqrt{\left(\alpha_\rho(x_i)-\frac1{N_2}\sum_{j=1}^{N_2}m_\rho(i,j)\right)^2+\left(\alpha_\sigma(x_i)-\frac1{N_2}\sum_{j=1}^{N_2}m_\sigma(i,j)\right)^2}<2\sqrt\epsilon_2
\end{multline}
for all $i\in I$ with with probability at least $1-4N_1\exp(-N_2\epsilon_2^2/2)$. Equivalently,
\begin{align}
{\Pr}\left(\frac{1}{N_1}\sum_{i\in I}\left|\,G(\alpha_\rho(x_i),\alpha_\sigma(x_i))-G\left(\frac1{N_2}\sum_{j=1}^{N_2}m_\rho(i,j),\frac1{N_2}\sum_{j=1}^{N_2}m_\sigma(i,j)\right)\,\right|>2\sqrt{\epsilon_2}\right)<4N_1\exp(-N_2\epsilon_2^2/2)\,,
\end{align}
and the proof is complete.
\end{proof}

\subsection{Proof of \Cref{th:asymm}}\label{app:proof-asym}
Note that \Cref{cor:assym} below corresponds to \Cref{th:asymm} from the main text. In this section, we will prove this theorem. It follows immediately from the following theorem:

\begin{theorem}\label{thm:new-asymm}
    Let $p_{\rho}(x)=\alpha_\rho(x)^2/2^n$ and let $\{x_1,\ldots,x_{N_{1}}\}$ be sampled from a distribution $q$ such that $||p_\rho - q||_{\rm TV}\le\Delta$, and let $\lambda >0$. Consider the estimator
    \begin{equation}
    f(N_1,N_\sigma,N_\rho)=\frac{1}{N_1}\sum_{i=1}^{N_1}\frac{\hat\alpha_\sigma(x_i,N_\sigma)}{s_\lambda(\hat\alpha_\rho(x_i,N_\rho))}
    \end{equation}
    where the function $s_\lambda(\cdot)$ is defined as
    \begin{equation}
    s_\lambda(z)=\begin{cases}
        z,&|z|>\lambda\\
        \text{sign}(z)\,\lambda,&|z|\le\lambda
    \end{cases}
    \end{equation}
    and, $\hat\alpha_\sigma(x,N_\sigma)$ and $\hat\alpha_\rho(x,N_\rho)$ are estimates of $\alpha_\sigma(x)=\langle P_x\rangle_\sigma$ and $\alpha_\rho(x)=\langle P_x\rangle_\rho$, respectively, estimated using $N_\sigma$ and $N_\rho$ measurements each. Then, for arbitrary $\epsilon_\sigma>0$, $\epsilon_\rho>0$, and $\gamma>0$ one has
    \begin{multline}
            {\Pr}\left(\Big|f(N_1,N_\sigma,N_\rho)-\tr(\rho\sigma)\Big|>\frac1{\lambda}\left(\epsilon_\sigma+\frac{\epsilon_\rho}{\lambda}\right)+\gamma+\frac{\Delta}{\lambda}+\frac{F_\rho(\lambda^2)}{\lambda}+2^{M_0(\rho)}\lambda\right)<\\
            2N_1\exp\left(-\epsilon_\sigma^2 N_\sigma/2\right)+2N\exp\left(-\epsilon_\rho^2 N_\rho/2\right)+(1+\Delta/\lambda^2)/(N_1\gamma^2).
    \end{multline} 
\end{theorem}

\begin{corollary}[\Cref{th:asymm} restated]\label{cor:assym}
With the same notation, one has
    \begin{equation}
    \Big|f(N_1,N_\sigma,N_\rho)-\tr(\rho\sigma)\Big|\le\frac1{\lambda}\left(\epsilon_\sigma+\frac{\epsilon_\rho}{\lambda}\right)+\gamma+\frac{\Delta}{\lambda}+\frac{F_\rho(\lambda^2)}{\lambda}+2^{M_0(\rho)}\lambda
    \end{equation}
    with probability at least $1-\delta$ provided that
    \begin{eqnarray}
        N_1&\geq&\frac{3(1+\Delta/\lambda^2)}{\gamma^2\delta},\\
        N_\sigma&\geq&\frac{2}{\epsilon_\sigma^2}\ln\left(\frac{6N_1}{\delta}\right),\\
        N_\rho&\geq&\frac{2}{\epsilon_\rho^2}\ln\left(\frac{6N_1}{\delta}\right).
    \end{eqnarray}
\end{corollary}

\begin{corollary}[\Cref{cor:main-result-IP-low-magic-asymmetric} restated]
    With the same notation, one has
    \begin{equation}
    \Big|f(N_1,N_\sigma,N_\rho)-\tr(\rho\sigma)\Big|\le \epsilon
    \end{equation}
    with probability at least $1-\delta$ provided that
    \begin{eqnarray}
        N_1&\geq&\frac{216}{\epsilon^2\delta},\\
        N_\sigma&\geq&\frac{2592\cdot16^{M_0(\rho)}}{\epsilon^4}\ln\left(\frac{6N_1}{\delta}\right),\\
        N_\rho&\geq&\frac{93312\cdot256^{M_0(\rho)}}{\epsilon^6}\ln\left(\frac{6N_1}{\delta}\right).
    \end{eqnarray}
\end{corollary}
\begin{proof}
    Note that $F_\rho(\lambda^2)\le2^{M_0(\rho)}\lambda^2$ (\cref{lem:bounding_probabilities_with_M0}) and take 
    \begin{align}
    \gamma &= \frac{\epsilon}{6},\\
    \lambda &= \frac{\epsilon}{6\cdot2^{M_0(\rho)}},\\
    \Delta &= \lambda^2 = \frac{\epsilon^2}{36\cdot4^{M_0(\rho)}},\\
    \epsilon_\sigma &= \frac{\lambda\epsilon}{6} = \frac{\epsilon^2}{36\cdot2^{M_0(\rho)}},\\
    \epsilon_\rho &= \frac{\lambda^2\epsilon}{6} = \frac{\epsilon^3}{216\cdot16^{M_0(\rho)}}.
    \end{align}
\end{proof}

\begin{proof}[Proof of Theorem~\ref{thm:new-asymm}]
    We expand the error in four terms as
    \begin{multline}
        \left|f(N_1,N_\sigma,N_\rho)-\tr(\rho\sigma)\right| \le \underbrace{\left|\frac{1}{N_1}\sum_{i=1}^{N_1}\frac{\hat\alpha_\sigma(x_i,N_\sigma)}{s_\lambda(\hat\alpha_\rho(x_i,N_\rho))}-\frac{1}{N_1}\sum_{i=1}^{N_1}\frac{\alpha_\sigma(x_i)}{s_\lambda(\alpha_\rho(x_i))}\right|}_\text{(i)}+\\
        \underbrace{\left|\frac{1}{N_1}\sum_{i=1}^{N_1}\frac{\alpha_\sigma(x_i)}{s_\lambda(\alpha_\rho(x_i))}-\sum_x q(x)\frac{\alpha_\sigma(x)}{s_\lambda(\alpha_\rho(x))}\right|}_\text{(ii)}+ \underbrace{\left|\sum_x q(x)\frac{\alpha_\sigma(x)}{s_\lambda(\alpha_\rho(x))}-\sum_x p(x)\frac{\alpha_\sigma(x)}{s_\lambda(\alpha_\rho(x))}\right|}_\text{(iii)}\\
        +\underbrace{\left|\sum_x p(x)\frac{\alpha_\sigma(x)}{s_\lambda(\alpha_\rho(x))}-\sum_x p(x)\frac{\alpha_\sigma(x)}{\alpha_\rho(x)}\right|}_\text{(iv)}
    \end{multline}
    and we analyze the right hand side of the previous inequality term by term.
    
    We start with the term (i) and first bound the probability that all the $2N$ estimates $\hat\alpha_\sigma(x_i,N_\sigma),\hat\alpha_\rho(x_i,N_\rho)$ are within additive error $\epsilon_\sigma$ and $\epsilon_\rho$, respectively, as a function of $N_\sigma$ and $N_\rho$. By Hoeffding's inequality, we have
    \begin{equation}
    \begin{aligned}
        {\Pr}\Big(|\hat\alpha_\sigma(x_i,N_\sigma)-\alpha_\sigma(x_i)|>\epsilon_\sigma\Big)&<2\exp\left(-N_\sigma\epsilon_\sigma^2/2\right)\\
        {\Pr}\Big(|\hat\alpha_\rho(x_i,N_\rho)-\alpha_\rho(x_i)|>\epsilon_\rho\Big)&<2\exp\left(-N_\rho\epsilon_\rho^2/2\right)
    \end{aligned}
    \end{equation}
    for each $i\in[N]$. Combined with the union bound, we get
    \begin{align}\label{eq:union-estimates}
        {\Pr}&\left(\bigcup_{i=1}^{N_1}\Big\{|\hat\alpha_\sigma(x_i,N_\sigma)-\alpha_\sigma(x_i)|>\epsilon_\sigma\Big\}\cup\Big\{|\hat\alpha_\rho(x_i,N_\rho)-\alpha_\rho(x_i)|>\epsilon_\rho\Big\}\right)\nonumber\\
        &\quad\quad\quad\quad\quad\quad\le\sum_{i=1}^{N_1}{\Pr}\Big(|\hat\alpha_\sigma(x_i,N_\sigma)-\alpha_\sigma(x_i)|>\epsilon_\sigma\Big)+{\Pr}\Big(|\hat\alpha_\rho(x_i,N_\rho)-\alpha_\rho(x_i)|>\epsilon_\rho\Big)\nonumber\\
        &\quad\quad\quad\quad\quad\quad<2N_1\exp\left(-N_\sigma\epsilon_\sigma^2/2\right)+2N_1\exp\left(-N_\rho\epsilon_\rho^2/2\right).
    \end{align}
    Now, note that, whenever $\hat\alpha_\sigma(x_i,N_\sigma),\hat\alpha_\rho(x_i,N_\rho)$ are within additive error $\epsilon_\sigma$ and $\epsilon_\rho$, we have:
    \begin{align}\label{eq:error-(i)}
        \left|\frac{\hat\alpha_\sigma(x_i,N_\sigma)}{s_\lambda(\hat\alpha_\rho(x_i,N_\rho))}-\frac{\alpha_\sigma(x_i)}{s_\lambda(\alpha_\rho(x_i))}\right|&\le\left|\frac{\epsilon_\sigma}{s_\lambda(\hat\alpha_\rho(x_i,N_\rho))}\right|
        +|\alpha_\sigma(x_i)|\left|\frac{s_\lambda(\alpha_\rho(x_i))-s_\lambda(\hat\alpha_\rho(x_i,N_\rho))}{s_\lambda(\hat\alpha_\rho(x_i,N_\rho))\,s_\lambda(\alpha_\rho(x_i))}\right|\nonumber\\
        &\le\frac{\epsilon_\sigma}{\lambda}+\frac{\epsilon_\rho}{\lambda^2},
    \end{align}
    where we have used the fact that $s_\lambda(\cdot)\ge\lambda$ and $|\alpha_\sigma(\cdot)|\le1$.
    Therefore, by combining \cref{eq:union-estimates} and \cref{eq:error-(i)}, we get
    \begin{align}
    {\Pr}&\left(\,\left|\frac{1}{N_1}\sum_{i=1}^{N_1}\frac{\hat\alpha_\sigma(x_i,N_\sigma)}{s_\lambda(\hat\alpha_\rho(x_i,N_\rho))}-\frac{1}{N_1}\sum_{i=1}^{N_1}\frac{\alpha_\sigma(x_i)}{s_\lambda(\alpha_\rho(x_i))}\right|>\frac1{\lambda}\left(\epsilon_\sigma+\frac{\epsilon_\rho}{\lambda}\right)\right)\nonumber\\
    &\quad\quad\quad\quad\quad\quad\leq {\Pr}\left(\bigcup_{i=1}^{N_1}\Big\{|\hat\alpha_\sigma(x_i,N_\sigma)-\alpha_\sigma(x_i)|>\epsilon_\sigma\Big\}\cup\Big\{|\hat\alpha_\rho(x_i,N_\rho)-\alpha_\rho(x_i)|>\epsilon_\rho\Big\}\right)\nonumber\\
    &\quad\quad\quad\quad\quad\quad\leq 2N_1\exp\left(-N_\sigma\epsilon_\sigma^2/2\right)+2N_1\exp\left(-N_\rho\epsilon_\rho^2/2\right),
    \end{align}
    
Now, we move to the term (ii). We use Chebyshev's inequality on the random variable $X_\lambda(x)=\alpha_\sigma(x)/s_\lambda(\alpha_\rho(x))$, distributed according to $q$. To use this inequality, we bound the variance of the random variable:
\begin{align}
    \text{Var}_q[X_\lambda(x)] &\leq \sum_x q(x)\left( \frac{\alpha_\sigma(x)}{s_\lambda(\alpha_\rho(x))} \right)^2\\
    &\leq \sum_x p(x) \left( \frac{\alpha_\sigma(x)}{s_\lambda(\alpha_\rho(x))} \right)^2 + \sum_x \abs{q(x)-p(x)} \left( \frac{\alpha_\sigma(x)}{s_\lambda(\alpha_\rho(x))} \right)^2 \nonumber\\
    &\leq \mathbb{E}_p[X_\lambda(x)^2] + \frac{||p-q||_{\rm TV}}{\lambda^2} \nonumber\\
    &\leq 1 + \frac{\Delta}{\lambda^2}\nonumber
\end{align}
where we have used the positivity of $X_\lambda(x)^2$ in the second line,  $s_\lambda(\cdot) \geq \lambda$ in the third line, and $\mathbb{E}_p[X_\lambda(x)^2] \leq \mathbb{E}_p[X(x)^2] = \sum_x \frac{\alpha_\sigma(x)^2}{2^n} = 1$ in the last line, for $X(x) = \frac{\alpha_\sigma(x)}{\alpha_\rho(x)}$, by noting that $X_\lambda(x)^2\leq X(x)^2, \forall x$.
Now, for any $\gamma$, we have
\begin{equation}
{\Pr}\left(\left|\frac{1}{N_1}\sum_{i=1}^{N_1}\frac{\alpha_\sigma(x_i)}{s_\lambda(\alpha_\rho(x_i))}-\sum_x q(x)\frac{\alpha_\sigma(x)}{s_\lambda(\alpha_\rho(x))}\right|>\gamma\right)<\frac{1+\Delta/\lambda^2}{N_1\gamma^2}.
\end{equation}

For the term (iii), we have
\begin{equation}
\left|\sum_x q(x)\frac{\alpha_\sigma(x)}{s_\lambda(\alpha_\rho(x))}-\sum_x p(x)\frac{\alpha_\sigma(x)}{s_\lambda(\alpha_\rho(x))}\right|\le\sum_x|p(x)-q(x)|\left|\frac{\alpha_\sigma(x)}{s_\lambda(\alpha_\rho(x))}\right|\le\frac{\Delta}{\lambda},
\end{equation}
where we have used again that $s_\lambda(\cdot)\ge\lambda$ and $|\alpha_\sigma(\cdot)|\le1$, and the assumption on the TV distance $||p-q||_{\rm TV}\le\Delta$.

Finally, the term (iv) is
\begin{align}
    \left|\sum_x p(x)\frac{\alpha_\sigma(x)}{s_\lambda(\alpha_\rho(x))}-\sum_x p(x)\frac{\alpha_\sigma(x)}{\alpha_\rho(x)}\right|&=\left|\sum_{\{x:|\alpha_\rho(x)|\le\lambda\}} p(x)\frac{\alpha_\sigma(x)}{s_\lambda(\alpha_\rho(x))}-\sum_{\{x:|\alpha_\rho(x)|\le\lambda\}}  p(x)\frac{\alpha_\sigma(x)}{\alpha_\rho(x)}\right|\\
    &\le\left|\sum_{\{x:|\alpha_\rho(x)|\le\lambda\}} p(x)\frac{\alpha_\sigma(x)}{s_\lambda(\alpha_\rho(x))}\right|+\left|\sum_{\{x:|\alpha_\rho(x)|\le\lambda\}}  p(x)\frac{\alpha_\sigma(x)}{\alpha_\rho(x)}\right|.\nonumber
\end{align}
For clarity, let us deal with each term in the expression above independently. First,
\begin{equation}
\left|\sum_{\{x:|\alpha_\rho(x)|\le\lambda\}} p(x)\frac{\alpha_\sigma(x)}{s_\lambda(\alpha_\rho(x))}\right|\le\frac1{\lambda}\sum_{\{x:|\alpha_\rho(x)|\le\lambda\}} p(x)=\frac1{\lambda}\sum_{\{x:\alpha_\rho(x)^2\le\lambda^2\}} p(x)=\frac{F_\rho(\lambda^2)}{\lambda}\,,
\end{equation}
where we have used again that $s_\lambda(\cdot)\ge\lambda$ and $|\alpha_\sigma(\cdot)|\le1$, and the definiton of the CDF. Second,
\begin{multline}
    \left|\sum_{\{x:|\alpha_\rho(x)|\le\lambda\}}  p(x)\frac{\alpha_\sigma(x)}{\alpha_\rho(x)}\right|=\left|\sum_{\{x:|\alpha_\rho(x)|\le\lambda\}}\frac{\alpha_\rho(x)\,\alpha_\sigma(x)}{2^n}\right|
    \le\frac{\lambda}{2^n}\sum_{\{x:\alpha_\rho(x)\neq0\}}1=\frac{2^{M_0(\rho)+n}\lambda}{2^n}=2^{M_0(\rho)}\lambda\,.
\end{multline}
from $|\alpha_\sigma(\cdot)|\le1$ once more and the definition of $M_0(\rho)$ involving the support of the characteristic distribution, i.e., $M_0(\rho)=\log|\{x:\alpha_\rho(x)\neq0\}|-n$.
We can now put together the bounds on the terms (i), (ii), (iii), (iv) as
    \begin{align}
            {\Pr}&\left(\Big|f(N_1,N_\sigma,N_\rho)-\tr(\rho\sigma)\Big|>\frac1{\lambda}\left(\epsilon_\sigma+\frac{\epsilon_\rho}{\lambda}\right)+\gamma+\frac{\Delta}{\lambda}+\frac{F_\rho(\lambda^2)}{\lambda}+2^{M_0(\rho)}\lambda\right)\\
            \nonumber&\quad\quad\quad\quad\quad\quad<{\Pr}\left(\left\{\text{(i)}>\frac1{\lambda}\left(\epsilon_\sigma+\frac{\epsilon_\rho}{\lambda}\right)\right\} \cup \left\{ \text{(ii)}>\gamma\right\}\right)\\
            \nonumber&\quad\quad\quad\quad\quad\quad< {\Pr}\left(\text{(i)}>\frac1{\lambda}\left(\epsilon_\sigma+\frac{\epsilon_\rho}{\lambda}\right)\right) + {\Pr}\Big(\text{(ii)}>\gamma\Big)\\
            \nonumber&\quad\quad\quad\quad\quad\quad<2N_1\exp\left(-\epsilon_\sigma^2 N_\sigma/2\right)+2N_1\exp\left(-\epsilon_\rho^2 N_\rho/2\right)+(1+\Delta/\lambda^2)/(N_1\gamma^2).\nonumber
    \end{align} 
where the second line follows from the complements of these probabilities, and the third line follows from the union bound. This completes the proof.
\end{proof}

\section{Pauli distributions with large $M_1$ where the algorithm is efficient}\label{app:m1}
For any $0<\tau<1$, consider the vector $\ket{\phi_0}=\sqrt{\tau}\ket{0\cdots 0}+\sqrt{1-\tau}\ket{+\cdots+}$ and the corresponding normalized pure state $\ket{\phi}=K^{-1/2}\ket{\phi_0}$ where the normalization constant $K=\langle{\phi_0}|{\phi_0}\rangle$ can be expressed as
\begin{equation}
K=\langle{\phi_0}|{\phi_0}\rangle=1+2\sqrt{\frac{\tau(1-\tau)}{2^n}}=1+2m
\end{equation}
with $m=\sqrt{\tau(1-\tau)/2^n}$. We have the following result
\begin{lemma}
    Consider the CDF $F_{\phi_\tau}$ (see \Cref{def:cdf}) of the $n$-qubits, normalized pure state $\ket{\phi_\tau}=K^{-1/2}\ket{\phi_0(\tau)}$, where $\ket{\phi_0(\tau)}=\sqrt{\tau}\ket{0\cdots 0}+\sqrt{1-\tau}\ket{+\cdots+}$ and $K=1+2m$ with $m=\sqrt{\tau(1-\tau)/2^n}$. One has,
    \begin{equation}
    F_\phi(\eta^2)\le2\tau(1-\tau)\,,\qquad\forall\eta\le\tau
   \end{equation}
\end{lemma}

\begin{proof}
There will be four important subsets $S_1$, $S_2$, $S_3$ and $S_4$ that identify Pauli strings of the form $X_1^{a_1}Z_1^{b_1}\cdots X_n^{a_n}Z_n^{b_n}$ that we label, as usual, by the vector $(\vec a, \vec b)\in\{0,1\}^{2n}$. More precisely, we have
\begin{eqnarray}
    S_1&=&\{(\vec a,\vec b)=(\vec 0,\vec 0)\},\\
    S_2&=&\{(\vec a,\vec 0):\vec a\neq\vec 0\},\\
    S_3&=&\{(\vec 0,\vec b):\vec b\neq\vec 0\},\\
    S_4&=&\{(\vec a,\vec b):\vec a\neq\vec 0\,,\vec b\neq\vec 0\,,\langle\vec a,\vec b\rangle\equiv0\mod 2\}
    .
\end{eqnarray}
Note that $|S_1|=1$ as it only contains the identity. On the other hand, $|S_2|=|S_3|=2^n-1$ since $S_2$ and $S_3$ contain, respectively nontrivial Pauli strings of the form $X_1^{a_1}\cdots X_n^{a_n}$ and $Z_1^{b_1}\cdots Z_n^{b_n}$. Regarding the cardinality of $S_4$, consider the set $R=\{(\vec a,\vec b):\langle\vec a,\vec b\rangle\equiv0\mod 2\}$. It is clear that $S_4\subseteq R$ and, indeed, one can write
\begin{equation}
    |S_4|=|R|-|S_1|-|S_2|-|S_3|=|R|-2\cdot (2^n-1)-1=|R|-2\cdot 2^n+1\,.
   \end{equation}
    Now, note that $R$ contains Pauli strings of the form $P_1\otimes\cdots\otimes P_n$, with $P_i\in\{1,X,Y,Z\}$ where the number of $Y$ factors is even. Therefore
    \begin{equation}
    |R|=\sum_{k=0}^n\binom{n}{k}\frac{1+(-1)^k}{2}3^{n-k}=\frac{4^n+2^n}{2}\,.
   \end{equation}
    and $|S_4|=|R|-2\cdot 2^n+1=2^{2n-1}-3\cdot 2^{n-1}+1$. One can explicitly check the following expressions from the definition of $\ket{\phi_0}$
\begin{equation}
\alpha_0(\vec{a},\vec{b}):=\bra{\phi_0}X_1^{a_1}Z_1^{b_1}\cdots X_n^{a_n}Z_n^{b_n}\ket{\phi_0}=\begin{cases}
    1+2m,&(\vec{a},\vec b)\in S_1,\\
    1-\tau+2m,&(\vec{a},\vec b)\in S_2,\\
    \tau+2m,&(\vec{a},\vec b)\in S_3,\\
    2m\,&(\vec{a},\vec b)\in S_4\,.
\end{cases}
\end{equation}
from where we can compute the quantity we are interested in, namely
\begin{equation}
\alpha(\vec{a},\vec{b}):=\bra{\phi}X_1^{a_1}Z_1^{b_1}\cdots X_n^{a_n}Z_n^{b_n}\ket{\phi}=\frac{\alpha_0(\vec{a},\vec{b})}{1+2m}\,.
\end{equation}
Recalling that $m=\sqrt{\tau(1-\tau)/2^n}$, we assume now that $\tau\le1/2$ and $2m<\tau$. Note that for the important case of $\tau=1/\sqrt{n}$ one has indeed $2m<\tau$ for all $n$. Now, noticing the inequalities
\begin{equation}
\frac{1-\tau+2m}{1+2m}\ge\frac{\tau+2m}{1+2m}>\tau>2m>\frac{2m}{1+2m}
\end{equation}
it follows that the set $S_4$ can be characterized as
\begin{equation}
S_4=\left\{(\vec{a},\vec{b}):0<|\alpha(\vec{a},\vec{b})|\le\frac{2m}{1+2m}\le2m\right\}=\Big\{(\vec{a},\vec{b}):0<|\alpha(\vec{a},\vec{b})|\le\tau\Big\}\,.
\end{equation}
In other words, $0<|\alpha(\vec{a},\vec{b})|\le\tau$ if and only if $(\vec{a},\vec{b})\in S_4$ if and only if $0<|\alpha(\vec{a},\vec{b})|\le2m$. We have
\begin{equation}
    F_{\phi_\tau}(\tau^2)=\sum_{\{x:\alpha(x)^2\le\tau^2\}}p_\phi(x)=\sum_{(\vec a,\vec b)\in S_4}\frac{\alpha(\vec a,\vec b)^2}{2^n}\le\frac{4m^2}{2^n}\sum_{(\vec a,\vec b)\in S_4}1=\frac{4m^2}{2^n}|S_4|
    =\frac{4\tau(1-\tau)}{4^n}|S_4|<2\tau(1-\tau)
\end{equation}
where we have used the fact that $|S_4|<4^n/2$. Since the CDF $F_{\phi_\tau}$ satisfies $F_{\phi_\tau}(x)\le F_{\phi_\tau}(y)$ whenever $x\le y$, this concludes the proof.
\end{proof}

The previous result shows that there are states $\phi$ with $F_\phi(\tau^2)\le2\tau(1-\tau)\le2\tau$. Consider \Cref{thm:error-symmetric} with $\epsilon_1=\tau$ and $\epsilon_2=\tau^2$ assuming that $F_\rho(\tau^2)\le2\tau$ and $F_\sigma(\tau^2)\le2\tau$. One has
\begin{align}
\big|f(N_1,N_2)-\tr(\rho\sigma)\big|\le6\tau+2\Delta
\end{align}
with probability at least $1-\delta$ provided that $N_1\ge(2\tau^2)^{-1}\log(8/\delta)$ and $N_2\ge(2/\tau^4)\log(8N_1/\delta)$.

\subsection{Magic and stabilizer entropies}
Clearly we have $M_0(\phi)=\log|R|-n$, where $R=\{(\vec a,\vec b):\langle\vec a,\vec b\rangle\equiv0\mod 2\}$ and we have shown that its cardinality is $|R|=2^{n-1}(1+2^n)$. It follows that $M_0(\phi)$, which is between $0$ and $n$, is very close to its maximum possible value for all values of the parameter $\tau$. Indeed
\begin{equation}
M_0(\phi)=\log(2^{n-1}(1+2^{n}))-n=\log(1+2^{n})-1>\log(2^{n})-1=n-1\,.
\end{equation}
Now we fix $\tau=1/\sqrt n$ (since we are assuming $\tau\le 1/2$, let us assume $n\ge4$) and consider
\begin{equation}
p_\phi(\vec a,\vec b)=\frac{|\alpha(\vec a,\vec b)|^2}{2^n}=\begin{cases}
    \frac{1}{2^n},&(\vec{a},\vec b)\in S_1,\\
    \frac{1}{2^n}\left(\frac{1-\tau+2m}{1+2m}\right)^2,&(\vec{a},\vec b)\in S_2,\\
    \frac{1}{2^n}\left(\frac{\tau+2m}{1+2m}\right)^2,&(\vec{a},\vec b)\in S_3,\\
    \frac{1}{2^n}\left(\frac{2m}{1+2m}\right)^2\,&(\vec{a},\vec b)\in S_4\,.
\end{cases}
\end{equation}
Now recall that $|S_1|=1$, $|S_2|=|S_3|=2^n-1$ and $|S_4|=4^n/2-3\cdot 2^n/2+1$. Since the parameter $m$ is just $\sqrt{\tau(1-\tau)/2^n}$, the previous expressions allows one to straightforwardly evaluate $M_1(\phi)$ for any value of $\tau$ and $n$. In particular, it is not hard to see that $M_1(\phi)>\sqrt{n}$ when $\tau=1/\sqrt{n}$.
    
\section{Higher dimensional systems}
\label{highdim}

In this section, we finally note that several of the above statements carry over to systems beyond quantum systems consisting of qubits, but of higher dimensional quantum systems, to exemplify the generality of the approach pursued. Specifically, for any local dimension $d$, one can pick a basis of
unitaries $\{U_j| j=1,\dots,d\}$ with $U_j\in U(d)$ that are orthonormal 
with respect to the Hilbert-Schmidt scalar product
as
\begin{equation}
\tr(U_j^\dagger U_k) = d\delta_{j,k}
\end{equation}
for $j,k=1,\dots, d$. Such basis actually exist for any integer 
dimension $d$ and generalize Pauli operators to arbitrary dimensions \cite{wernerAllTeleportationDense2001},
albeit not necessarily as Hermitian operators. Again denoting with 
\begin{equation}
\ket{\Omega}:=\frac{1}{\sqrt{d}}\sum_{v=0}^{d-1} \ket{v,v}
\end{equation}
a fiducial maximally entangled state vector in the $d\times d$-dimensional
bi-partite system, one finds a basis of maximally entangled state vectors
as
\begin{equation}
\{\ket{\psi_j}:=  \{(U_j\otimes I) \ket{\Omega}| j=0,\dots, d-1\}
\end{equation}
as a basis in $\mathbb{C}^{d}\otimes\mathbb{C}^{d}$. It is easy to see that the orthonormality of the unitary operator basis is inherited by the orthonormality of this basis of state vectors.

In order to estimate the purity via maximally entangled state sampling for pure states $\rho=\ket{\psi}\bra{\psi}$ with real state vectors $\ket{\psi}$,
prepare $\rho\otimes \rho$. Then measure in
the maximally entangled basis $\{\ket{\psi_j}\}$,
giving rise to outcomes $x\in\{0,\dots, d-1\}$. 
It takes a moment of thought to see that the statistics of the maximally entangled measurements is given
\begin{equation}
	p_\rho (x) = \frac{1}{d} |\bra{\psi} U_x\ket{ \psi^*}|^2
\end{equation}
in terms of a complex conjugation arising from transposition,
which for real state vectors is
\begin{equation}
	p_\rho (x) = \frac{1}{d} |\tr (\rho U_x)|^2.
\end{equation}
That is to say, by preparing two copies and performing suitable measurements in a maximally entangled basis, for any local dimension $d$, one can sample from the distribution $p_\rho$ in this case. This means that many of the above statements carry over to this situation. If a multi-qubit system is seen as a single higher dimensional one, the measurements involve entangled measurements over several qubits.

It is also worth noting that all eigenvectors of 
$U_j\otimes U_j$ for all $j=1,\dots, d$ with eigenvalue $+1$ are all maximally entangled. 
Since for all $j$
\begin{equation}
(U_j\otimes U_j) |\psi\rangle\langle\psi| (U_j\otimes U_j)^\dagger =   |\psi\rangle\langle\psi| ,
\end{equation}
we have that
\begin{equation}
U_j {\rm tr}_2(  |\psi\rangle\langle\psi| ) U_j^\dagger =   {\rm tr}_2(|\psi\rangle\langle\psi| ),
\end{equation}
by partial trace. Invoking the unitary design property of the collection of unitaries \cite{grossEvenlyDistributedUnitaries2007a}, we find
\begin{equation}
\frac{1}{d^2}
\sum_j U_j \tr_2(  |\psi\rangle\langle\psi| ) U_j^\dagger =   
\frac{I}{d} = \tr_2(|\psi\rangle\langle\psi| ),
\end{equation}
which means that $|\psi\rangle$ is maximally entangled.
    \end{widetext}
\end{document}